\documentclass[twocolumn,twocolappendix]{aastex631} 
\usepackage[utf8]{inputenc}
\usepackage{enumerate}
\usepackage{graphicx}
\usepackage{hyperref}
\usepackage{mathtools}
\usepackage{comment}
\usepackage{tabularx}
\usepackage{amsmath}
\usepackage{changepage}
\usepackage{ amssymb }
\usepackage{lipsum} 
\usepackage{float}
\usepackage{commath}
\usepackage{enumerate}
\usepackage{longtable}
\usepackage{stackengine}

\usepackage{cleveref}
\usepackage[caption=false]{subfig}
\usepackage{placeins}
\usepackage[ruled,lined]{algorithm2e}

\AfterEndEnvironment{figure}{\noindent\ignorespaces}

\usepackage{xcolor}

\begin{document}
\title{\texttt{StreamSculptor}: Hamiltonian Perturbation Theory for Stellar Streams \\ in Flexible Potentials with Differentiable Simulations
}

\author[0000-0001-8042-5794]{Jacob Nibauer}
\altaffiliation{NSF Graduate Research Fellow}
\affiliation{Department of Astrophysical Sciences, Princeton University, 4 Ivy Ln, Princeton, NJ 08544, USA}
\author[0000-0002-7846-9787]{Ana Bonaca}
\affiliation{The Observatories of the Carnegie Institution for Science, 813 Santa Barbara Street, Pasadena, CA 91101, USA}
\author{David N. Spergel}
\affiliation{Center for Computational Astrophysics, Flatiron Institute, 162 5th Avenue, New York, NY, 10010, USA}
\affiliation{Department of Astrophysical Sciences, Princeton University, 4 Ivy Ln, Princeton, NJ 08544, USA}
\author[0000-0003-0872-7098]{Adrian M. Price-Whelan}
\affiliation{Center for Computational Astrophysics, Flatiron Institute, 162 5th Avenue, New York, NY, 10010, USA}
\author[0000-0002-5612-3427]{Jenny E. Greene}
\affiliation{Department of Astrophysical Sciences, Princeton University, 4 Ivy Ln, Princeton, NJ 08544, USA}
\author[0000-0003-3954-3291]{Nathaniel Starkman}
\affiliation{Department of Physics and Kavli Institute for Astrophysics and Space Research, Massachusetts Institute of Technology, 77 Massachusetts Ave, Cambridge,
MA 02139, USA}
\author[0000-0001-6244-6727]{Kathryn V. Johnston}
\affiliation{Department of Astronomy, Columbia University, New York, NY 10027, USA}

\correspondingauthor{Jacob Nibauer}
\email{jnibauer@princeton.edu}

\begin{abstract}
\noindent
Stellar streams retain a memory of their gravitational interactions with small-scale perturbations. While perturbative models for streams have been formulated in action-angle coordinates,
a direct transformation to these coordinates is only available for static and typically axisymmetric models for the galaxy. The real Milky Way potential is in a state of disequilibrium, complicating the application of perturbative methods around an equilibrium system. Here, we utilize a combination of differentiable simulations and Hamiltonian perturbation theory to model the leading-order effect of dark matter subhalos on stream observables. To obtain a perturbative description of streams, we develop a direct and efficient forward mode differentiation of Hamilton's equations of motion. Our model operates in observable coordinates, allowing us to treat the effects of arbitrary subhalo potentials on streams perturbatively, while simultaneously capturing non-linear effects due to other substructures like the infalling LMC or the rotating bar. The model predicts the velocity dispersion of streams as a function of subhalo statistics, allowing us to constrain the low-mass range of subhalos down to $\sim 10^5~M_\odot$. We forecast the velocity dispersion of the GD-1 stream, and find that observations are in agreement with a CDM subhalo population, with a slight preference for more dense subhalos. The method provides a new approach to characterize streams in the presence of substructure, with significantly more modeling flexibility compared to previous works.
\vspace{1cm}
\end{abstract}

\section{Introduction}
The cold-dark-matter (CDM) paradigm predicts that galaxies should contain a significant amount of substructure within their halos, specifically in the form of dark-matter subhalos (e.g., \citealt{1999ApJ...522...82K, 1999ApJ...524L..19M, 2008MNRAS.391.1685S}). While the more massive subhalos ($\gtrsim 10^8 M_\odot$) retain a stellar component, less massive subhalos are also predicted and would not retain any stars. Measuring the statistics of a population of dark matter subhalos---including their mass, central concentration, and radial number counts---would provide a stringent test of CDM and probe the particle nature of dark matter, since distinct dark matter candidates differ in their predictions for these distributions at low subhalo masses ($\lesssim 10^6~M_\odot$; \citealt{2000PhRvL..84.3760S,2000PhRvL..85.1158H,2001ApJ...556...93B}).

While subhalos devoid of any stars cannot be directly observed, if they are gravitationally bound as predicted in CDM then it is expected that they will impart gravitational signatures on their environment. For instance, strong gravitational lensing provides one pathway to measuring completely dark subhalos down to masses of roughly $10^7~M_\odot$ using flux anomalies in an otherwise smooth lensed image (e.g., \citealt{1998MNRAS.295..587M}). This method is applicable to external galaxies. Another approach for detecting dark  subhalos is to consider the dynamical effect of the subhalos on their surrounding environment. For low mass subhalos without any stars, some fraction of bound subhalos are expected to have encounters with kinematically cold stellar systems. The details of this approach depend on the tracer population. Perhaps the most promising dynamical tracer of subhalo perturbations are stellar streams (e.g., \citealt{2002ApJ...570..656J, 2002MNRAS.332..915I, 2009ApJ...705L.223C}).  

Stellar streams, or tidal tails, are the result of a satellite galaxy or globular cluster (GC) disrupting in the tidal field of a more massive host. As stars are lost around the lagrange points of a cluster, they extend along a series of similar orbits producing long filamentary tails that connect through the dissolving progenitor. Crucially, stellar streams that form in a smooth gravitational potential have predictable surface densities and will generally appear continuous in their density and kinematic distribution. The effect of small-scale perturbations (e.g., due to subhalos) is to distort the stream, producing gaps, bifurcations, broadening, and kinematic heating of its tidal tails (e.g., \citealt{2002ApJ...570..656J,2002MNRAS.332..915I,2009ApJ...705L.223C,2011ApJ...731...58Y, 2012ApJ...748...20C, 2015MNRAS.450.1136E,2016MNRAS.457.3817S,2017MNRAS.466..628B,2019ApJ...880...38B,2024arXiv240518522C}).

There are already several kinematically cold streams whose morphology and kinematic distribution seem to suggest that they encountered a subhalo flyby. The most well-studied of these is the GD-1 stream, which has a gap and bifurcation that can be reproduced in models with a $\sim 10^6~M_\odot$ subhalo with a central density that is higher than expected under CDM \citep{2019ApJ...880...38B}, and consistent with predictions from self interacting dark matter \citep{2024arXiv240919493Z}. Recently, \citet{2024arXiv240402953H} finds that the ``forked" morphgology of the ATLAS-Aliqa Uma stream can be explained by a recent encounter with a $\sim 10^7~M_\odot$ subhalo. The tidal tails of the GC Palomar 5 (Pal-5) are rich with density variations, which are similarly compatible with some contribution from dark matter subhalos, though also with perturbations from the galactic bar \citep{2017MNRAS.470...60E, 2017NatAs...1..633P, 2020ApJ...889...70B}. In fact, to date, almost all streams that have been observed at an appreciable depth reveal density variations that would be unexpected in a smooth, nearly spherical halo (e.g., \citealt{2019MNRAS.485.4726K,2022AJ....163...18F, 2023A&A...669A.102W, 2024arXiv240519410B}).

There are several methods for modeling streams in the presence of substructure. While $N-$body simulations provide the highest degree of accuracy in modeling the disruption of a satellite, these simulations are prohibitively expensive to run for different combinations of subhalo impacts, masses, and orbits. Alternatively, semi-analytic models for stellar streams have been successful in reproducing the features of tidal tails from $N-$body simulations at a fraction of the computational cost \citep{2014ApJ...795...95B, 2015MNRAS.452..301F,2015MNRAS.450..575A,2023MNRAS.525.3662A, 2024arXiv240801496C}. Virtually all semi-analytic models for streams operate under the same principle: particles are released near the progenitor, extending along a series of similar, albeit slightly different orbits. The detailed mechanics of how this approach is carried out, however, varies depending on the method. 

Angle-action $(\theta, J)$ based models for stream generation exploit the simple structure of streams in this space (e.g., \citealt{1999MNRAS.307..877T, 2013MNRAS.433.1813S, 2014ApJ...795...95B}). In $(\theta, J)$ coordinates, cold stellar streams are approximately 1D linear features, extending along a single eigenvector in (e.g.) frequency-angle space. By generating linear features in this space and transforming to the observable space of positions and velocities, streams can be forward modeled provided that this transformation can be estimated accurately. This method has the advantage of computational speed and is appealing theoretically, since the dynamics of streams are straightforward in $(\theta, J)$ coordinates.

Another semi-analytic approach for stream generation that does not rely on action-angle coordinates is a class of particle-spray based algorithms \citep{2015MNRAS.452..301F, 2023MNRAS.525.3662A,2024arXiv240801496C}. This method operates under the same principle as angle-action models, but is formulated entirely in observable coordinates (positions and velocities: labeled generically as $\boldsymbol{q}$ and $\boldsymbol{p}$ throughout this work). Under this approach, particles are released near the lagrange points of the progenitor and integrated forward as test-particle orbits. This method is also fast, and enjoys the benefit of working with any differentiable potential, since transformation to the unobserved space of actions and angles is never required.

Both methods for generating streams have been applied in modeling stream--subhalo interactions. For example, \citet{2016MNRAS.457.3817S} used action-angle coordinates to explain the dynamics of streams undergoing collisions with subhalos, and \citet{2017MNRAS.466..628B} extended this formalism to model streams in the regime of many subhalo impacts in frequency-angle space. These works assume that subhalo impacts are impulsive (i.e., instantaneous), and use perturbative methods in frequency-angle space, $\left(\Omega(J), \theta\right)$, to model the evolution of a stream post-impact. Because this approach relies on action-angle modeling, the method of \citet{2017MNRAS.466..628B} has been applied in static axisymmetric models for the galactic potential, since this is the class of potentials for which the transformation to observable coordinates is most straightforward and accurate. Particle-spray models for streams and subhalos have also been explored (e.g., \citealt{2019ApJ...880...38B,2024arXiv240402953H}). While this approach can relax assumptions about the galactic potential and the impulsive nature of the subhalo impacts, particle-spray techniques are still too computationally expensive when attempting to forward model the $\mathcal{O}(100)$ unknown impacts that are expected for a GD-1 like stream down to a subhalo mass of $10^5~M_\odot$ \citep{2011ApJ...731...58Y,2016MNRAS.463..102E}.

In this work we seek to bridge the gap between pertubrative stream models in action-angle coordinates, and fully non-linear simulations based on particle-spray techniques. We focus on the ability to model streams in realistic or user-defined galactic potentials (i.e., time-evolving, highly aspherical, etc.) with both baryonic and non-baryonic sources of perturbation. This is a crucial step for not only capturing the orbital dynamics of a stream, but also for including the known sources of baryonic perturbation that may act non-linearly on MW streams. These include (e.g.) the infalling LMC \citep{2021ApJ...923..149S,2023MNRAS.518..774L,2024arXiv241002574B}, the halo's response \citep{2019ApJ...884...51G,2021ApJ...919..109G, 2021Natur.592..534C}, the galactic bar \citep{2016MNRAS.460..497H,2017NatAs...1..633P,2020ApJ...889...70B}, disk evolution \citep{2023MNRAS.518.2870D,2024ApJ...969...55N},  etc., all of which are capable of shaping the morphology and kinematics of MW streams. Failure to account for time-dependence in the potential can lead to bias when inferring properties of the host halo using streams (e.g., \citealt{2021MNRAS.502.4170R,2022ApJ...939....2A, 2024MNRAS.532.2657B, 2024ApJ...969...55N}).

To support arbitrary models for the galactic potential, we apply methods from Hamiltonian perturbation theory to the scenario where the ``unperturbed" system is not necessarily one that is in equilibrium, but actively time-evolving or analytically intractable to solve. The astrophysical literature on perturbation theory is almost entirely concerned with another class of problems, where the unperturbed system is static and analytically tractable to solve. In our case, the unperturbed system consists of the entire galaxy model without low mass dark matter subhalos, and we apply perturbation theory around this complicated system to account for the plethora of low mass subhalos that are expected in a CDM universe. To accomplish this, we develop analytic and numerical methods to perturb the solution to a dynamical system (i.e., the galaxy) at linear order in the perturbation strength parameters.

The paper is organized as follows. In \S\ref{sec: method} we introduce the method. In \S\ref{sec: grav_pots_and_subhalos} we specify gravitational potentials for the galaxy and perturbations, along with a prescription for sampling subhalo impacts. We demonstrate the method in \S\ref{sec: demonstration}, and consider applications to MW streams in \S\ref{sec: applications}. We discuss the method in \S\ref{sec: discussion}, and conclude in \S\ref{sec: conclusion}.

\section{Method}\label{sec: method}

The basic principle of our method is to model a stream as an ensemble of orbits with physically motivated initial conditions (i.e., generated from a particle-spray type model or a $N-$body model). Perturbation theory is then applied to the orbit ensemble, to model the collective response of a stream to external perturbations. Perturbation theory is applied in observable coordinates, so that transformation to angles and actions is never required. The reader can safely skip this section if they wish to see demonstration, validation, and applications of the model.

In \S\ref{sec: perturbative_orbit_model} we introduce our perturbative orbit model, in \S\ref{sec: intuition} we provide intuition for our perturbative treatment, in \S\ref{sec: boundary_conditions} we specify boundary conditions and the stream model, in \S\ref{sec: self_grav_method} we discuss how self-gravity of the progenitor is implemented in our model, in \S\ref{sec: varying_structural_params} we describe how perturbations are applied to the subhalo structural parameters (i.e., scale-radius), in \S\ref{sec: convergence_criterion} we discuss a convergence criterion for our method, and in \S\ref{sec: code_implementation} we specify our code implementation.   

\subsection{Perturbative Orbit Model}\label{sec: perturbative_orbit_model}
Our approach is to model the galaxy in terms of a smooth, possibly time-evolving component plus a series of local time-dependent fluctuations in the potential. In terms of a Hamiltonian, we consider a system of the form
\begin{multline}\label{eq: Hamiltonian}
    H\left(\boldsymbol{q},\boldsymbol{p},t; \boldsymbol{\theta}\right) = H_{\rm base}\left(\boldsymbol{q},\boldsymbol{p},t ;\boldsymbol{\theta}_{\rm base}\right) \\ + \sum\limits_{\alpha=1}^N \epsilon_\alpha \Phi_\alpha\left(\boldsymbol{q},t; \boldsymbol{\theta}_\alpha\right),
\end{multline}
where $H_{\rm base}$ encodes the smooth, global mass distribution of the galaxy, while the summation captures local fluctuations in the potential. The $\boldsymbol{\theta}$ vectors represent parameter sets for $H_{\rm base}$ and the series of perturbing potentials. We also introduce the notation $\boldsymbol{\epsilon}=\left(\epsilon_1, \epsilon_2, ..., \epsilon_N\right)$, representing a vector of perturbation parameters. Importantly, $H_{\rm base}$ does not need to represent an analytically tractable dynamical system in our analysis. Motivated by global disequilibrium in the Milky Way, we allow for the possibility of non-trivial spatial and time dependence in $H_{\rm base}$.

The base Hamiltonian is the sum of the specific kinetic energy and the base potential,
\begin{equation}
    H_{\rm base}\left( \boldsymbol{q}, \boldsymbol{p}, t;\boldsymbol{\theta}_{\rm base}\right) = \frac{\boldsymbol{p}^2}{2} + \Phi_{\rm base}\left(\boldsymbol{q},t;\boldsymbol{\theta}_{\rm base}\right).
\end{equation}

Without approximation, the equations of motion (EOM) for Eq.~\ref{eq: Hamiltonian} are given by Hamilton's equations
\begin{equation}\label{eq: EOM}
\begin{split}
    \dot{\boldsymbol{q}} &= \frac{\partial H_{\rm base}}{\partial \boldsymbol{p}} = \boldsymbol{p} \\
    \dot{\boldsymbol{p}} 
    &= -\left(\frac{\partial \Phi_{\rm base}}{\partial \boldsymbol{q}} +  \sum\limits_{\alpha=1}^N \epsilon_\alpha \frac{\partial \Phi_\alpha}{\partial \boldsymbol{q}}\right).
\end{split}
\end{equation}
We now restrict ourselves to scenarios where the summation terms are small compared to the base Hamiltonian, $H_{\rm base}$, in the sense that the gradients of $\Phi_\alpha$ are comparable in magnitude to the gradients of $\Phi_{\rm base}$, and $\epsilon_\alpha<<1$ for all $\alpha$.\footnote{Assuming that the perturbing potentials reach their peak gradients with respect to the unperturbed trajectory at roughly non-overlapping times---as will be the case for subhalo potentials---requiring that each $\epsilon_\alpha$ is individually small is a sufficient condition for a perturbative treatment.} This allows us to solve the EOM perturbatively, as a power series in $\epsilon_\alpha$. In canonical perturbation theory, the solution to Eq.~\ref{eq: EOM} takes the form
\begin{equation}\label{eq: expansion_series}
\begin{split}
    \boldsymbol{q}\left(t, \boldsymbol{\epsilon}\right) &= \boldsymbol{q}_0\left(t\right) + \sum\limits_{\alpha=1}^N\sum\limits_{\beta=1}^{\infty} \epsilon_\alpha^\beta \boldsymbol{q}_{\alpha\beta}\left(t\right) + \mathcal{O}\left( \epsilon_\alpha \epsilon_{\alpha^\prime} I_q(\alpha, \alpha^\prime) \right)\\
    \boldsymbol{p}\left(t, \boldsymbol{\epsilon} \right) &= \boldsymbol{p}_0\left(t\right) + \sum\limits_{\alpha=1}^N\sum\limits_{\beta=1}^{\infty} \epsilon_\alpha^\beta \boldsymbol{p}_{\alpha\beta}\left(t\right) + \mathcal{O}\left( \epsilon_\alpha \epsilon_{\alpha^\prime} I_p(\alpha, \alpha^\prime) \right)
\end{split}
\end{equation}
where $\boldsymbol{q}_0, \boldsymbol{p}_0$ represent solutions to the base EOM with $\boldsymbol{\epsilon} = 0$, while $\boldsymbol{q}_{\alpha\beta}, \boldsymbol{p}_{\alpha\beta}$ represent corrections to the EOM for perturbation label $\alpha$ at order $\epsilon^\beta$ in the perturbation parameter. The interaction function, $I_{\{q,p\}}(\cdot)$, characterizes the covariant effect between different perturbations. We will discuss interaction terms of order $\mathcal{O}(\epsilon_\alpha \epsilon_{\alpha^\prime})$ below. Dependence on the parameter vectors $\boldsymbol{\theta}$ have been suppressed for notational simplicity. We apply perturbation theory to the parameters $\boldsymbol{\theta}$ in \S\ref{sec: varying_structural_params}.

The terms in the summation of Eq.~\ref{eq: expansion_series} do not include the covariant effect of multiple perturbations. For example, if a single massive subhalo impacts a stream, it modifies the stream's subsequent orbit. Additional subhalo impacts affect the already perturbed set of orbits. The dynamics of these cascading interactions are captured by terms of $\mathcal{O}\left(\epsilon_\alpha \epsilon_{\alpha^\prime}\right)$ with $\alpha\neq \alpha^{\prime}$. These are the interaction terms in Eq.~\ref{eq: expansion_series}. In this work we will make (and test) the simplifying assumption that interaction terms are subdominant, and can be neglected. Additionally, we will neglect all terms of $\mathcal{O}(\epsilon^2)$. This is a valid assumption, provided that the fractional change in any conserved quantity is small under the perturbation. Neglecting interaction terms greatly simplifies the complexity of the perturbation theory overall, because at linear order each perturbation is independent of the others. This massively simplifies the combinatorics of, e.g., stream-subhalo interactions, since different combinations of perturbations can be sampled independently, and the collective effect of a series of perturbations is just the sum of the individual perturbations. This simplification means that adding in new perturbations does not require any additional computation involving the existing perturbations.

At linear order in the perturbation parameter $\epsilon_\alpha$, the solution to the EOM (Eq.~\ref{eq: EOM}) takes the form
\begin{equation}\label{eq: linearized_qp}
\begin{split}
    \boldsymbol{q}\left(t, \boldsymbol{\epsilon}\right) &\approx \boldsymbol{q}_0\left(t\right) + \sum\limits_{\alpha=1}^N \epsilon_\alpha \boldsymbol{q}_{\alpha 1}\left(t\right) \\
    \boldsymbol{p}\left(t, \boldsymbol{\epsilon}\right) &\approx \boldsymbol{p}_0\left(t\right) + \sum\limits_{\alpha = 1}^N \epsilon_\alpha \boldsymbol{p}_{\alpha 1}\left( t\right),
\end{split}
\end{equation}
where we have truncated the series expansion in Eq.~\ref{eq: expansion_series} at $\mathcal{O}\left(\epsilon_\alpha^2\right)$. When this truncated expansion is valid (i.e., for small perturbations), the response of each tracer particle to the perturbation is linear in the perturbation parameter, $\epsilon_\alpha$, and a series of perturbations can be added together algebraically once $\{(\boldsymbol{q}_{\alpha 1}, \boldsymbol{p}_{\alpha 1})\}$ are obtained. Furthermore, the presence, absence, and strength of the perturbation are all controlled by $\epsilon_\alpha$, allowing one to scale up and down the effect of the perturbation rapidly once the leading-order correction terms have been obtained.

To obtain $(\boldsymbol{q}_{\alpha 1}, \boldsymbol{p}_{\alpha1})$, we must first solve the EOM for the base Hamiltonian. This can be done in general by solving the equations
\begin{equation}\label{eq: base_EOM}
\begin{split}
    \dot{\boldsymbol{q}}_0 &= \frac{\partial H_{\rm base}}{\partial \boldsymbol{p}} = \boldsymbol{p}_0\\
    \dot{\boldsymbol{p}}_0 &= -\frac{\partial H_{\rm base}}{\partial \boldsymbol{q}} = -\frac{\partial \Phi_{\rm base}}{\partial \boldsymbol{q}},
\end{split}
\end{equation}
with some vector of initial conditions $\left(\boldsymbol{q}_{0,\rm{init}}, \boldsymbol{p}_{0,\rm{init}}\right)$. Once Eq.~\ref{eq: base_EOM} is solved numerically, we can now derive expressions for the correction terms to the base EOM at linear order in $\epsilon_\alpha$. Namely, $\left(\boldsymbol{q}_{\alpha1}(t), \boldsymbol{p}_{\alpha1}(t)\right)$ in Eq.~\ref{eq: expansion_series}. In canonical perturbation theory, the correction terms are usually derived by representing $\Phi_\alpha$ as a Fourier series in angles ${\theta}$ and actions $\boldsymbol{J}$ associated with $H_{\rm base}$. This can be done analytically when $H_{\rm base}$ has a transformation $\left(\boldsymbol{q}, \boldsymbol{p}\right) \xrightarrow[]{} \left({\theta}, \boldsymbol{J}\right)$. As stated, we would like to avoid such an assumption in order to retain flexibility in our modeling.

In general, the functions $\{(\boldsymbol{q}_{\alpha\beta}(t), \boldsymbol{p}_{\alpha\beta}(t))\}$ can be obtained by differentiating Eq.~\ref{eq: expansion_series} with respect to each order of $\epsilon_\alpha$, evaluated at $\boldsymbol{\epsilon}=0$. That is,
\begin{equation}\label{eq: expansion_terms}
    \boldsymbol{q}_{\alpha\beta}\left(t\right) = \frac{1}{\beta!}\frac{d^\beta\boldsymbol{q}}{d\epsilon^{\beta}_\alpha}\Big\vert_{\boldsymbol{\epsilon}=0}, \ \boldsymbol{p}_{\alpha\beta}\left(t\right) = \frac{1}{\beta!}\frac{d^\beta\boldsymbol{p}}{d\epsilon^{\beta}_\alpha}\Big\vert_{\boldsymbol{\epsilon}=0}.
\end{equation}

We are free to take a time-derivative of Eq.~\ref{eq: expansion_terms} on both sides, giving us
\begin{equation}\label{eq: expansion_terms_timederiv}
    \dot{\boldsymbol{q}}_{\alpha\beta}\left(t\right) = \frac{1}{\beta!}\frac{d^\beta\dot{\boldsymbol{q}}}{d\epsilon^{\beta}_\alpha}\Big\vert_{\boldsymbol{\epsilon}=0}, \ \dot{\boldsymbol{p}}_{\alpha\beta}\left(t\right) = \frac{1}{\beta!}\frac{d^\beta\dot{\boldsymbol{p}}}{d\epsilon^{\beta}_\alpha}\Big\vert_{\boldsymbol{\epsilon}=0}.
\end{equation}
This is a useful form for the perturbation terms, since $\dot{\boldsymbol{q}}$ and $\dot{\boldsymbol{p}}$ on the right-hand-sides are provided by Hamilton's equations, Eq.~\ref{eq: EOM}.

We will now specialize to the linear order correction, with $\beta = 1$. The time-derivative of the linear terms are given by $d\dot{\boldsymbol{q}}/d\epsilon_\alpha$ and $d\dot{\boldsymbol{p}}/d\epsilon_\alpha$. If we can obtain these derivatives, then they provide an update rule to pass an initial state $d\left(\boldsymbol{q},\boldsymbol{p}\right)/d\epsilon_\alpha$ to a final state. We derive this update rule below. For the leading order correction, we have
\begin{equation}\label{eq: basic_vector_form}
    \begin{pmatrix}
        \dot{\boldsymbol{q}}_{\alpha 1} \\
        \dot{\boldsymbol{p}}_{\alpha 1}
    \end{pmatrix} = \left[\frac{d}{d\epsilon_\alpha}\begin{pmatrix}
        \dot{\boldsymbol{q}} \\
        \dot{\boldsymbol{p}}
    \end{pmatrix}\right]_{\boldsymbol{\epsilon}=0}.
\end{equation}
We now substitute in Hamilton's equations (Eq.~\ref{eq: EOM}), giving us
\begin{equation}\label{eq: dot_q_alpha_1}
    \dot{\boldsymbol{q}}_{\alpha 1} = \frac{d\boldsymbol{p}}{d\epsilon_\alpha}\Big\vert_{\boldsymbol{\epsilon}=0}
\end{equation}
for the $\boldsymbol{q}$ element. The second element, $\dot{\boldsymbol{p}}_{\alpha 1}$, involves an application of the chain rule:
\begin{small}
\begin{equation}\label{eq: dot_p_alpha_1_init}
\begin{split}
    \dot{\boldsymbol{p}}_{\alpha 1} 
    &=\frac{d\dot{\boldsymbol{p}}}{d\epsilon_\alpha}\Big\vert_{\boldsymbol{\epsilon}=0}\\
    &= -\left [ \frac{d}{d\epsilon_\alpha}\left( \frac{\partial \Phi_{\rm base}}{\partial \boldsymbol{q}} +  \sum\limits_{{\alpha^\prime}=1}^N \epsilon_{\alpha^\prime} \frac{\partial \Phi_{\alpha^\prime}}{\partial \boldsymbol{q}}\right) \right]_{\boldsymbol{\epsilon}=0} \\
    &= -\left[ \frac{\partial^2 \Phi_{\rm base}}{\partial \boldsymbol{q}^2} \frac{d\boldsymbol{q}}{d\epsilon_{\alpha}} \right. \\ & \hspace{1cm}+ \left. \sum\limits_{{\alpha^\prime}=1}^N \left( \epsilon_{\alpha^\prime} \frac{\partial^2\Phi_{\alpha^\prime}}{\partial\boldsymbol{q}^2} \frac{d \boldsymbol{q}}{d \epsilon_{\alpha^\prime}} + \frac{\partial \Phi_{\alpha^\prime}}{\partial \boldsymbol{q}} \delta_{\alpha^\prime \alpha}\right)\right]_{\boldsymbol{\epsilon}=0},
\end{split}
\end{equation}\end{small}
where $\delta_{\alpha^\prime \alpha}$ is the Kronecker delta-function. The term $\partial^2\Phi_{\rm base}/\partial \boldsymbol{q}^2$ is the tidal tensor of the base potential. We will label the tidal tensor with $\mathbf{T}$, whose elements are given by
\begin{equation}\label{eq: tidal_tensor_base}
    \mathbf{T}_{ij} = \frac{\partial^2\Phi_{\rm base}}{ \partial q_i \partial q_j}.
\end{equation}

With this notation, and evaluating at $\boldsymbol{\epsilon}=0$, Eq.~\ref{eq: dot_p_alpha_1_init} becomes
\begin{equation}\label{eq: dot_p_alpha1}
\dot{\boldsymbol{p}}_{\alpha 1} = - \frac{\partial \Phi_\alpha}{\partial \boldsymbol{q}}\Big\vert_{\boldsymbol{q}_0(t)} -\mathbf{T}\left( \boldsymbol{q}_0(t) \right) \frac{d\boldsymbol{q}}{d\epsilon_\alpha}\Big\vert_{\boldsymbol{\epsilon}=0}.
\end{equation}
Repackaging Eqs.~\ref{eq: basic_vector_form}-\ref{eq: dot_p_alpha1} in vector form gives us
\begin{equation}\label{eq: update_rule}
    \frac{d}{dt}\begin{pmatrix}
        d{\boldsymbol{q}}/d\epsilon_\alpha  \\
        d{\boldsymbol{p}}/d\epsilon_\alpha
    \end{pmatrix}_{ \boldsymbol{\epsilon}=0} =
    \begin{pmatrix}
        d\boldsymbol{p}/d\epsilon_\alpha \\
         - \frac{\partial \Phi_\alpha}{\partial \boldsymbol{q}} -\mathbf{T}\left(\boldsymbol{q}, t\right)\frac{d\boldsymbol{q}}{d\epsilon_\alpha}
    \end{pmatrix}_{ \boldsymbol{\epsilon}=0}
\end{equation}
where the quantities are evaluated at $\boldsymbol{\epsilon}=0$, i.e., the base trajectory $\left( \boldsymbol{q}_0(t), \boldsymbol{p}_0(t)\right)$. 

Eq.~\ref{eq: update_rule} is an integro-differential equation, and provides a means to evolve an initial state $\left( d\boldsymbol{q}_{\rm init}/d\epsilon_\alpha, d\boldsymbol{p}_{\rm init}/d\epsilon_\alpha\right)$ to a final state $\left( d\boldsymbol{q}_{\rm final}/d\epsilon_\alpha, d\boldsymbol{p}_{\rm final}/d\epsilon_\alpha\right)$, analogous to evolving the position and a velocity of a particle forward in time. Here, instead of position and velocity, we evolve \emph{derivatives} of these quantities forward in time. Provided that the proper initial conditions are chosen, these equations can be integrated numerically for any differentiable choice of $\Phi_{\rm base}$ and $\Phi_\alpha$. We discuss initial conditions in \S\ref{sec: boundary_conditions}.

\subsection{Intuition behind Linear Perturbations}\label{sec: intuition}
\begin{figure}
\centering\includegraphics[scale=.5]{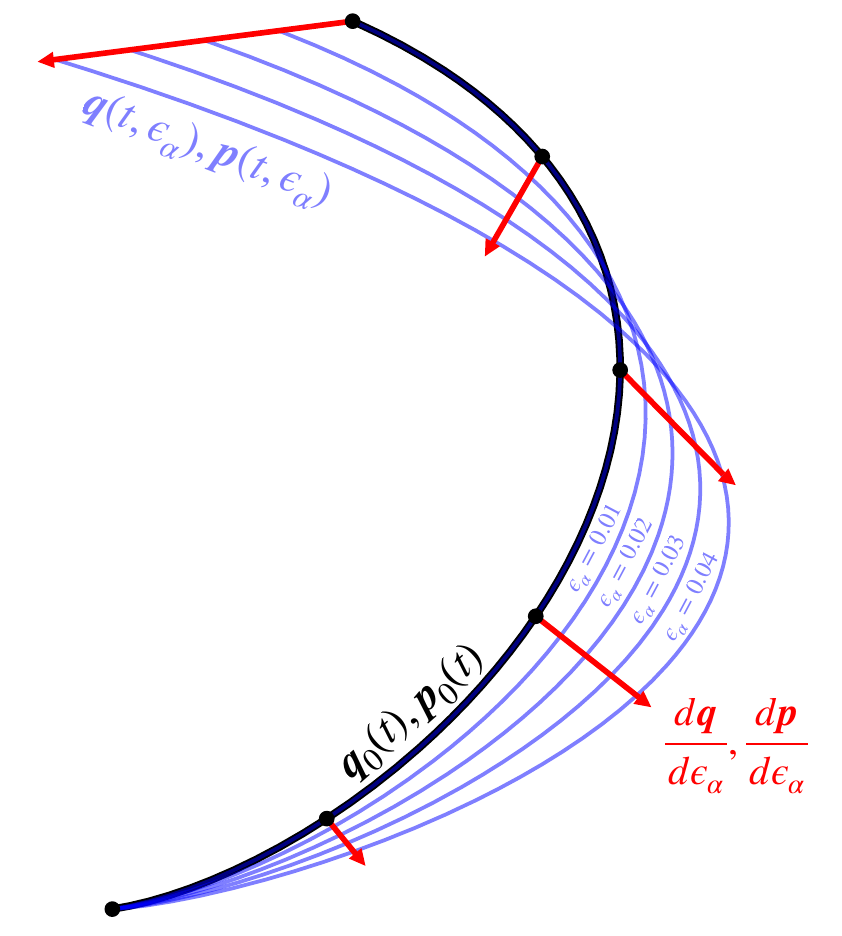}
    \caption{Illustration of the method in phase-space. The black curve is the base trajectory, $\boldsymbol{q}_0, \boldsymbol{p}_0$, governed by $H_{\rm base}$. How does this trajectory deform as the perturbation parameter, $\epsilon_\alpha$, increases? The blue curves illustrate a continuous deformation of the base trajectory due to increasing $\epsilon_\alpha$. If $\epsilon_\alpha$ is small, then the deformed trajectories at a fixed point in time are all colinear. The derivatives $d\boldsymbol{q}/d\epsilon_\alpha$ and $d\boldsymbol{p}/d\epsilon_\alpha$ provide the direction of the deformation to the orbit for small $\epsilon_\alpha$. Each blue curve can be thought of as a contour of constant perturbation. Scaling $\epsilon_\alpha$ up and down pierces through these contours, allowing us to algebraically sample the effect of different values for the perturbation parameters. }
    \label{fig: method_illustration}
\end{figure}

We now comment on the physical intuition behind Eq.~\ref{eq: update_rule}, since this expression provides the leading order effect of a perturbing (i.e., subhalo) potential on a tracer particle. An illustration of linear perturbation theory in phase-space is provided in Fig.~\ref{fig: method_illustration} for an arbitrary choice of $H_{\rm base}$ and $\Phi_\alpha$. When Eq.~\ref{eq: update_rule} is integrated along the base trajectory, one obtains the perturbation vectors $(d\boldsymbol{q}/d\epsilon_\alpha, d\boldsymbol{p}/d\epsilon_\alpha)$ (red). At each point along the trajectory, the perturbation vector pierces through contours of constant perturbation strength. Scaling up and down the perturbation parameter, $\epsilon_\alpha$, corresponds to continuously deforming the phase-space trajectory. Importantly, the perturbation vector can have components both along and perpendicular to the base trajectory. Thus, the orbit and frequencies of particles can change in our linear treatment. When the linear approximation is valid, contours of constant perturbation are colinear with the local perturbation vectors in Fig.~\ref{fig: method_illustration}. 

The physical meaning of the perturbation vectors are further developed below. Eq.~\ref{eq: update_rule} is a differential equation that is valid for any time $t$. Realistically, we only have access to the phase-space position of a star at the time of observation, labeled $t_f$. At this time, we can ask, how does the perturbed phase-space position of the particle differ from the base trajectory without perturbations, $\left(\boldsymbol{q}_0(t_f), \boldsymbol{p}_0(t_f)\right)$? To first order in the perturbation parameter, the answer is given by Eq.~\ref{eq: linearized_qp}. Re-arranging this expression and assuming only a single perturbation term (i.e., $N=1$ for simplicity---we will therefore suppress the index $\alpha$ for now), we have
\begin{equation}\label{eq: delta_q_p}
\begin{split}
    \Delta \boldsymbol{q}\left(t_f, \epsilon\right) &= \epsilon \boldsymbol{q}_{1}\left(t_f\right) \\
    \Delta \boldsymbol{p}\left(t_f, {\epsilon}\right) &= \epsilon \boldsymbol{p}_{1}\left(t_f\right),
\end{split}
\end{equation}
where $\Delta$ is the difference between the perturbed and unperturbed phase-space location up to linear order in $\epsilon$. Unsurprisingly, as $\epsilon\xrightarrow[]{} 0$ the difference between the perturbed state and base state is $0$. Eq.~\ref{eq: delta_q_p} also provides a clear interpretation for the perturbation functions $\boldsymbol{q}_1$ and $\boldsymbol{p}_1$: these are simply the rate of change of $\Delta\boldsymbol{q}$ and $\Delta\boldsymbol{p}$ with respect to $\epsilon$ (e.g., $\boldsymbol{q}_1 = d\Delta \boldsymbol{q}/d\epsilon$). Physically, $\boldsymbol{q}_1$ and $\boldsymbol{p}_1$ provide the rate at which the position and velocity of a particle will scatter with increasing $\epsilon$. 

Now we discuss the actual terms that constitute $\boldsymbol{q}_1$ and $\boldsymbol{p}_1$. As mentioned, we only measure a star at the time of observation, $t_f$. To obtain an expression for $\boldsymbol{q}_1(t_f)$ and $\boldsymbol{p}_1(t_f)$ in Eq.~\ref{eq: delta_q_p}, we can integrate Eq.~\ref{eq: update_rule} from an initial time, $t_{\rm init}$, to the final time, $t_f$. Starting with the position term, Eq.~\ref{eq: update_rule} states that $d\dot{\boldsymbol{q}}/d\epsilon = d\boldsymbol{p}/d\epsilon$. This means that $d\boldsymbol{q}/d\epsilon$, or $\boldsymbol{q}_1$, is just the time-integral of the velocity term. From this relation, the integral expression for $\Delta\boldsymbol{q}$ is
\begin{equation}\label{eq: Delta_q_int}
    \Delta\boldsymbol{q}\left( t_f, \epsilon\right) = \epsilon \int\limits_{t_{\rm init}}^{t_f} \frac{d\boldsymbol{p}}{d\epsilon}\Big\vert_{\epsilon=0} dt.
\end{equation} Similarly for $\Delta\boldsymbol{p}$, its integral expression is
\begin{multline}\label{eq: Delta_p_int}
    \Delta\boldsymbol{p}\left(t_f, \epsilon\right) \\= \epsilon \int \limits_{t_{\rm init}}^{t_f} dt\left[-\frac{\partial \Phi_{1}}{\partial\boldsymbol{q}}\Big\vert_{\boldsymbol{q}_0(t)} - \mathbf{T}\left(\boldsymbol{q}_0(t)\right) \int \limits_{t_{\rm init}}^{t}  \frac{d\boldsymbol{p}}{d\epsilon}\Big\vert_{\epsilon=0} dt^\prime \right],
\end{multline}
where $\Phi_1$ is the potential of the perturbation. 

We can interpret Eq.~\ref{eq: Delta_p_int} as follows: the first bracketed term, $-\partial \Phi_1/\partial\boldsymbol{q}$, is the force due to the perturbation, evaluated along the base trajectory. The integral of this term has units of velocity, representing the \emph{velocity injected} by the perturbation. Unlike the impulse approximation which injects a velocity kick at a single time, Eq.~\ref{eq: Delta_p_int} distributes the force throughout the time-evolution of a test particle. The second bracketed term is a product between the tidal tensor, $\mathbf{T}$, and the integral of the velocity term from the initial time up to a generic time $t$. This integral gives the spatial displacement of the particle at time $t$ due to the perturbation. The appearance of the tidal tensor in this expression is also intuitive: if a particle is perturbed by the force $-\partial \Phi_1/\partial\boldsymbol{q}$ throughout its evolution, it becomes displaced in position by an amount $\delta \boldsymbol{q} = \int \left(d\boldsymbol{p}/d\epsilon\right) dt$. This displacement means that the particle no longer feels the same force in the base potential, $\Phi_{\rm base}$, since it has traversed to a new location. As a result of the displacement $\delta \boldsymbol{q}$, the acceleration of the particle due to the base potential must be modified. This modification is captured by the tidal tensor of the base potential, multiplied by the displacement of the particle up to time $t$. The (negative) integral of this product has units of velocity, representing the change in velocity due to the perturbation knocking the particle off of its base trajectory, thereby modifying where the base force is evaluated.

The change in velocity due to these two effects (injected force and correction to the base force) must then be integrated forward to obtain the total change in position due to the perturbation, $\Delta\boldsymbol{q}$. This integration is carried out in Eq.~\ref{eq: Delta_q_int}. 

While we have considered only a single perturbation term in this section, the same intuition holds for $N$ perturbations up to linear order in $\epsilon_\alpha$. The only difference is that the total change in position and velocity, $\Delta\boldsymbol{q}$ and $\Delta\boldsymbol{p}$, will consist of a sum of individual changes due to each perturbation:
\begin{equation}\label{eq: change_q_change_p}
    \Delta\boldsymbol{q} = \sum\limits_{\alpha = 1}^N \Delta\boldsymbol{q}_\alpha,  \ \ 
    \Delta\boldsymbol{p} = \sum\limits_{\alpha = 1}^N \Delta\boldsymbol{p}_\alpha
\end{equation}
where $\Delta\boldsymbol{q}_\alpha$ is the total change in position due to perturbation $\alpha$ (and similar for $\Delta \boldsymbol{p}_\alpha$).

The computational benefit of Eq.~\ref{eq: change_q_change_p} is that the strength of each perturbation, $\epsilon_\alpha$, can be scaled up and down algebraically once the integrals terms in Eq.~\ref{eq: Delta_q_int} and Eq.~\ref{eq: Delta_p_int} are calculated. This allows for an extremely rapid sampling of different sets of perturbations, once integral terms have been pre-computed.

\subsection{Stream Model and Boundary Conditions}\label{sec: boundary_conditions}
We now discuss how the perturbative approach developed above can be generalized to orbit ensembles, i.e., stellar streams. The basic recipe involves first obtaining initial conditions for every stream particle, computing (and saving) orbits, and then obtaining the linear perturbation derivatives for each orbit.

The central equation of this work is Eq.~\ref{eq: update_rule}, which provides a velocity and acceleration term to update a state $\left(d\boldsymbol{q}/d\epsilon_\alpha, d\boldsymbol{p}/d\epsilon_\alpha\right)$ at time $t$ to some later time $t_f$. As with any update rule, one needs to specify the initial conditions. As we will discuss below, the initial conditions that one adopts for each particle depend on the underlying stream model, and what assumptions the modeler would like to make.

Throughout this work we will utilize the particle-spray technique as a starting point \citep{2015MNRAS.452..301F}, though the formalism we develop here is more general and can be applied to any differentiable model for particles leaving a cluster. Specifically, we will assume that the initial condition in $\boldsymbol{q}$ and $\boldsymbol{p}$ for each particle is any  
differentiable function of the progenitor's phase-space position at the particle's release time. Let $\boldsymbol{q}_{\rm init}$ and $\boldsymbol{p}_{\rm init}$ represent the initial position and velocity of a stream particle. The position and velocity of the progenitor at the release time, $t_{\rm rel}$, is $(\boldsymbol{q}_{\rm prog}(t_{\rm rel}, \boldsymbol{\epsilon}), \boldsymbol{p}_{\rm prog}(t_{\rm rel}, \boldsymbol{\epsilon}))$, where there is an $\boldsymbol{\epsilon}$ dependence since the progenitor is subject to the perturbation terms just like any other particle. In mathematical terms, each stream particle is initialized according to a generic vector-valued release function
\begin{equation}\label{eq: F_rel}
    \left( \boldsymbol{q}_{\rm init}, \boldsymbol{p}_{\rm init} \right) = \mathbf{F}_{\rm rel}\left( \boldsymbol{q}_{\rm prog}(t_{\rm rel}), \boldsymbol{p}_{\rm prog}(t_{\rm rel})\right).
\end{equation}
From this expression, we can obtain the initial state $(d\boldsymbol{q}_{\rm init}/d\epsilon_\alpha, d\boldsymbol{p}_{\rm init}/d\epsilon_\alpha)$, by differentiating both sides with respect to $\epsilon_\alpha$, and evaluating at $\boldsymbol{\epsilon}=0$.

Using the chain rule, and defining $\boldsymbol{w}_{\rm prog} = (\boldsymbol{q}_{\rm prog}, \boldsymbol{p}_{\rm prog})$, the initial conditions for Eq.~\ref{eq: update_rule} are given by
\begin{small}
\begin{multline}\label{eq: init_cond_general}
    \begin{pmatrix}
        d\boldsymbol{q}_{\rm init}/d\epsilon_\alpha \\
        d\boldsymbol{p}_{\rm init}/d\epsilon_\alpha
    \end{pmatrix}_{\boldsymbol{\epsilon}=0} = \\\frac{\partial\mathbf{F}_{\rm rel}}{\partial\boldsymbol{w}}\Big\vert_{\boldsymbol{w}_{\rm prog}(t_{\rm rel}, \boldsymbol{\epsilon}=0)} 
    \begin{pmatrix}
        d\boldsymbol{q}_{\rm prog}/d\epsilon_\alpha \\
        d\boldsymbol{p}_{\rm prog}/d\epsilon_\alpha
    \end{pmatrix}_{(t_{\rm rel}, \boldsymbol{\epsilon}=0)},
\end{multline}\end{small}
where $\partial \mathbf{F}_{\rm rel}/\partial\boldsymbol{w}$ is the $6\times 6$ Jacobian
\begin{equation}\label{eq: JacobianF}
    \frac{\partial \mathbf{F}_{\rm rel}}{\partial\boldsymbol{w}} = \begin{pmatrix} 
        \frac{\partial \mathbf{F}_{\rm rel}}{\boldsymbol{\partial{q}}} & \frac{\partial \mathbf{F}_{\rm rel}}{\partial{\boldsymbol{p}}} 
    \end{pmatrix},\end{equation}
which is evaluated at the location of the progenitor in the base potential only (with $\boldsymbol{\epsilon}=0$). This Jacobian is computed using automatic differentiation (discussed in \S\ref{sec: code_implementation}). Note that the Jacobian does not change as we introduce different perturbation terms: it can be precomputed for each particle based on whatever one adopts for the progenitor's orbit and the base potential, $\Phi_{\rm base}$.

From Eq.~\ref{eq: init_cond_general}, we have offloaded the problem of finding a generic initial condition for every particle to calculating the response of the progenitor to each perturbation. That is, the vector $(d\boldsymbol{q}_{\rm prog}/d\epsilon_\alpha, d\boldsymbol{p}_{\rm prog}/d\epsilon_\alpha)$ still needs to be dealt with. One simple choice is that the progenitor's orbit is independent of any $\epsilon_\alpha$. That is, the effect of any perturbation term on the progenitor is negligible, amounting to $\frac{d}{d\epsilon_\alpha}(\boldsymbol{q}_{\rm init}, \boldsymbol{p}_{\rm init}) = 0$. In \citet{2017MNRAS.466..628B}, this assumption is made (effectively), since they assume that the progenitor's orbit is unaffected by the perturbations. 
If the progenitor's orbit itself is perturbed by a subhalo, then particles released well-after the perturbation will have new initial positions and velocities, and therefore orbits. Below we develop a simple framework to account for this effect.

To anchor our stream model,  we would like to require that the final phase-space location of the progenitor (i.e., today) remains the same, independent of $\boldsymbol{\epsilon}$.
This provides a boundary condition for our problem, of the form
\begin{equation}\label{eq: prog_ic}
\frac{d\left(\boldsymbol{q}_{\rm prog}, \boldsymbol{p}_{\rm prog} \right)}{d\epsilon_\alpha} \Big\vert_{t_f, \boldsymbol{\epsilon}=0} = 0,
\end{equation}
enforcing that the progenitor's final location is invariant to the perturbations. 

Eq.~\ref{eq: prog_ic} is a boundary condition for the progenitor, which can then be integrated \emph{backwards} in time according to the update rule Eq.~\ref{eq: update_rule} (up to a minus sign for the backwards integration). Saving the values of this integration at each particle release time then gives us the necessary information to evaluate the released particle's initial conditions, from Eq.~\ref{eq: init_cond_general}. With this choice, the initialization for each particle's derivatives $d\boldsymbol{q}_{\rm init}/d\epsilon_\alpha$ and $d\boldsymbol{p}_{\rm init}/d\epsilon_\alpha$ are specified, allowing us to numerically integrate Eq.~\ref{eq: update_rule} forward from the release time, $t_{\rm rel}$, for each particle up to a common final time, $t_{f}$.

\subsection{Incorporating the Progenitor's Self-Gravity}\label{sec: self_grav_method}
We now discuss how self-gravity of the progenitor can be incorporated in our perturbative stream model. Self-gravity can play an important role in determining the release conditions of stars leaving the progenitor cluster, and therefore the epicyclic structure of tidal tails (e.g.,  \citealt{2008MNRAS.387.1248K, 2017MNRAS.470...60E,2020ApJ...891..161I}). 

Let $\tilde{\Phi}_{\rm prog}$ represent the (possibly time-dependent) gravitational potential of the progenitor. We will utilize an analytic potential for the progenitor, centered on the progenitor's spatial location labeled $\boldsymbol{q}_{\rm prog}\left(t\right)$. The potential felt by a test-particle at location $\boldsymbol{q}$ due to the self-gravity of the progenitor is ${\Phi}_{\rm prog}(\boldsymbol{q},t) \equiv \tilde{\Phi}_{\rm prog}(\boldsymbol{q} - \boldsymbol{q}_{\rm prog}(t))$. This term can be added to the base Hamiltonian in Eq.~\ref{eq: Hamiltonian}. The only complication of this treatment is that the progenitor's orbit itself can be perturbed by fluctuations in the potential, $\Phi_\alpha$. Therefore, the derivative $d\dot{\boldsymbol{p}}/d\epsilon_\alpha$ must now self-consistently account for perturbations to the test particle at $\boldsymbol{q}$, and perturbations to the progenitor's orbit at $\boldsymbol{q}_{\rm prog}$. The derivation proceeds identically to \S\ref{sec: perturbative_orbit_model}, though Eq.~\ref{eq: dot_p_alpha1} now becomes
\begin{multline}\label{eq: acceleration_term_self_grav}
    \dot{\boldsymbol{p}}_{\alpha 1} = - \frac{\partial \Phi_\alpha}{\partial \boldsymbol{q}}\Big\vert_{\boldsymbol{q}_0(t)} -\mathbf{T}\left( \boldsymbol{q}_0(t) \right) \frac{d\boldsymbol{q}}{d\epsilon_\alpha}\Big\vert_{\boldsymbol{\epsilon}=0} \\
    -\mathbf{T}_{\rm prog}\left(\boldsymbol{q}_0(t)\right)\left[\frac{d\boldsymbol{q}}{d\epsilon_\alpha} - \frac{d\boldsymbol{q}_{\rm prog}}{d\epsilon_\alpha} \right]_{\boldsymbol{\epsilon}=0},
\end{multline}
where the term $\mathbf{T}_{\rm prog}$ is the $3\times 3$ tidal tensor of the progenitor's potential, with elements
\begin{equation}\label{eq: tidal_prog}
    \mathbf{T}_{{\rm prog}, ij} \equiv \frac{\partial^2 \Phi_{\rm prog}}{\partial q_i \partial q_j},
\end{equation}
and the bracketed term in Eq.~\ref{eq: acceleration_term_self_grav} is the relative distance between the perturbed test particle and the perturbed progenitor's orbit. We separate these terms since they are calculated individually. The bracketed term governs how the perturbation displaces a test particle relative to the progenitor, while $\mathbf{T}_{\rm prog}$ characterizes the local acceleration field created by the progenitor's self-gravity. The derivative $d\boldsymbol{q}_{\rm prog}/d\epsilon_\alpha$ in Eq.~\ref{eq: acceleration_term_self_grav} can be calculated using Eq.~\ref{eq: update_rule}, since the progenitor's orbit does not feel its own self-gravity. Eq.~\ref{eq: acceleration_term_self_grav} can then be integrated numerically. For the case of including the progenitor's self-gravity, initial conditions for the \emph{derivatives} of every particle can be obtained following the same procedure outlined in \S\ref{sec: boundary_conditions}, though adopting Eq.~\ref{eq: acceleration_term_self_grav} for the acceleration term.

\subsection{Varying the Structural Parameters of the Perturbing Potentials}\label{sec: varying_structural_params}
In this section we incorporate the effect of varying the structural parameters of the perturbing potentials. These parameters might include (e.g.) the scale-radius of a dark matter subhalo, its inner-density slope, or other intrinsic characteristics. We use the term structural parameters, since these are internal properties of the perturbing potentials rather than orbital properties like its velocity.

We will proceed perturbatively, and again obtain correction terms to the motion of a particle due to varying the structural parameters of the perturbations. Following the notation adopted in Eq.~\ref{eq: Hamiltonian}, let us assume that $\Phi_\alpha$ has the structural parameters labeled $\boldsymbol{\theta}_\alpha$. This means that all quantities involving $\Phi_\alpha$ must also be a function of $\boldsymbol{\theta}_\alpha$. The terms $\boldsymbol{q}_{\alpha \beta}$ and $\boldsymbol{p}_{\alpha \beta}$ in Eq.~\ref{eq: expansion_series} are functions of $\Phi_\alpha$, and therefore $\boldsymbol{\theta}_\alpha$ as well.

We now seek the leading order corrections to the EOM in the structural parameters $\boldsymbol{\theta}_\alpha$. To derive this dependence, we only need to take derivatives of $\boldsymbol{q}$ and $\boldsymbol{p}$ in Eq.~\ref{eq: expansion_series} with respect to $\boldsymbol{\theta}_{\alpha}$. Because our treatment is perturbative, all derivatives are evaluated at $\boldsymbol{\epsilon}=0$. This has the immediate repercussion that $\partial (\boldsymbol{q}, \boldsymbol{p})/\partial\boldsymbol{\theta}_\alpha$ evaluated at $\boldsymbol{\epsilon}=0$ is itself $0$. This makes intuitive sense: changing the internal properties of perturbations at zeroth order in the perturbation parameter has no effect on the EOM. 

The leading non-zero derivatives are the mixed partials, 
\begin{equation}\label{eq: first_surviving_theta}
\begin{split}
    \left[\frac{\partial }{\partial \boldsymbol{\theta}_\alpha} \left( \frac{d \boldsymbol{q}}{\partial\epsilon_{\alpha}}\right)\right]_{\left(\boldsymbol{\epsilon}=0, \boldsymbol{\theta}_\alpha = \boldsymbol{\theta}_\alpha^*\right)} &= \frac{\partial \boldsymbol{q}_{\alpha 1}}{\partial \boldsymbol{\theta}_\alpha}\Big\vert_{\boldsymbol{\theta}_\alpha = \boldsymbol{\theta}_\alpha^*} \\
    \left[\frac{\partial}{\partial \boldsymbol{\theta}_\alpha}\left(\frac{d \boldsymbol{p}}{\partial\epsilon_{\alpha}}\right)\right]_{\left(\boldsymbol{\epsilon}=0, \boldsymbol{\theta}_\alpha = \boldsymbol{\theta}_\alpha^*\right)} &= \frac{\partial \boldsymbol{p}_{\alpha 1}}{\partial \boldsymbol{\theta}_\alpha}\Big\vert_{\boldsymbol{\theta}_\alpha = \boldsymbol{\theta}_\alpha^*}, 
\end{split}
\end{equation}
where on the left-hand-side we evaluate the derivatives at a root expansion point $\boldsymbol{\theta}_{\alpha}^*$, and $\boldsymbol{\epsilon}=0$. On the right-hand-side, we have utilized Eq.~\ref{eq: expansion_series} to evaluate the derivatives. The root expansion point is the value for the structural parameters that we expand around at linear order. For streams and subhalos, we will expand around a value of the subhalo scale-radius, $r_s$, that is drawn from the subhalo mass-radius relation that will be defined in \S\ref{sec: sampling_mass_func}.

Eq.~\ref{eq: first_surviving_theta} are the first derivative terms in a Taylor expansion for $\boldsymbol{q}_{\alpha 1}$ and $\boldsymbol{p}_{\alpha 1}$ in the parameter $\boldsymbol{\theta}_\alpha$. Up to linear order in $\boldsymbol{\theta}_\alpha$, we can write $\boldsymbol{q}_{\alpha 1}$ and $\boldsymbol{p}_{\alpha 1}$ as a linear function of $\boldsymbol{\theta}_{\alpha}$ using
\begin{equation}\label{eq: qp_alpha1_theta}
\begin{split}
    \boldsymbol{q}_{\alpha 1}\left( \boldsymbol{\theta}_\alpha \right) &\approx \boldsymbol{q}_{\alpha 1}\left( \boldsymbol{\theta}_\alpha^* \right) + \frac{\partial \boldsymbol{q}_{\alpha 1}}{\partial \boldsymbol{\theta}_\alpha}\Big\vert_{\boldsymbol{\theta}^*_\alpha} \cdot \left( \boldsymbol{\theta}_\alpha - \boldsymbol{\theta}_\alpha^*\right) \\
    \boldsymbol{p}_{\alpha 1}\left( \boldsymbol{\theta}_\alpha \right) &\approx \boldsymbol{p}_{\alpha 1}\left( \boldsymbol{\theta}_\alpha^* \right) + \frac{\partial \boldsymbol{p}_{\alpha 1}}{\partial \boldsymbol{\theta}_\alpha}\Big\vert_{\boldsymbol{\theta}^*_\alpha} \cdot \left( \boldsymbol{\theta}_\alpha - \boldsymbol{\theta}_\alpha^*\right). 
\end{split}
\end{equation}

With the leading order response to varying $\boldsymbol{\theta}_\alpha$ obtained, we can now substitute Eq.~\ref{eq: qp_alpha1_theta} into Eq.~\ref{eq: linearized_qp} to obtain an algebraic expression for how each particle responds to changes in the perturbation strength ($\epsilon_\alpha$) and structural parameters ($\boldsymbol{\theta}_\alpha$):
\begin{multline}\label{eq: qp_theta_dependence}
    \boldsymbol{q}\left( t, \boldsymbol{\epsilon}, \{\boldsymbol{\theta}_{\alpha}\}\right) \approx \boldsymbol{q}_0\left(t\right)   \\+ \sum\limits_{\alpha=1}^N \epsilon_\alpha\left[ \boldsymbol{q}_{\alpha 1}\left( \boldsymbol{\theta}_\alpha^* \right) + \frac{\partial \boldsymbol{q}_{\alpha 1}}{\partial \boldsymbol{\theta}_\alpha}\Big\vert_{\boldsymbol{\theta}^*_\alpha} \cdot \left( \boldsymbol{\theta}_\alpha - \boldsymbol{\theta}_\alpha^*\right) \right] \\
    \hspace{-4.5cm}\boldsymbol{p}\left( t, \boldsymbol{\epsilon}, \{\boldsymbol{\theta}_{\alpha}\}\right) \approx \boldsymbol{p}_0\left(t\right) 
    \\ +\sum\limits_{\alpha=1}^N \epsilon_\alpha\left[ \boldsymbol{p}_{\alpha 1}\left( \boldsymbol{\theta}_\alpha^* \right) + \frac{\partial \boldsymbol{p}_{\alpha 1}}{\partial \boldsymbol{\theta}_\alpha}\Big\vert_{\boldsymbol{\theta}^*_\alpha} \cdot \left( \boldsymbol{\theta}_\alpha - \boldsymbol{\theta}_\alpha^*\right) \right].
\end{multline}
The terms $\boldsymbol{q}_{\alpha 1}$ and $\boldsymbol{p}_{\alpha 1}$ are obtained by numerically integrating Eq.~\ref{eq: update_rule}. These are simply the first order correction terms we have already accounted for. The other derivative terms are developed below.

In order to obtain
$\partial\left(\boldsymbol{q}_{\alpha 1}, \boldsymbol{p}_{\alpha 1} \right)/\partial \boldsymbol{\theta}_\alpha$, we will use Eq.~\ref{eq: first_surviving_theta}, which connects these derivatives to $\boldsymbol{q}$ and $\boldsymbol{p}$. We can substitute Hamilton's EOM into Eq.~\ref{eq: first_surviving_theta} by taking a time derivative on both sides. This gives us
\begin{equation}\label{eq: first_surviving_theta_time_deriv}
\begin{split}
    \frac{\partial \dot{\boldsymbol{q}}_{\alpha 1}}{\partial \boldsymbol{\theta}_\alpha}\Big\vert_{\boldsymbol{\theta}_\alpha = \boldsymbol{\theta}_\alpha^*} &= \left[\frac{\partial }{\partial \boldsymbol{\theta}_\alpha} \left( \frac{d \dot{\boldsymbol{q}}}{d\epsilon_{\alpha}}\right)\right]_{\left(\boldsymbol{\epsilon}=0, \boldsymbol{\theta}_\alpha = \boldsymbol{\theta}_\alpha^*\right)}  \\
    \frac{\partial  \dot{\boldsymbol{p}}_{\alpha 1}}{\partial \boldsymbol{\theta}_\alpha}\Big\vert_{\boldsymbol{\theta}_\alpha = \boldsymbol{\theta}_\alpha^*} &= \left[\frac{\partial}{\partial \boldsymbol{\theta}_\alpha}\left(\frac{d \dot{ \boldsymbol{p}}}{d\epsilon_{\alpha}}\right)\right]_{\left(\boldsymbol{\epsilon}=0, \boldsymbol{\theta}_\alpha = \boldsymbol{\theta}_\alpha^*\right)}. 
\end{split}
\end{equation}
We have already obtained expressions for $d\dot{\boldsymbol{q}}/d\epsilon_\alpha$ and $d\dot{\boldsymbol{p}}/d\epsilon_\alpha$ evaluated at $\boldsymbol{\epsilon}=0$, this is the update rule from Eq.~\ref{eq: update_rule}. Substituing Eq.~\ref{eq: update_rule} into Eq.~\ref{eq: first_surviving_theta_time_deriv}, and using the notation $\boldsymbol{q}_{\alpha 1} = d\boldsymbol{q}/d\epsilon_\alpha$ and $\boldsymbol{p}_{\alpha 1}=d\boldsymbol{p}/d\epsilon_\alpha $, we can obtain a new matrix differential equation whose solution (numerically integrated) gives us the leading order effect of varying $\boldsymbol{\theta}_\alpha$ on the phase-space position of a particle. 
The expression is
\begin{equation}\label{eq: update_rule_theta}
    \frac{d}{dt}\begin{pmatrix}
        \partial \boldsymbol{q}_{\alpha 1}/\partial \boldsymbol{\theta}_\alpha \\
        \partial{\boldsymbol{p}_{\alpha 1}}/\partial\boldsymbol{\theta}_\alpha
    \end{pmatrix}_{ \boldsymbol{\epsilon}=0} =
    \begin{pmatrix}
        \partial\boldsymbol{p}_{\alpha 1}/\partial \boldsymbol{\theta}_\alpha\\
         - \frac{\partial }{\partial\boldsymbol{\theta}_\alpha}\frac{\partial \Phi_\alpha}{ \partial \boldsymbol{q}} -\mathbf{T}\left(\boldsymbol{q}, t\right)\frac{\partial \boldsymbol{q}_{\alpha 1}}{\partial \boldsymbol{\theta}_\alpha}
    \end{pmatrix}_{\boldsymbol{\epsilon}=0},
\end{equation}
where all derivatives are evaluated at $\boldsymbol{\epsilon} = 0$ and $\boldsymbol{\theta}_\alpha = \boldsymbol{\theta}_\alpha^*$. Because Eq.~\ref{eq: update_rule_theta} is evaluated at $\boldsymbol{\epsilon} = 0$, this means that the equation is integrated along the base trajectory, $(\boldsymbol{q}_0(t), \boldsymbol{p}_0(t))$.

Eq.~\ref{eq: update_rule_theta} provides another update rule, to time-evolve some initial state $(\partial\boldsymbol{q}_{\alpha 1}/\partial \boldsymbol{\theta}_\alpha, \partial\boldsymbol{p}_{\alpha 1}/\partial \boldsymbol{\theta}_\alpha)_{\rm initial}$ to a final state $(\partial\boldsymbol{q}_{\alpha 1}/\partial \boldsymbol{\theta}_\alpha, \partial\boldsymbol{p}_{\alpha 1}/\partial \boldsymbol{\theta}_\alpha)_{\rm final}$. Starting from the proper initial conditions, Eq.~\ref{eq: update_rule_theta} is solved numerically to obtain the $\boldsymbol{\theta}_\alpha$ derivatives in Eq.~\ref{eq: qp_theta_dependence}. To obtain initial conditions, we follow an identical procedure to \S\ref{sec: boundary_conditions}, and provide the derivation in Appendix~\ref{app: structural_boundary_conds}.

\subsection{Criterion for the Linear Approximation to Hold}\label{sec: convergence_criterion}
An important limitation of any linear approximation is that for some $\epsilon > \epsilon_{\rm max}$, the linear approximation will fail. Here we discuss a criterion for convergence.

For our linearized treatment to remain valid, it is required that the gradient of the perturbing potential, $\vert \epsilon_\alpha \nabla \Phi_\alpha \vert$, is much smaller than $\vert \nabla \Phi_{\rm base} \vert$. This condition amounts to assuming that the perturbation only slightly deforms an orbit in phase-space rather than completely changing its conserved quantities (i.e., energy, action, etc). Intuitively, $\epsilon_\alpha$ should never be so large to change any conserved quantity by of order itself (i.e., a 100\% change). Otherwise, we have potentially changed the orbital family that a particle belongs to and our linear approximation will surely fail.

A major benefit of our model is that we allow for time-dependent potentials. Then, the energy of any single particle is not strictly conserved, but we may still use the energy of the particle at a final time $t_f$ as an approximate constant of motion. Even for time-dependent systems, energy can still provide a heuristic indicator of the orbit family that any given particle belongs to, and can therefore be used to determine a domain of validity for the linear approximation. Let $\Delta E$ represent the energy difference between the perturbed particle and unperturbed particle at time $t$. If the unperturbed particle has energy $E_{\rm base}$, then at the final time, $t_f$, we require 
\begin{equation}\label{eq: nrg_convergence_criterion}
    \Big\vert \frac{\Delta E\left(t_f, \boldsymbol{\epsilon} \right)}{E_{\rm base}} \Big\vert << 1
\end{equation}
in order for  our linear approximations to remain valid.

For any vector of perturbation parameters, $\boldsymbol{\epsilon}$, we can compute the convergence criterion using the left-hand-side of Eq.~\ref{eq: nrg_convergence_criterion} and check if the quantity is less than 1. For inference of the perturbation strength parameters, we can place (e.g.) top-hat priors on the convergence ratio and never sample an $\boldsymbol{\epsilon}$ vector that violates Eq.~\ref{eq: nrg_convergence_criterion}. This is equivalent to enforcing that the sum of coefficients in a linear regression should be bounded above by some fixed quantity.

\begin{figure}
\centering\includegraphics[scale=.6]{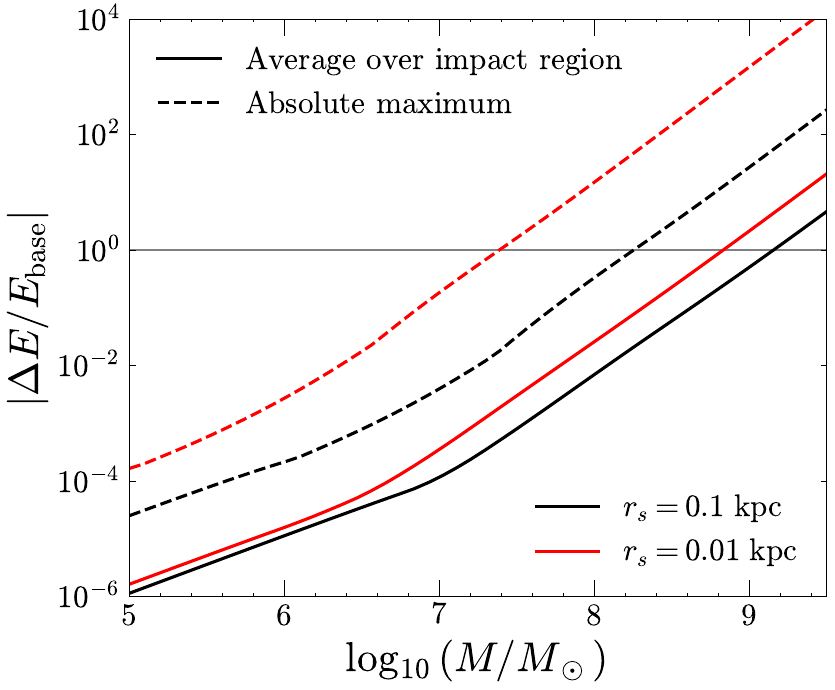}
    \caption{The fractional change in energy between a stream perturbed by a subhalo, versus the same stream without a subhalo perturbation. Increasing the perturbation strength (proportional to the subhalo mass) produces larger changes in energy, with an order unity change indicating a departure from the linear regime. More dense subhalos (e.g., those with smaller scale-radii) achieve an order unity change in energy more rapidly than less dense subhalos.    }
    \label{fig: energy_error}
\end{figure} 

An illustration of Eq.~\ref{eq: nrg_convergence_criterion} is provided in Fig.~\ref{fig: energy_error}. In this figure, we apply our perturbative formalism to a GD-1 like stream that suffers a single subhalo impact $1~\rm{Gyr}$ ago. The x-axis shows the mass of the subhalo (proportional to the perturbation strength, $\epsilon_\alpha$), while the y-axis is the fractional change in energy relative to the base case without any subhalo (Eq.~\ref{eq: nrg_convergence_criterion}). Solid lines represent an average over the particles in a $\sim 40~\rm{deg}$ window around the impacted region of the stream. The dashed line represent the absolute maximum of the convergence criterion over the length of the stream. We consider two scale radii for the subhalo: $r_s = 0.1~\rm{kpc}$ and $r_s = 0.01~\rm{kpc}$. As expected, all curves show a monotonic increase in the fractional change in energy as a function of subhalo mass. At fixed mass, subhalos with smaller scale radii (i.e., higher density) induce a larger change in energy. Still, for all curves in Fig.~\ref{fig: energy_error} the energy error remains less than 1 for all particles up to a subhalo mass of $\approx 10^7 M_\odot$. In practice, even if a single particle experiences an order unity change in energy this does not indicate an overall failure of our perturbation theory. It is only when the local average of particles in the impacted region suffer a change of order unity in energy that the perturbation theory fails. This is true since we are not interested in the dynamics of any individual star in a stream, but the ensemble behavior of the stars in spatial and kinematic spaces (i.e., the solid curves in Fig.~\ref{fig: energy_error}).

\subsection{Code Implementation}\label{sec: code_implementation}
We implement our code for this project in the python library \texttt{Jax} \citep{jax2018github}, using the differential equation solvers from \texttt{Diffrax} \citep{kidger2021on}. The \texttt{Jax} framework enables GPU acceleration of our code-base, and allows for all tracer-particle computations to be parallelized across several GPUs or CPUs. All dynamics calculations presented in the main text (e.g., Eq.~\ref{eq: update_rule}) will be carried out on a single GPU. The runtime of our method is discussed in Appendix~\ref{app: efficiency}, where we also distribute computation across two GPUs.

Besides computational efficiency, the major advantage of utilizing \texttt{Jax} is its native support for automatic differentiation (AD). AD provides exact derivatives of user-defined functions, up to numerical precision. It accomplishes this by iterative application of the chain-rule, and essentially breaking down complicated functions into their smaller elementary components. This means that once a gravitational potential is supplied, AD can be used to rapidly calculate the force, tidal tensor, or even higher order derivatives. Additionally, we use AD to compute the Jacobian of the particle-spray release function, discussed in \S\ref{sec: boundary_conditions}. 

While AD can even be used to obtain the perturbative corrections $\boldsymbol{q}_{\alpha 1}$ and $\boldsymbol{p}_{\alpha 1}$, doing so is computationally (and memory) intensive, in large part due to the recomputation of base trajectories for each new perturbation. A crucial detail of the perturbation equations we have derived is that all quantities are computed along the base trajectory, without any subhalo perturbations (e.g., Eq.~\ref{eq: Delta_p_int}). We exploit this feature of the problem, and compute unperturbed stream trajectories once, and interpolate to obtain continuous functions $(\boldsymbol{q}_0(t), \boldsymbol{p}_0(t))$. These continuous functions are then used to evaluate the $\boldsymbol{\epsilon=0}$ terms in Eq.~\ref{eq: update_rule}, so that the perturbation equations are integrated along a pre-saved trajectory. This routine reduces the runtime of derivative computations by a factor of $\approx 2$. AD is still used in this procedure to compute derivatives of the potential, and of the stream release function (\S\ref{sec: boundary_conditions}). The usage of AD for these functions makes our code highly modular, in that any differentiable potential and stream release function can be easily implemented without having to refactor the codebase or adjust the methodology. We expand on the discussion of AD and our usage of the \texttt{Jax} library in Appendix~\ref{app: autodiff}.

\section{Gravitational Potentials and Subhalo Statistics}\label{sec: grav_pots_and_subhalos}
Here we specify the gravitational potentials that will be utilized in subsequent sections where our model is demonstrated. We will also specify our method for sampling subhalos impacts. Importantly, any differentiable function for the galaxy's potential or the subhalo potentials can work using our method.

\subsection{Base Galaxy Model}\label{sec: galaxy_model}
All of our potential models will, at minimum, consist of a Miyamoto-Nagai disk \citep{1975PASJ...27..533M} and a spherical NFW profile \citep{1997ApJ...490..493N} to represent the dark matter halo of the galaxy. The disk mass is chosen to be $5\times 10^{10}~M_\odot$, with a scale-length of $3~\rm{kpc}$ and a scale-height of $0.25~\rm{kpc}$. The NFW halo has a scale-mass of $6\times 10^{11}~M_\odot$ and a scale-radius of $14~\rm{kpc}$. These values for the disk and the halo are similar to those adopted in $\texttt{MilkyWayPotential2022}$ from Gala \citep{gala}. 

For the time-dependent potentials, we will consider the effects of a rotating bar at the Galactic center, and the infalling LMC. For the rotating bar, we adopt the same model used in \citet{2016MNRAS.461.3835M}, which is a 3D version of the pure quadrupole model utilized in \citet{1994ApJ...420..597W,2000AJ....119..800D}. Its functional form is
\begin{equation}\label{eq: bar}
    \Phi_{\rm b}\left( R, \phi, z, t\right) = \alpha_{b} \frac{v^2_0}{3}
    \left(\frac{R_0}{R_b}\right)^3 U(r) \frac{R^2}{r^2} \cos\left(\gamma_{b}(\phi, t)\right),
\end{equation}
where $r^2 = R^2 + z^2$ is the (squared) spherical radius, and $\phi$ is the azimuthal cylindrical angle. The other quantities are $\gamma_b(\phi,t) = 2\left( \phi - \phi_b - \Omega_b t\right)$, and 
\begin{equation}
    U\left(r\right) = \begin{cases} 
      \left(r/R_b\right)^{-3}, & r \geq R_b \\
      \left(r/R_b\right)^3 - 2, & r < R_b. 
   \end{cases}
\end{equation}
For the bar parameters, we use similar values to those adopted in \citet{2016MNRAS.461.3835M} based on constraints from the Milky Way's bar. We set $v_0 = 224~\rm{km/s}, R_0 = 8~\rm{kpc}, \ \alpha = 0.1, \ R_b = 2~\rm{kpc}, \phi_b = 25~\rm{deg}$. We adopt a constant bar pattern speed of $\Omega_b = 38~\rm{km/s/kpc}$, roughly in agreement with constraints for the Milky Way's bar pattern speed (e.g., \citealt{2019MNRAS.488.4552S}). We only seek to demonstrate our method with a bar, so the exact value we adopt of $\Omega_b$ is unimportant for this work.

We will also test our model in the presence of a time-evolving LMC component. The LMC is modeled as a NFW potential, with a mass and radius relation set by observational constraints from \citet{2021MNRAS.501.2279V}. The LMC scale-mass is set to $10^{11}~M_\odot$, and the scale-radius is $8.5~\rm{kpc}$. These values are approximate, but in agreement with \citet{2021MNRAS.501.2279V,2024MNRAS.527..437V}. The LMC is not fixed, but follows its own orbit in the Galactic potential. While our perturbative model is able to handle both the motion of the LMC and the MW's response when supplied with a differentiable time-dependent potential for both components, here we will model the motion of the LMC but not the MW's response. The latter could be incorporated in our base model using, for example, basis function expansions, which are differentiable.
For the LMC's orbit, we first fix the 6D phase-space location of the LMC's center of mass today and integrate it backwards in time. For the present-day location of the LMC, we utilize the values conveniently tabulated in a script from the \texttt{AGAMA} dynamics package \citep{2019MNRAS.482.1525V}, and reproduced in Appendix~\ref{app: prog_and_LMC}. We have adjusted the MW halo mass iteratively until the LMC is on its first passage (we adopt a halo mass of $6\times 10^{11}~M_\odot$ as previously stated), though we find that it is easy to produce models where the LMC has $\sim 2$ pericentric passages when integrating back for $\sim 5~\rm{Gyr}$ (see \citealt{2024MNRAS.527..437V} for a discussion of the second-infall scenario). To evolve the LMC's potential in time, we utilize a cubic spline interpolation to adjust the center of the LMC's NFW potential. Future work can use a similar approach to account for the response of the MW disk to the infalling LMC in our base model. Regardless of the level of realism we have achieved in our potential modeling, the combination of a time-dependent bar and LMC potential will provide a demonstration that our perturbative method can be applied in aspherical, out-of-equilibrium systems.

\subsection{Subhalo Potentials}
We now specify the potential of the perturbations used throughout this work. For the subhalo potentials, we adopt a Hernquist potential of the form
\begin{equation}\label{eq: subhalo_potential}
    \Phi(\boldsymbol{r}) = -\frac{G M}{r + r_s},
\end{equation}
where $M$ is the mass of the subhalo and $r_s$ is the scale-radius. Other subhalo potentials are also applicable, including potentials with time-dependent mass and radius terms, though we use the Hernquist profile for simplicity in this work. 

Eq.~\ref{eq: subhalo_potential} has two structural parameters: scale-mass ($M$) and scale-radius ($r_s$). Recall that in our perturbative treatment, the potential of the perturbation with index $\alpha$ is the product $\epsilon_\alpha \Phi_\alpha$. Therefore, the choice of the scale-mass is somewhat arbitrary, since it is really the product $\epsilon_\alpha M$ that controls the characteristic amplitude of the force imparted by a perturbation. However, for the perturbative series in Eq.~\ref{eq: expansion_series} to remain valid, we would like to ensure that each increase in the order of $\epsilon_\alpha$ represents successively smaller contributions to the dynamics. That is,
$\epsilon_\alpha > \epsilon^2_\alpha > ... > 0$. It is therefore standard to choose a scale-mass $M$ such that the gradients of $\Phi_\alpha$ are comparable in magnitude to the gradients of $\Phi_{\rm base}$ (e.g., \citealt{f2381f32-aadb-3c8a-a473-53334cac15e7}). Then, it is valid to have $\epsilon_\alpha <<1$. For instance, if we choose $M = 10^{10} M_\odot$ in Eq.~\ref{eq: subhalo_potential} and $\epsilon_\alpha = 0.001$, then the effective scale-mass of the perturbation is $\epsilon_\alpha M = 10^7 M_\odot$. 

If we are only expanding to linear order in $\epsilon_\alpha$ then the choice of $M>0$ is in fact completely arbitrary. This is because the EOM for $q_{\alpha 1}$ and $p_{\alpha 1}$ (Eq.~\ref{eq: update_rule}) are directly proportional to $M$, so only the product $\epsilon_\alpha M$ carries any physical meaning (i.e., there is a degeneracy between $\epsilon_\alpha$ and $M$). Therefore, 
we define the effective mass
\begin{equation}
    M \xrightarrow[]{} \epsilon_\alpha M,
\end{equation}
which characterizes the amplitude of the perturbing potential and carries the same physical units as $M$. When we discuss the mass of a subhalo, we are referring to the product $\epsilon_\alpha M$. The reader does not need to do any extra work to convert the stated mass to the actual mass by multiplying by the perturbation parameter.

\subsubsection{Treatment of Subhalo Dynamics}
We now discuss our treatment of the subhalo's motion in the base potential, $\Phi_{\rm base}$. Our model assumes that each potential (both the base and perturbations) are possibly time-dependent. Implementing subhalo dynamics is therefore a user-defined choice. To remain in general terms for now, suppose that subhalo $\alpha$ follows the 3D parametric curve $\boldsymbol{r}_\alpha\left(t\right)$. For a particle at position $\boldsymbol{r}$, it feels the perturbing potential $\Phi_\alpha\left(  \boldsymbol{r} - \boldsymbol{r}_\alpha(t)  \right)$. 

Therefore we must specify the trajectory of each subhalo, $\boldsymbol{r}_\alpha(t)$. One straightforward choice is to integrate the subhalo's orbit in the base potential, and fit a spline function to its orbit. This method has the benefit of physical subhalo orbits, though wastes compute since each subhalo is only relevant for a stream when it has a close approach (within a few scale radii, $r_s$). An alternative approach, then, entails only turning the subhalo ``on" in the simulation when it approaches the stream. That is, computing only a segment of the subhalo's orbit. This method is effective, though still overly complex for most orbital segments since the majority of interactions occur over short timescales ($\approx 100~\rm{Myr}$; \citealt{2016MNRAS.457.3817S}). We therefore take the simplest, non-impulsive approach possible and model the subhalo dynamics as straight lines with constant velocity. We refer to this as the \emph{orbit tangent method}. Because the majority of stream-subhalo interactions are short, we only turn on the subhalo's potential in a $250~\rm{Myr}$ window around the impact time (to be specified in \S\ref{sec: coord_sys}). We adopt this time-scale for the same reasoning as in \citet{2016MNRAS.457.3817S}. Namely, the interaction time of any given subhalo is approximately $r_s / v_{\rm rel}$, where $v_{\rm rel}$ is the velocity of the subhalo in the stream's local frame of rest. For example, a subhalo with scale-radius of $0.5~\rm{kpc}$ and a relative velocity of $100~\rm{km/s}$ has an interaction time of $\approx 5~\rm{Myr} << 250~\rm{Myr}$. 

The approach for modeling subhalo dynamics outlined above is agnostic to the choice of $\Phi_{\rm base}$, which means that a single subhalo that impacts a stream multiple times would require multiple subhalos in the treatment we explore here. We have compared stream perturbations generated from subhalos following orbit-tangents to the same perturbations generated from subhalos following orbits. We have verified that with the exception of a few edge-cases (like the repeat-encounter discussed above), the orbit-tangent approximation does not produce quantitatively different streams compared to the spline-based approach of modeling the full orbit of the subhalo.

\subsection{Sampling Subhalo Impacts}\label{sec: sampling_impacts}
In this section we specify how subhalo impacts are sampled. In \S\ref{sec: coord_sys} we introduce a coordinate system for specifying the orbital parameters of the impacts. In \S\ref{sec: sampling_mass_func} we discuss a prescription for sampling the number, mass, and scale-radius of subhalos when considering a CDM subhalo population, and departures from CDM.

\subsubsection{Coordinate System}\label{sec: coord_sys}
\begin{figure}
\centering\includegraphics[scale=.9]{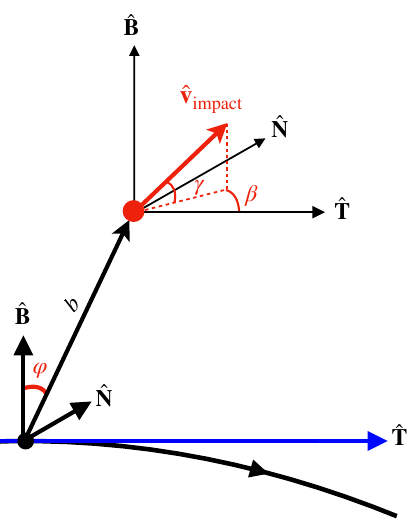}
    \caption{Illustration of the coordinate system used for sampling subhalo impacts. The black curve is a local orbit segment of a phase-space patch, and the coordinate system is anchored at the black point along this curve at time $t_{\rm impact}$. The basis vectors ($TNB$) depend on the impact time, and are determined by the velocity and angular momentum direction of the phase-space patch. The impact parameter of a subhalo is $b$, at an angle $\varphi$ from the binormal vector $\boldsymbol{\hat B}$. The other two angles $\beta$ and $\gamma$ determine the direction of the subhalo's velocity at $t_{\rm impact}$.}\label{fig: coord_sys}
\end{figure}

To specify a subhalo impact (or more generally, a fly-by), we require an impact time, impact parameter, impact velocity, impact angles, and structural parameters for the subhalo (i.e., mass and radius). To sample an impact, we start by generating an unperturbed stream, selecting the average phase-space location of particles in one region of the stream (determined according to the $\phi_1$ values of the particles, where $\phi_1$ is the angle along the stream), and integrating this average 6D coordinate back to the impact time, $t_{\rm impact}$. For a GD-1 like stream, we adopt a typical $\phi_1$ window of order $0.5~\rm{deg}$ in our tests. At $t_{\rm impact}$, the patch of stream we have selected has a mean 3D velocity. We label the direction of the velocity vector with ${\mathbf{\hat{T}}}$ (i.e., tangent to the motion). The angular momentum vector of the selected patch points in the direction ${\mathbf{\hat B}}$, which is orthogonal to ${\mathbf{\hat T}}$. The third orthogonal vector at the impact site is the normal vector, ${\mathbf{\hat N}} = {\mathbf{\hat B}} \times {\mathbf{\hat T}}$. These so-called Frenet-Serret coordinates can be constructed at any impact site, and the basis vectors $TNB$ depend on the impact time. An illustration of the coordinate system is provided in Fig.~\ref{fig: coord_sys}, where the black curve is the local orbit of the phase-space patch, and the $TNB$ vectors are labeled at a particular time. 

We label the impact parameter with $b$ in the TNB frame. This informs the distance between the subhalo and the phase-space patch at $t_{\rm impact}$. A direction is also needed, which we specify with the parameter $\varphi$ in Fig.~\ref{fig: coord_sys}, defined as the angle between the binormal ($\mathbf{\hat B}$) and the impact vector (of length $b$). The direction of the subhalo velocity at $t_{\rm impact}$ is specified by the unit vector $\mathbf{\hat v}_{\rm impact}$, which points at an angle $\beta$ from the tangent axis $\mathbf{\hat T}$, and an angle $\gamma$ above the $T-N$ plane. The speed of the subhalo at $t_{\rm impact}$ is $v_{\rm impact}$. With the impact time, distance, speed, and direction angles specified the impact is ready to be simulated once the subhalo structural parameters have been specified. 

When we randomly sample impacts, we randomly distribute impact times from the present day to an earlier time, usually the age of the stream (unless otherwise specified). For impact velocities, we sample these uniformly from $0- 2$ times the speed of the phase-space patch at $t_{\rm impact}$. This ensures sensible orbital velocities for the subhalo fly-by. We have tested uniformly sampling between $0-400~\rm{km/s}$ and find that the statistical distributions we will present in \S\ref{sec: heating} remain unchanged. We sample all impact angles uniformly as well, so that $\mathbf{\hat v}_{\rm impact}$ can point in any direction. Sampling of the impact parameter is discussed in the following section.

\subsubsection{Subhalo Mass Function and the Impact Rate}\label{sec: sampling_mass_func}
For sampling subhalo masses, we utilize a subhalo mass function (SHMF) of the form
\begin{equation}\label{eq: SHMF}
    \frac{dN}{dM} = a_0 \left(\frac{M}{m_0}\right)^{\alpha_{\rm SHMF}},
\end{equation}
where $M$ is the scale-mass of a subhalo. We adopt parameters from the Aquarius simulations, which are characterized by $\alpha_{\rm SHMF} = -1.9, \ a_0 = 3.26\times 10^{-5} M^{-1}_\odot$ and $m_0 = 2.52\times10^7 M_\odot$ \citep{2008MNRAS.391.1685S}. The number density of subhalos from the same suite of simulations is well-described by an Einasto profile, allowing us to write the differential number of subhalos per mass bin as
\begin{equation}\label{eq: subhalo_number_density}
    \frac{dn}{dM} = c_0 \left(\frac{M}{m_0} \right)^{\alpha_{\rm SHMF}}\exp\left[-\frac{2}{\alpha_r} \left( \frac{r^\alpha}{r_{-2}^\alpha} - 1 \right) \right],
\end{equation}
with $c_0 = 2.02 \times 10^{-13} M^{-1}_\odot \rm{kpc}^{-3}$, $\alpha_r = 0.678$, and $r_{-2} = 0.81 r_{\rm 200} = 199~\rm{kpc}$ \citep{2008MNRAS.391.1685S}. The variable $r$ is a galactocentric position. The normalization $c_0$ is rescaled to the virial mass of the MW (roughly $10^{12}~M_\odot$) as discussed in \citet{2016MNRAS.463..102E}.

For the scale-radius of the subhalos, we adopt the same deterministic mass-radius relation (though we will allow for variations around this relation) used in \citet{2016MNRAS.463..102E}, derived from fitting subhalos with Hernquist profiles in the Via Lactea II simulations \citep{2008Natur.454..735D}. The fit is
\begin{equation}\label{eq: r_of_m}
    r_s = 1.05~\rm{kpc} \left( \frac{M}{10^8 M_\odot}\right)^{0.5}.
\end{equation}

Eq.~\ref{eq: subhalo_number_density} provides the key formula to determine the expected number of subhalos in a given mass and radius bin at a specific spatial location. Under this model, the number of impacts for a given stream can be treated as a Poisson process, with a rate that was derived in \citet{2011ApJ...731...58Y} by treating a stream as a cylinder with radius $b_{\rm max}$. The impact formalism was revisited in \citet{2016MNRAS.463..102E} under Gaussian assumptions for the relative velocity distribution of subhalo encounters in the stream's frame of rest. The Poisson impact rate for a given stream is
\begin{equation}\label{eq: Num_impacts}
    N_{\rm impact} = \sqrt{\frac{\pi}{2}} l_{\rm obs} b_{\rm max} n(r_{\rm avg}) \sigma t,
\end{equation}
where $l_{\rm obs}$ is the angular length of the stream at the present-day, $n(r_{\rm avg})$ is the number density of subhalos at the average spherical radius of the stream's orbit (Eq.~\ref{eq: subhalo_number_density}), $\sigma$ is the velocity dispersion of the subhalo population, and $t$ is the dynamical age of the tidal tails. For the subhalo velocity dispersion, we use the same value adopted in \citet{2016MNRAS.463..102E} of $180~\rm{km/s}$, derived from the Via Lactea II simulation suite \citep{2008Natur.454..735D}. While the parameter $b_{\rm max}$ is a cylindrical radius around the evolving stream, it can also be thought of as the maximum impact parameter. Following \citet{2016MNRAS.463..102E, 2017MNRAS.466..628B}, we sample impacts parameters uniformly from 0 to $\sim 5$ times the scale radius of the subhalo (before applying perturbation theory to the scale-radius). \citet{2017MNRAS.466..628B} provides a series of convergence tests for this choice. While the small-scale structure of a stream is well-converged under the sampling range of $[0,5]r_s$, the track of a stream can shift in response to more distant perturbations. It is beyond the scope of this work to develop a different sampling scheme for distant subhalo encounters, though future work could consider the effect of a large number of long-range perturbations.  

In \citet{2011ApJ...731...58Y}, several approximations are made to derive Eq.~\ref{eq: Num_impacts}. The most important, perhaps, is the assumption of roughly circular orbits. For streams on appreciably radial orbits, Eq.~\ref{eq: Num_impacts} will tend to under-estimate the number of subhalos entering the stream's cylinder. However, Eq.~\ref{eq: Num_impacts} provides a good measure of the average impact rate since most streams will spend the majority of their time away from the inner galaxy where the subhalo number density is expected to be higher \citep{2011ApJ...731...58Y,2016MNRAS.463..102E, 2017MNRAS.466..628B}. Additionally, Eq.~\ref{eq: subhalo_number_density} does not account for the disruption of subhalos in the inner galaxy. More physically motivated statistics for the subhalo population (e.g., \citealt{2024arXiv240611989M}) can be incorporated in future work.

\begin{figure*}
\centering\includegraphics[scale=.6]{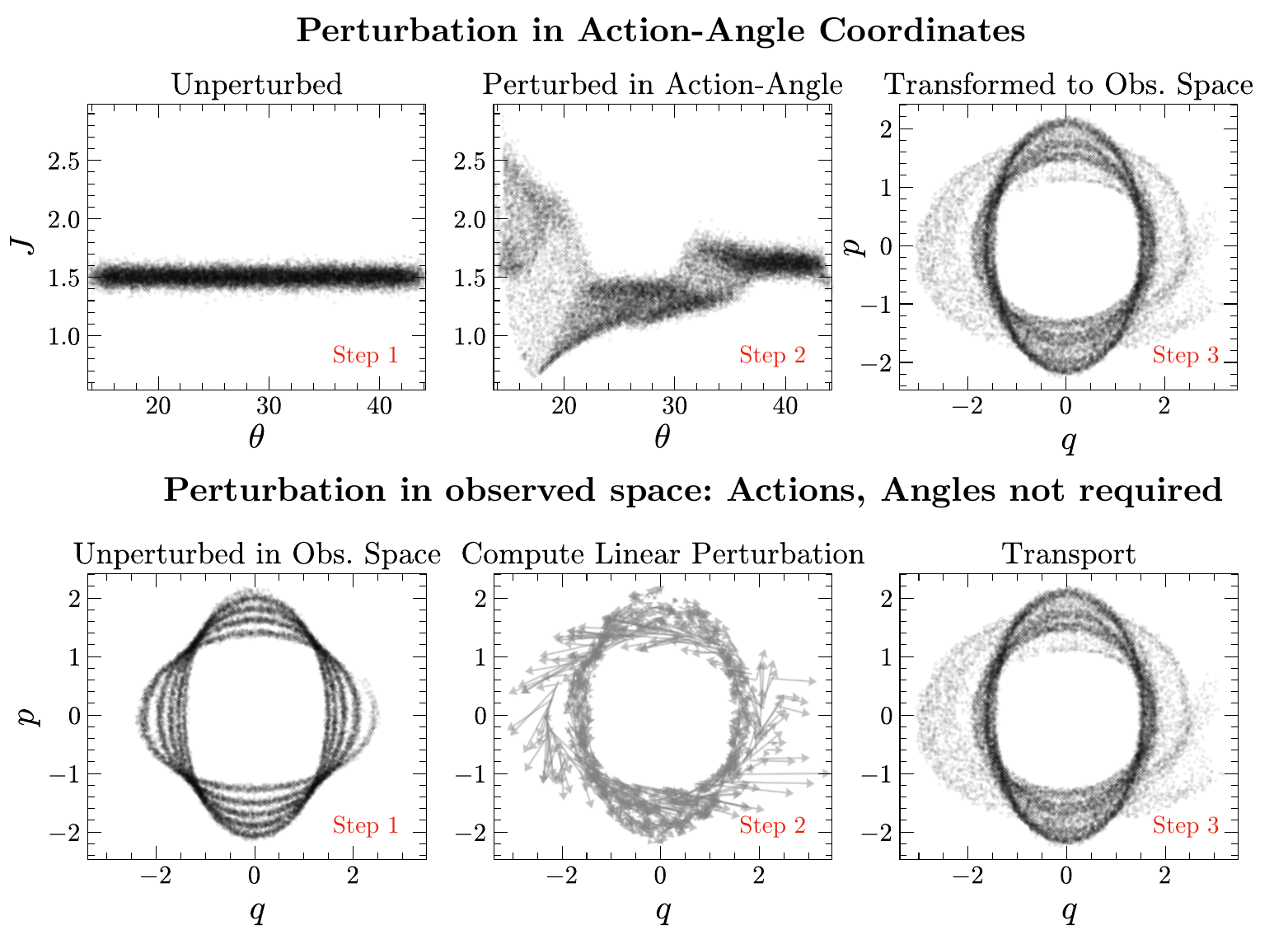}
    \caption{Schematic illustration of perturbation theory in action-angle coordinates (top) versus applying the same perturbation theory entirely in observable coordinates (bottom). In action-angle coordinates, one may start with a distribution of particles (step 1) that then receives an impulse $\Delta J$, thereby distorting the action and angle distribution (step 2). If the transformation from $(\theta, J)$ to the observable $(q,p)$ can be estimated, then one obtains the phase-wrapped particle distribution in the top right panel. This final step hinges on a transformation that is often approximate, or completely inaccessible for more complicated potentials. In the bottom row, we start with the same initial particle distribution shown in the top left, though this time it is expressed in observable coordinates $(q,p)$. We use Eq.~\ref{eq: update_rule} to obtain the linear response of the particles to the perturbation (step 2). The particles are then transported along the linear response vectors (step 3). We obtain the same final perturbed system without ever requiring a transformation to the unobserved latent space of actions and angles.     }
    \label{fig: method_workflow}
\end{figure*}

\section{Demonstration of the Method}\label{sec: demonstration}
In this section we demonstrate our method, and compare predictions from our model to simulated streams with subhalo impacts. We first provide a conceptual illustration of the method in \S\ref{sec: conceptual_example} for a simple harmonic oscillator potential. In \S\ref{sec: single_impacts} we demonstrate the method on a Pal-5 like stream with a single subhalo impact. The case of multiple impacts is demonstrated in \S\ref{sec: multiple_impacts}. 

We validate our method by comparing the predictions from our model to a direct simulation. For clarity, we define what we mean by simulation and model below:
\begin{itemize}
    \item \textbf{Simulation:} Test-particle stream (particle-spray based) with subhalo perturbations. Orbits are integrated directly in the joint potential of the Galaxy and subhalos, without approximating the dynamics. 

    \item \textbf{Model:} Linear perturbation theory applied to a test-particle stream (developed in \S\ref{sec: method}). The dynamical effect of each subhalo on the stream is modeled at linear order in the subhalo mass. 
\end{itemize}

\subsection{Conceptual Example}\label{sec: conceptual_example}
Here we will demonstrate our perturbative model by applying the method to the simple harmonic oscillator (SHO). We carry out this exercise to demonstrate our methodology on a simple, analytically tractable simple. Additionally, we will demonstrate that the end result of our method is identical to canonical perturbation theory in action-angle space, but without requiring the transformation $(\boldsymbol{q}, \boldsymbol{p}) \leftrightarrow (\boldsymbol{\theta}, \boldsymbol{J})$. For the SHO discussed in this section, we provide an analytic derivation of the perturbation equations in Appendix~\ref{app: SHO}.

We start by showing the result of canonical perturbation theory in actions and angles, and then compare to perturbation theory in real-space. This exercise is illustrated in Fig.~\ref{fig: method_workflow}. We start with an unperturbed distribution of particles that is reminiscent of a stream: a small spread in action distributed over a range in angle. The action-angle representation of the stream is shown in the top left panel, while the real-space representation is shown in the bottom left panel. A cubic perturbation of the form $\Phi_{\rm pert} \propto q^3$ is temporarily applied to the system, representing a global perturbation. In the top row (step 2), we show the perturbed stream which has suffered a change in actions due to the perturbing potential. We may transform the perturbed particle distribution to real-space when the mapping $(\theta, J) \xrightarrow{} (q,p)$ is known, giving us the real-space picture in the top right panel (step 3).

The workflow of our method is shown in the bottom row of Fig.~\ref{fig: method_workflow}. In our case, we start with the particle distribution in real-space (step 1). We then use Eq.~\ref{eq: update_rule} to estimate the linear response of the system to the perturbation, by applying perturbation theory to the individual particles. The linear response vectors, $dq/d\epsilon$ and $dp/d\epsilon$, are shown as gray arrows in step 2. The direction of each vector is determined by the properties of the perturbing potential, and the length of each vector quantifies the strength of the perturbation. To obtain the perturbed particle distribution, we simply transport the unperturbed particles along the linear response vectors to obtain step 3. Crucially, the result of perturbation theory in actions and angles is the same as in real-space positions and velocities: step 3 in the top row is identical to step 3 in the bottom row. This discussion is generalized in mathematical terms in Appendix~\ref{app: canonical_theory}. 

It is a useful feature that perturbation theory in action-angle coordinates can be equivalently expressed entirely in real-space. By applying perturbative methods in real-space, we never require transformation to action-angle coordinates, allowing us to consider significantly more complex dynamical systems for which transformation to action-angle coordinates is not straightforward with current methods. Importantly, it is not expected that the galaxy, nor streams, can be accurately characterized entirely in terms of actions and angles, in large part due to the impact of the infalling LMC and the MW's response to the merger \citep{2022ApJ...939....2A, 2023MNRAS.518..774L, 2024MNRAS.532.2657B}. Because our method supports time-dependent and arbitrary mass distributions, the flexibility of our framework allows for substantial freedom in modeling streams, the galaxy, and perturbations.

\subsection{A Single Subhalo Impact}\label{sec: single_impacts}
 We now demonstrate our method using a Palomar 5 (Pal 5) like stream with a single subhalo impact. We focus on relatively massive and dense perturbers. If these can be modeled successfully, then perturbations due to lower density subhalos will also be captured accurately by our model (i.e., the linear approximation only improves for lower mass subhalos). For the perturber, we utilize a spherical Hernquist profile with a scale-mass of $10^7 M_\odot$ and a scale-radius of $0.14~\rm{kpc}$ ($\sim 50\%$ smaller $r_s$ than CDM expectations from Eq.~\ref{eq: r_of_m}). The impact occurs $\approx 530~\rm{Myr}$ ago, with an impact angle that is 45~\rm{deg} to the stream's track and in the stream's local orbital plane. The base potential in this section consists of a disk and spherical NFW potential (parameters defined in \S\ref{sec: galaxy_model}). We model the joint effects of the bar and subhalo impacts on a Pal-5 like stream in \S\ref{sec: Pal5_and_Bar}.

 \begin{figure*}
\centering\includegraphics[scale=.56]{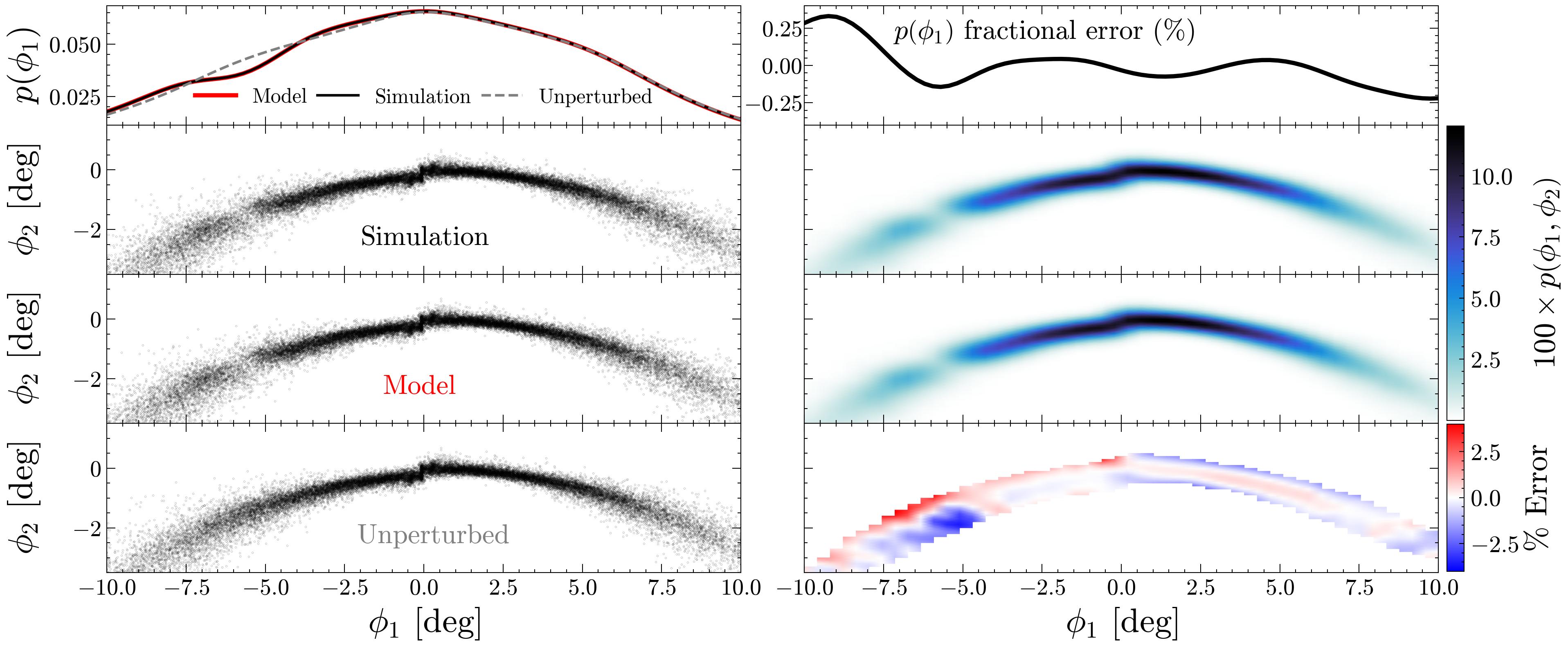}
    \caption{Demonstration of the method for a Pal 5 like stream. We inject a $10^7~M_\odot$ subhalo with $r_s = 0.14~\rm{kpc}$, and model the interaction as a linear perturbation. We compare the resulting stream to the equivalent simulation generated without approximation. Left column: the subhalo clears out a gap around $\phi_1\approx -5~\rm{deg}$. The $\phi_1$ density of the linear perturbation stream (model, red) and the simulation (black) are shown in the top left panel, with the unperturbed density provided for reference (dashed). The particle distributions of the streams are shown in $(\phi_1,\phi_2)$ on-sky coordinates in the left column. Right column: the top right panel shows the fractional difference in $\phi_1$ densities between the linear perturbation model  and the simulation. Below this panel, we show 2D densities for each stream, with the $\%$ error between the model and simulation (in 2D) shown in the bottom right panel. Errors are at the 1\% level and below.}
    \label{fig: single_impact}
\end{figure*}

Fig.~\ref{fig: single_impact} provides an illustration of the present-day stream in a Pal 5 frame. The left column shows three models: the first is the simulated stream with a subhalo impact, second is a stream with our linear perturbation applied, and the third panel is an unperturbed stream for reference (no subhalo impact). The top panel shows the particle density in $\phi_1$ along the stream for the different models, represented with a kernel density estimate. Qualitatively, the morphology of the linear model stream and the simulation stream appear very similar.

A quantitative determination of errors is provided in the second column, where the fractional error in the $\phi_1$ density is shown in the top panel, and a kernel density estimate (KDE) is provided for both the simulation and model streams in 2D.\footnote{Throughout this work, the KDE bandwidth is determined using a cross-validation technique: we try different trial bandwidths and evaluate the likelihood of a held-out sample. We choose the bandwidth that maximizes the likelihood. } The bottom right panel illustrates the difference between the 2D density estimate for the model and simulation. Spatially resolved residuals are at the $\lesssim 1$\% level (bottom), and even lower when considering only the $\phi_1$ density. While this is only one example, the magnitude of the fractional errors seen here is typical in our tests, as we will demonstrate.

\begin{figure*}
\centering\includegraphics[scale=.6]{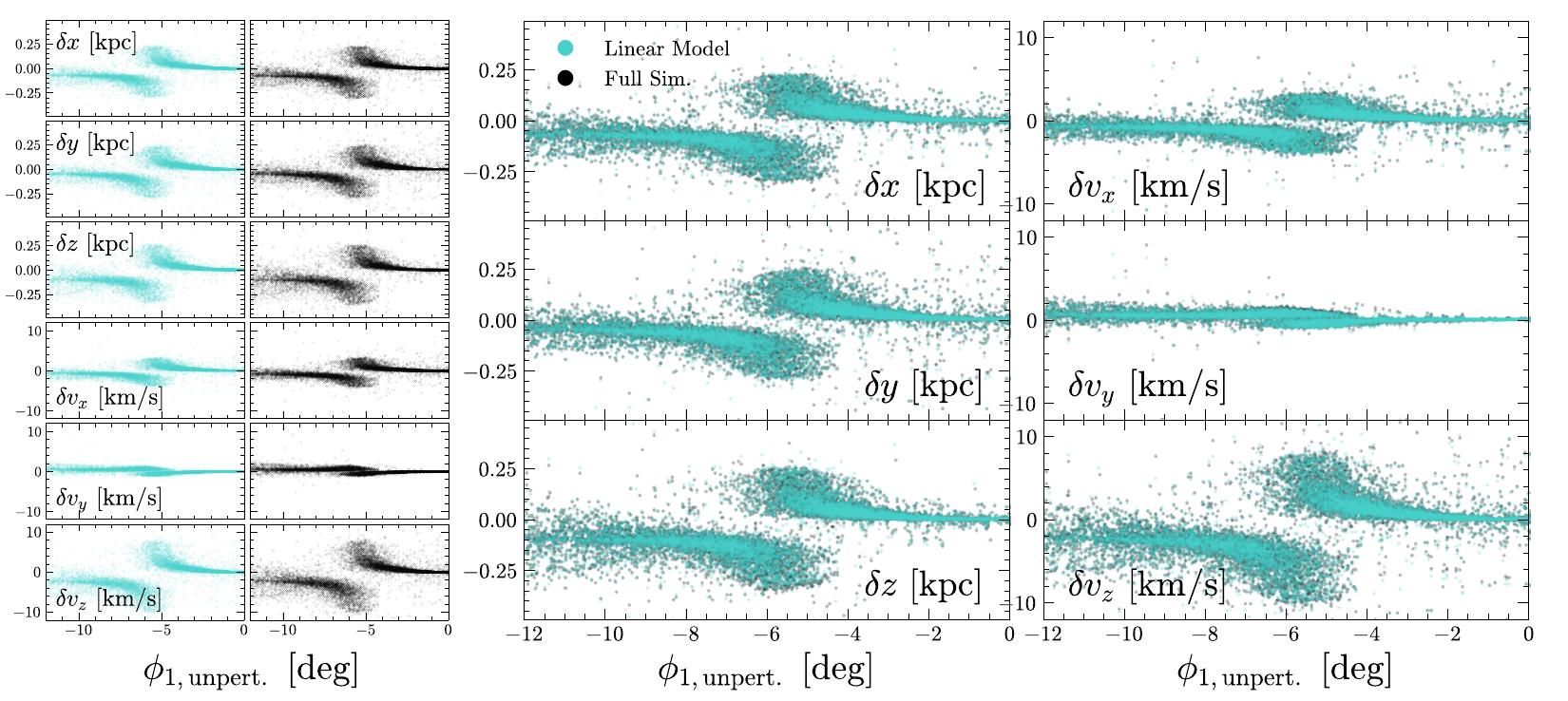}
    \caption{Here we provide a more rigorous illustration of how the perturbative linear model (cyan) compares to the simulation (black) in 6D cartesian phase-space. Each panel shows a vector component of $\delta\left(\boldsymbol{x}, \boldsymbol{v}\right)$, where $\delta$ is the difference---on a per-particle basis---between the perturbed stream and the unperturbed stream. This figure therefore shows the response of the stream to the $10^7~M_\odot$ subhalo, since any non-zero value for each particle is entirely due to the stream's interaction with the subhalo. The x-axis is the unperturbed $\phi_1$ angle for each particle. The small panels on the left show a side-by-side view of the linearly perturbed and simulation streams in each dimension, while the right larger panels overplot the two particle distributions.  }
    \label{fig: delta_Kinematic_single}
\end{figure*}

In Fig.~\ref{fig: delta_Kinematic_single} we visualize the performance of our model in 6D phase-space. In this figure, we plot the difference, on a per-particle basis, between a perturbed stream and unperturbed stream in each of the 6 phase-space dimensions. Differences between the simulations are labeled with the symbol $\delta$. We perform this exercise both for our stream model (cyan points) and the simulation (black points). A distinctive ``1/x" feature is seen in the different phase-space projections. The shape of the perturbation signal is due to some particles getting boosted to higher energies (lower frequencies) and others being dragged to lower energies (higher frequencies) due to the subhalo fly-by \citep{2011ApJ...731...58Y, 2016MNRAS.457.3817S, 2015MNRAS.450.1136E}. In all dimensions, the linear model performs extremely well. The only appreciable difference between the model and simulation is the location of particles in the outskirts of the bulk of the stream. These outlier particles represent stars that suffered the largest amount of scattering due to the subhalo impact, or those that experience the largest $\Vert \nabla \Phi_\alpha \Vert$. Our model successfully ejects these particles from the stream, since the derivative $d\boldsymbol{p}/d\epsilon_\alpha$ is proportional to the integral of the subhalo's force, $\Vert \nabla \Phi_\alpha \Vert$ (Eq.~\ref{eq: update_rule}). However, because the force imparted by the subhalo is large for these particles, their long-term orbital evolution is not reliably captured. This is because the orbital evolution of perturbed debris is governed by the tidal tensor of the base potential in our linear treatment. The tidal tensor provides an excellent approximation for stars that remain near the stream, but for those that are perturbed to appreciably different orbits the base tidal tensor no longer provides an accurate characterization of the acceleration field felt by the star. Still, we make the argument that our results are robust to the misrepresentation of these stellar obits, since the flux of stars \emph{leaving} the stream is reliably captured in our model, and most analyses use stars still associated with the stream to constrain substructure.   

\subsubsection{Time-Evolution of Perturbed Stream}

In order to demonstrate how our model evolves a perturbed stream forward in time, in this section we compare the time evolution of a stream in our model to the simulation. This exercise highlights our model's ability to deal with the realistic phase-mixing of perturbed debris.

We again utilize a Pal 5 like stream for this test, and evolve it for $5~\rm{Gyr}$. The stream encounters a subhalo at a lookback time of 2.1~\rm{Gyr}. This is an older interaction than the one we consider in Fig.~\ref{fig: delta_Kinematic_single}, allowing us to visualize the phase-mixing process over $\rm{Gyr}$ time-scales. We use a massive and dense subhalo to demonstrate that our model remains accurate even in this regime. The subhalo has a mass of $5\times 10^7 M_\odot$ and a scale-radius of $0.05~\rm{kpc}$ (the CDM $r_s(M)$ expectation is $\sim 0.7~\rm{kpc}$ for a subhalo of this mass). The impact parameter is $b = 4~\rm{pc}$, and the impact geometry is $\varphi = \gamma = -23~\rm{deg}, \ \beta = 83~\rm{deg}$. The angle parameters were randomly chosen to generate a grid of test-case streams, though we set $\varphi$ equal to $\gamma$ for simplicity. 

\begin{figure*}
\centering\includegraphics[scale=.56]{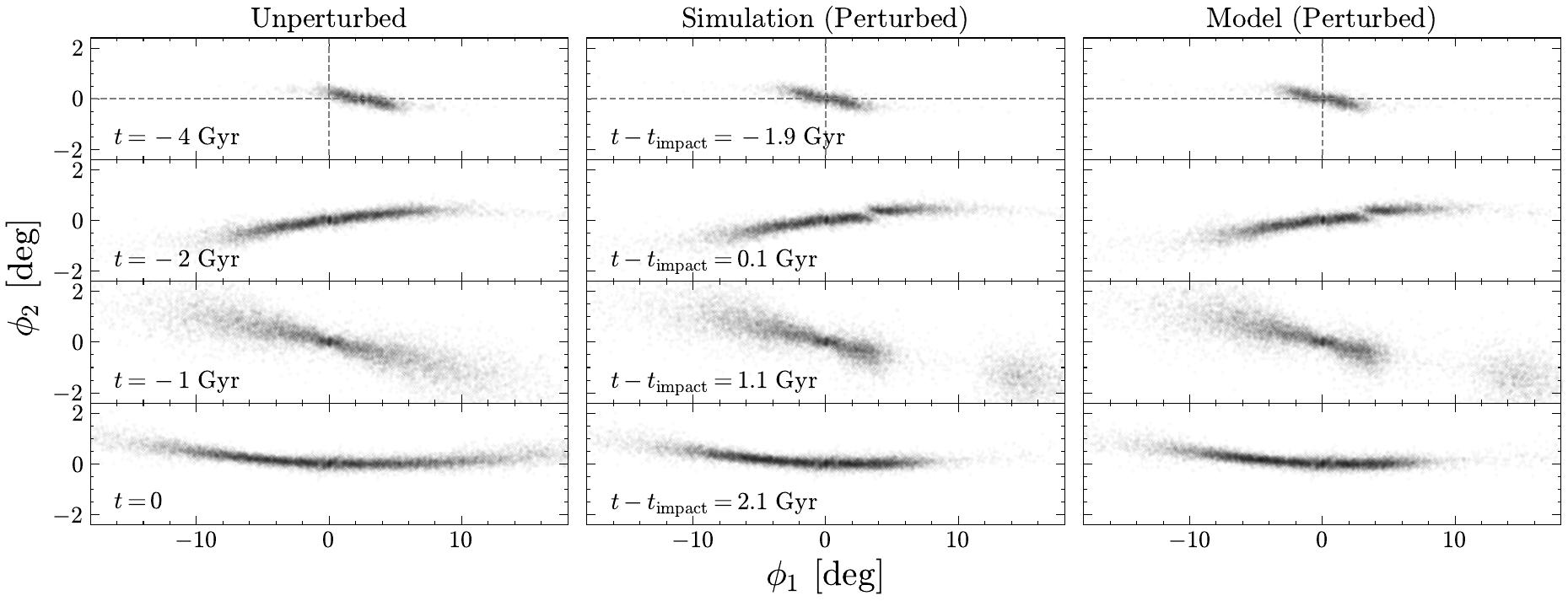}
    \caption{Time-evolution of a Pal 5 like stream with a $5\times 10^7~M_\odot$ subhalo impact. The left column shows the unperturbed stream at four equally spaced snapshots from $4~\rm{Gyr}$ ago to the present-day. The middle column shows a simulation of the stream and subhalo interaction. The time since the subhalo impact is listed. Around $ 100~\rm{Myr}$ after the impact a small spur-like feature begins to form, and  rapidly phase-mixes over the course of a Gyr to clear out a wide gap. Around $2~\rm{Gyr}$ after the impact particles from the impacted region have departed from the main-body of the stream.
    The right column shows the same information as the middle, but as predicted by our linear model. The subhalo perturbs the progenitor's orbit, 
    so the unperturbed and perturbed models do not share the same progenitor location (dashed lines) at $t=-4~\rm{Gyr}$. Perturbations to the progenitor's orbit are approximated with linear perturbation theory in our model. }
    \label{fig: time_evo_Pal5}
\end{figure*}

Four snapshots of the unperturbed and perturbed stream are provided in Fig.~\ref{fig: time_evo_Pal5}. The left column shows the unperturbed stream, with the time of the snapshots listed. The middle column shows the perturbed stream from the simulation, generated without the use of perturbation theory. The time since the impact listed. The right column shows the perturbed stream from our linear model. This figure clearly demonstrates the ability of our model to capture the phase-mixing process of a perturbed stream. Fig.~\ref{fig: time_evo_Pal5} also shows that the progenitor location (marked with dashed lines) of a perturbed stream differs from that of the unperturbed stream at $t = -4~\rm{Gyr}$. This is because we impose a boundary condition on the present-day location of the progenitor, and fix its final phase-space location (\S\ref{sec: boundary_conditions}). Subhalo impacts can perturb the progenitor's orbit, so the initial location of the progenitor is adjusted from the unperturbed case at linear order in the perturbation parameter. We find that failure to account for perturbations to the progenitor's trajectory can lead to significant disagreements between the model and simulation, since the initial condition of every stream particle depends on the progenitor's orbit \citep{2024arXiv240402953H}.

Fig.~\ref{fig: time_evo_Pal5} also illustrates the performance of our model in capturing the change in morphology of a stream over time. In particular, the initial bifurcation of the stream $0.1~\rm{Gyr}$ after the impact is driven by the velocity injected by the subhalo fly-by, while the subsequent growth of the gap is governed by the tidal tensor of the base potential ($\mathbf{T}$ in Eq.~\ref{eq: update_rule}). Even for a significant perturbation like the one we consider here, a comparison between the model and simulation shows no appreciable disagreement. For such a massive perturbation, an entire segment of the stream at $\phi_1 \gtrsim 8~\rm{deg}$ phase mixes away from the main stream, producing truncated asymmetric tidal tails in both the model and simulation.

\subsubsection{Incorporating the Progenitor's Self-Gravity}
In \S\ref{sec: self_grav_method} we provide a formalism to incorporate self-gravity of the progenitor in our model. The progenitor's self-gravity can redistribute mass along tidal tails \citep{2012MNRAS.420.2700K}, and can therefore be a useful feature to incorporate in realistic stream models. Here we validate this component of our formalism.

We consider a GD-1 like stream disrupting for $4~\rm{Gyr}$, with a progenitor mass of $10^4 M_\odot$. The progenitor is modeled as a Plummer sphere with a scale-radius of $10~\rm{pc}$. Any differentiable function for the mass, $M(t)$, can be easily incorporated through the tidal tensor of the progenitor (Eq.~\ref{eq: tidal_prog}), though here we elect to use a constant mass for the progenitor since we are only aiming to illustrate the effect of self-gravity. For simplicity, the base potential consists of a static disk and NFW component (parameters defined in \S\ref{sec: galaxy_model}). We apply our model to more realistic potentials in \S\ref{sec: applications}.

\begin{figure}
\centering\includegraphics[scale=.45]{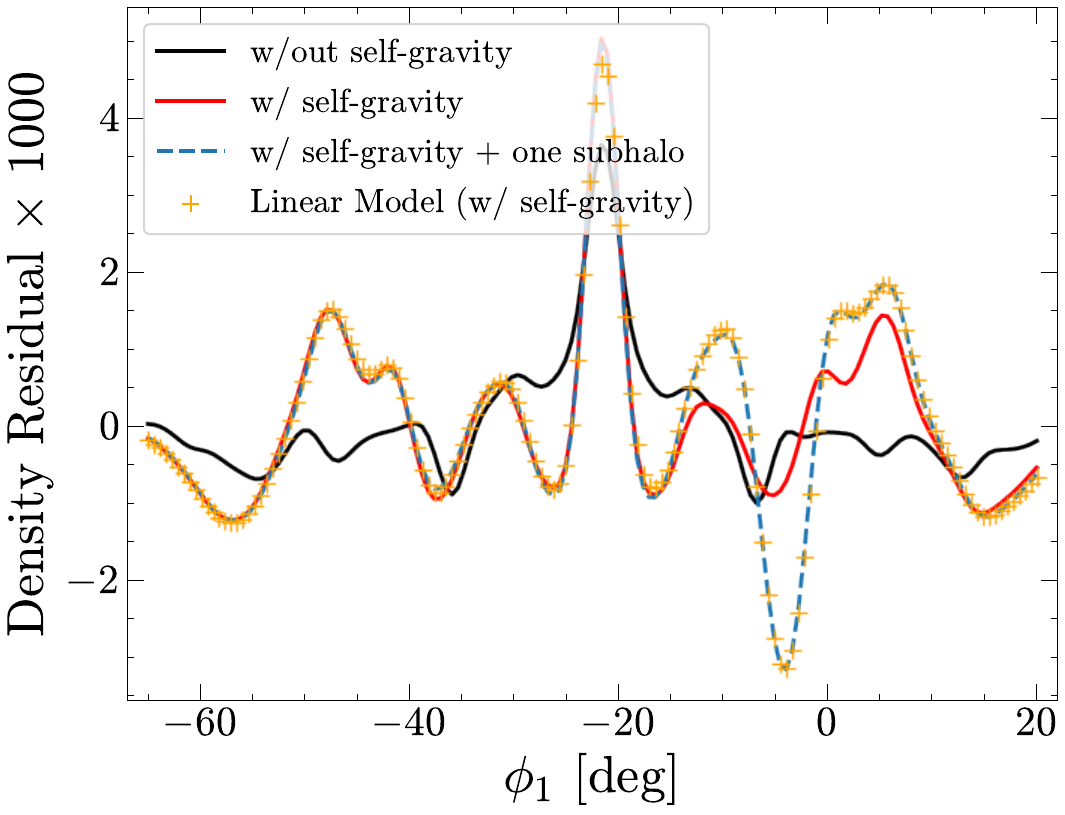}
    \caption{Impact of the progenitor's self-gravity on the density structure of a GD-1 like stream. The y-axis is the density residual, which shows small-scale density fluctuations for models with and without self-gravity (red and black, respectively). The spike around $\phi_1 = -20~\rm{deg}$ is the location of the progenitor. The other density fluctuations are due to epicyclic over and under densities. The dashed blue line and orange plus symbols show stream models with a subhalo impact at $\phi_1=-5~\rm{deg}$, with the former generated by a full simulation, and the latter by our linear approximation.
    }
    \label{fig: self_gravity}
\end{figure}

For comparison, we test  three scenarios: the first is a GD-1 like stream generated without self-gravity, the second is the same stream with self-gravity, and the third is the stream with self-gravity plus a subhalo impact. For the third case, we validate our linear model with self-gravity against the equivalent simulation generated without the use of linear perturbation theory. The density structure of the three streams is shown in Fig.~\ref{fig: self_gravity}. To highlight density fluctuations along each stream, we use a density residual. The density residual is defined as the difference between a sharp KDE model of the stream and a coarse KDE model with a few times the bandwidth (i.e., an unsharp filter).

The black curve in Fig.~\ref{fig: self_gravity} shows the simplest of the three scenarios, where we generated a GD-1 like stream without accounting for the progenitor's self-gravity after particles are released. Importantly, the progenitor mass is still used to determine the initial velocity dispersion that particles are released with (\citealt{2015MNRAS.452..301F}). There are density variations even in this scenario, with the largest at $\phi_1 \approx -20~\rm{deg}$ near the progenitor's location due to the larger number of particles near the progenitor. Other density variations are due to particles oscillating around their guiding centers with similar, albeit slightly different, radial, azimuthal, and vertical frequencies. Patterns of epicyclic overdensities have been detected in the tidal tails of Pal 5 \citep{2016MNRAS.460.2711T, 2017MNRAS.470...60E}.

Incorporating self-gravity of the progenitor produces a non-linear response along the tidal tails of the stream, which is why we must incorporate the self-gravity of the progenitor in our base model, $H_{\rm base}$, rather than the linear model. The effect of the progenitor's self-gravity is illustrated by the red curve in Fig.~\ref{fig: self_gravity}. Here we can see that the epicycle spikes typically grow in amplitude, and shift in $\phi_1$ compared to the black curve without self-gravity. This is due to the tidal field of the progenitor itself changing the release conditions of the particles as they enter the tidal tails. In particular, self-gravity can have the effect of launching stars with higher release velocities, with larger amplitude oscillations around the orbital guiding center \citep{2012MNRAS.420.2700K, 2014MNRAS.445.3788G}. 

Finally, we take the model for the stream with self-gravity, and inject a $10^6 M_\odot$ subhalo around $\phi_1 \approx -5~\rm{deg}$, which intersected the stream around $0.8~\rm{Gyr}$ ago. We add the effect of the subhalo both using direct simulation (i.e., non-linear effects are accounted for, blue dashed line) and using our linear model with self-gravity of the progenitor (orange + symbols). The agreement between the model and simulation is excellent, with virtually no noticeable differences in the $\phi_1$ density distribution.

In this example, the subhalo induced gap at $\phi_1 = -5~\rm{deg}$ is a few times deeper than the epicycle ``gaps", but, subhalos with smaller masses, or large impact parameters, can produce similarly shallow density variations. However, epicycle density variations tend to be symmetrical with respect to the progenitor, or at least predictable, thus providing a way to identify their common origin. It may be possible to use kinematic signatures of subhalo impact to break degeneracies with epicycle variations, which we will explore in future work. 

In Appendix~\ref{app: N_body}, we compare particle spray streams with and without self-gravity to a $N-$body stream model. We find that the particle-spray model without self-gravity produces the most similar stream. This is because the release function for a particle spray model is tuned to $N-$body simulations that already incorporate self-gravity by construction. We therefore elect to use standard particle-spray-based models for the unperturbed stream for the remainder of this work, though methods for stream generation that depend on self-gravity are applicable with our model.

\subsection{Multiple Subhalo Impacts}\label{sec: multiple_impacts}
In this section we validate our model in the regime of many subhalo impacts. We consider two scenarios: the first consists of many impacts from subhalos that follow the fiducial CDM mass-radius relation from Eq.~\ref{eq: r_of_m} (\S\ref{sec: many_CDM_impacts}). The second scenario represents a departure from CDM, where we have more impacts than expected with denser subhalos (\S\ref{sec: deviations_from_CDM}). 
We explore the second case to demonstrate that our model can characterize streams outside of CDM expectations. In both cases we compare our linearized stream model to the full simulation. The smooth potential of the Galaxy in this section consists of a disk, spherical NFW halo, and time-dependent LMC model (all parameters are defined in \S\ref{sec: galaxy_model}).

\subsubsection{Many Impacts with CDM Subhalos}\label{sec: many_CDM_impacts}

\begin{figure}
\centering\includegraphics[scale=.52]{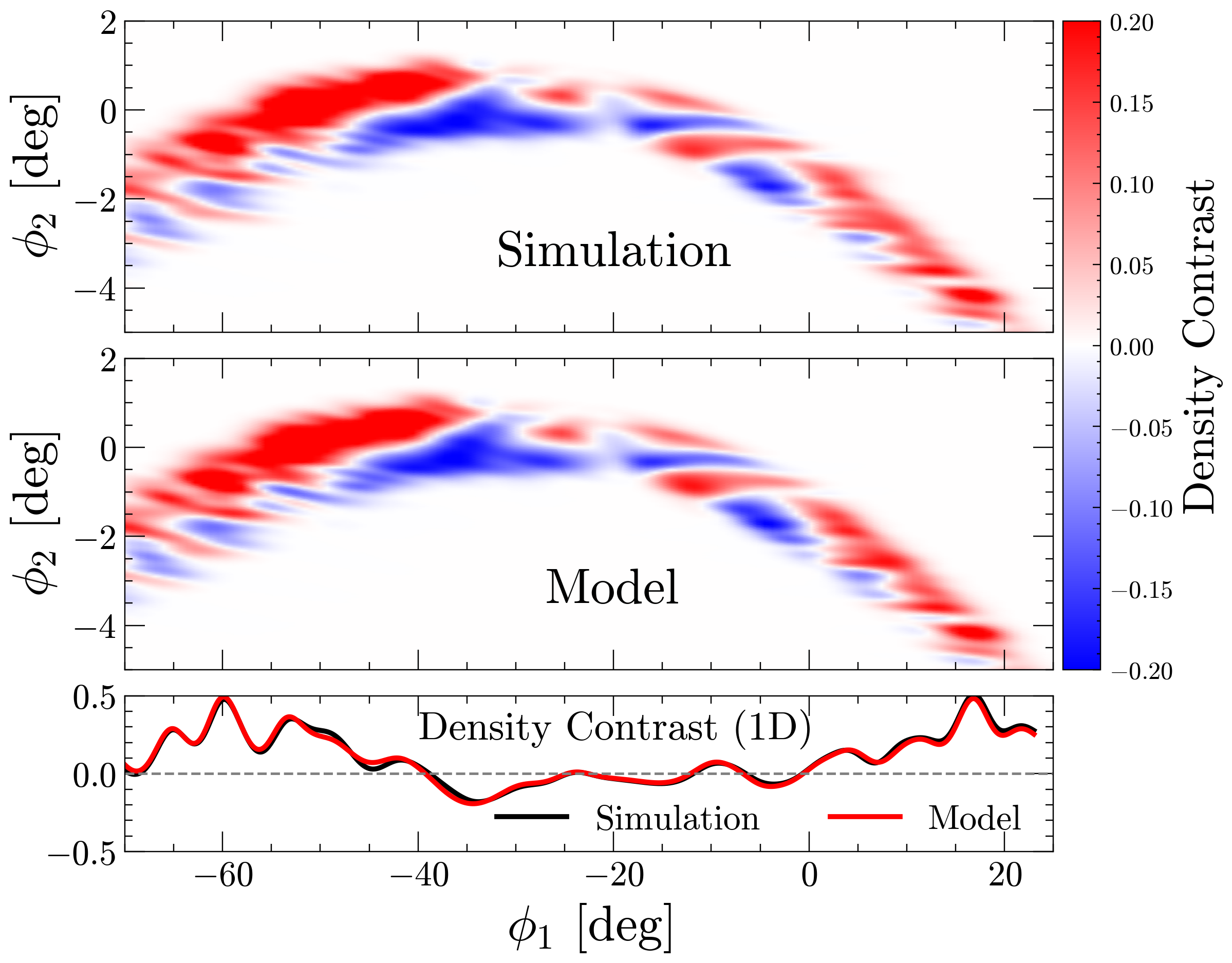}
    \caption{The top two panels show a 2D density contrast in $(\phi_1, \phi_2)$ coordinates for the simulation and model. Density fluctuations are induced by $67$ subhalo fly-bys generated from the CDM mass-spectrum. Red regions represent overdensities, blue regions are underdensities. Our model is able to characterize 2D changes in stream morphology. The bottom panel shows the 1D density contrast for the simulation (black) and model (red). }
    \label{fig: 2CDM_density_constrast}
\end{figure} 
We sample from the fiducial CDM SHMF (Eq.~\ref{eq: SHMF}) in a mass range of $[10^5-10^7]~M_\odot$. The Poisson rate for the number of impacts in this mass range is roughly $63$ \citep{2011ApJ...731...58Y,2016MNRAS.463..102E,2017MNRAS.466..628B}. The sampled number of impacts we visualize in this section is 67. The subhalo potentials are Hernquist profiles, and we use the fiducial CDM mass-radius relation from Eq.~\ref{eq: r_of_m}.
\begin{figure*}
\centering\includegraphics[scale=.61]{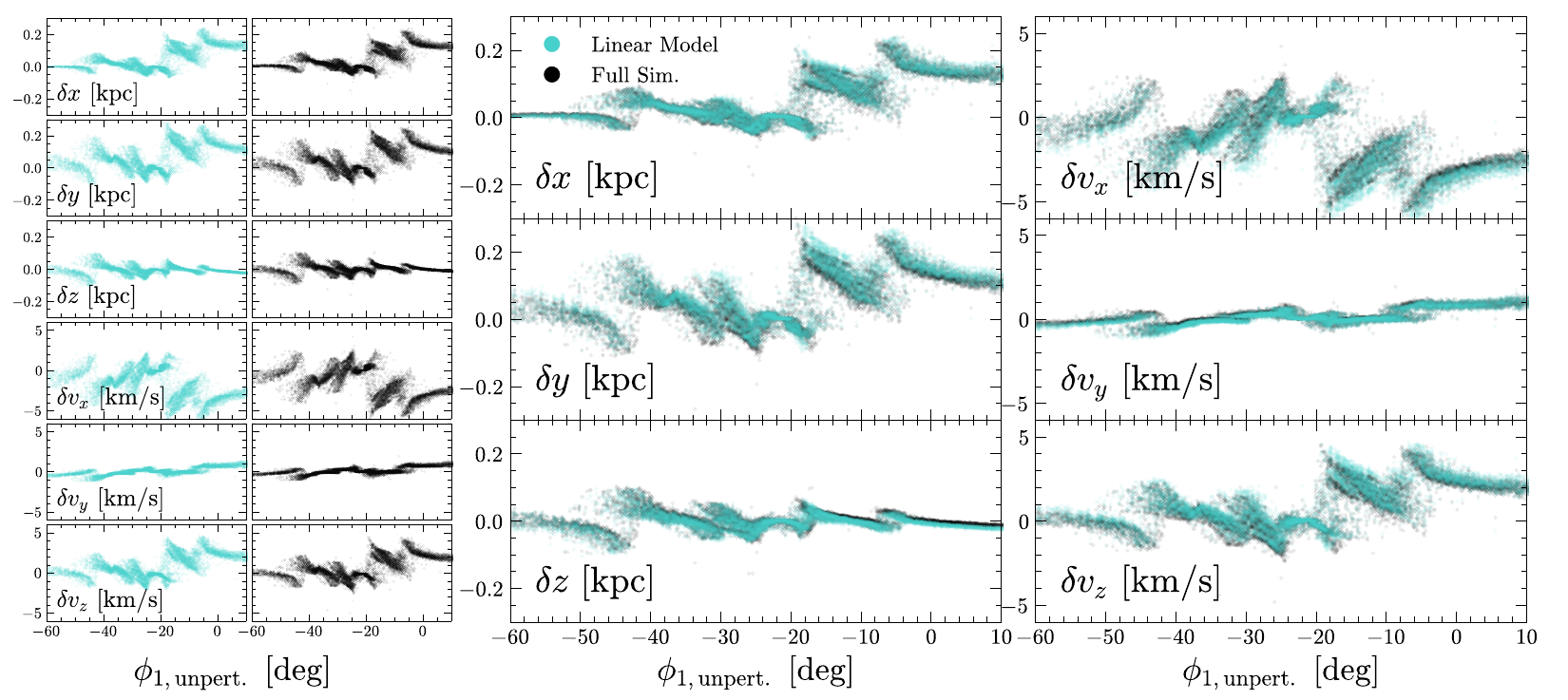}
    \caption{Same as Fig.~\ref{fig: delta_Kinematic_single}, but for the case of 15 massive and dense subhalos impacting a GD-1 like stream. To generate this figure, we apply perturbation theory in the subhalo mass \emph{and} radius (that is, the root subhalo radius is different from the truth. We apply a perturbation $\Delta r_s$ and compare the resulting particle distributions).}
    \label{fig: ManyMajorImpacts}
\end{figure*}

To assess the performance of our linear model we visualize the stream in observable coordinates in Fig.~\ref{fig: 2CDM_density_constrast}. The top two panels show a 2D version of the density contrast (relative again to an unperturbed model). The 2D contrast is fit similarly to the 1D contrast, using smooth KDEs of the perturbed and unperturbed streams (described in \S\ref{sec: single_impacts}). The 1D density contrast is shown in the bottom panel for the simulation (black) and model (red). From Fig.~\ref{fig: 2CDM_density_constrast}, the linear model does an excellent job of capturing the more informative summary statistic of a 2D density contrast, with the amplitude of peaks and troughs for the linear model showing strong agreement with the simulation. In Appendix~\ref{app: additional_validation} we provide a comparison between the model and simulation in all 6 phase-space dimensions. The spatial and kinematic distribution of the linear perturbation stream is in extremely good agreement with the full simulation.

Now that we have established that the model and simulation agree, we discuss the density features in Fig.~\ref{fig: 2CDM_density_constrast}. The red regions in Fig.~\ref{fig: 2CDM_density_constrast} (2D contrast) represent overdensities, while the blue regions are underdensities. The arm of the stream with $\phi_1 < -20$ has shifted in the positive $\phi_2$ direction. The blue under-dense region around of the stream around $\phi_1 = -15~\rm{deg}$ is notably most prominent as a $\phi_2$ under-density. That is, at every $\phi_1$ slice along the stream there is a $\phi_2$ density, $p(\phi_2 | \phi_1)$. Because our approach characterizes the $\phi_1$ and $\phi_2$ density distribution of a perturbed stream, we are able to recover 2D stream morphologies that are induced by substructure perturbations. Fig.~\ref{fig: 2CDM_density_constrast} indicates that subhalo impacts produce not only gaps, but also distortions to their tracks (e.g., small wiggles or overall shifts). In this regime, 1D density variations alone convey only a limited projection of the full morphology, indicating that informative summary statistics---like the 2D contrast considered here---can help place more stringent constraints on the stream's encounter with substructure.

\subsubsection{Deviations from CDM}\label{sec: deviations_from_CDM}
In order to use our model to test CDM, we must be able to allow for deviations around the CDM expectation. In this section we demonstrate our method beyond CDM, by injecting more subhalos of a higher density than expected under CDM considerations. We also show how observable quantities in our model (like density fluctuations) can be recomputed on-the-fly for different subhalo masses and scale-radii. We sample masses from the CDM SHMF (Eq.~\ref{eq: SHMF}), in the range $[10^6, 5\times 10^7]~M_\odot$. For a GD-1 like stream, we estimate an average impact rate of $\approx 6.7$ for the observable regions of the stream that have been evolving for $\approx 3~\rm{Gyr}$ (Eq.~\ref{eq: Num_impacts}). To consider departures from the CDM expectation, we sample 15 subhalo impacts in this mass range, representing a roughly $2\times$ increase in the number of subhalos in this higher mass interval over CDM expectations.

\begin{figure}
\centering\includegraphics[scale=.65]{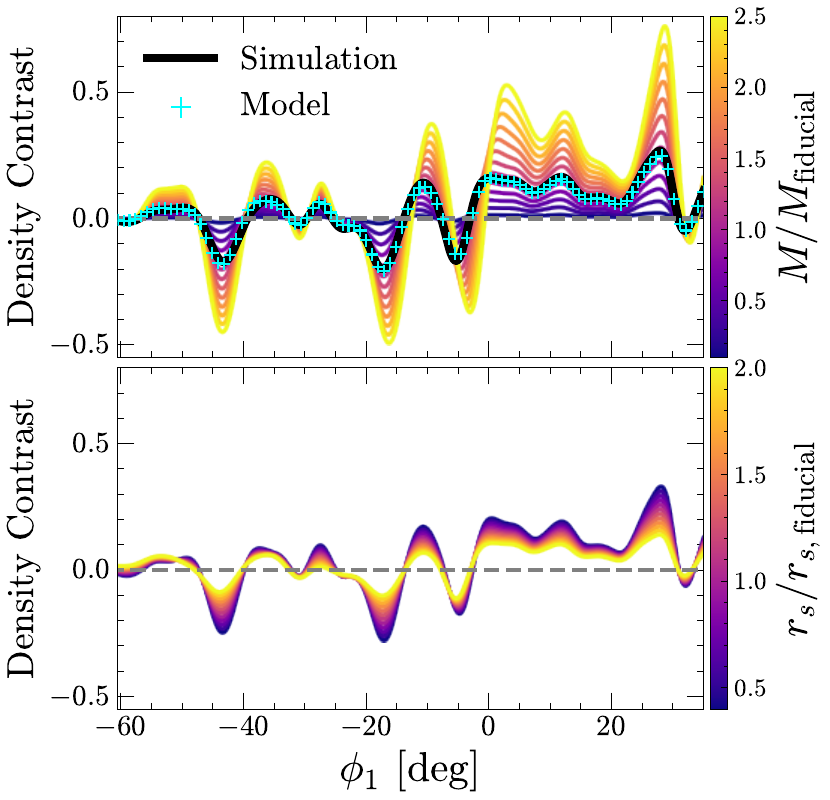}
    \caption{1D density structure of the stream with 15 major impacts. Top panel: the black curve shows the density contrast of the stream when generated with the simulation, while the blue symbols show the result of the linear model. Our model allows for efficient rescaling of the subhalo masses, shown as colored lines, with darker and lighter colors for less and more massive subhalos, respectively.
    Bottom panel: same as the top, but here we rescale the fiducial subhalo scale-radius using perturbation theory. }
    \label{fig: Contrast_Amplitudes}
\end{figure}

In addition to sampling more subhalos than expected in a mass range, we will also reduce the scale-radius of all subhalos by $30\%$ below the mass-radius relation from Eq.~\ref{eq: r_of_m}. That is, we will map $r_s(M) \xrightarrow[]{} 0.7\times r_s(M)$. We will apply this deviation with perturbation theory, starting from the CDM mass-radius relation from  Eq.~\ref{eq: r_of_m} and applying the appropriate $\Delta r_s$ perturbation.

The Cartesian phase-space comparison of the model and simulation is provided in Fig.~\ref{fig: ManyMajorImpacts} (cyan and black, respectively). Here we can see that the stream has several gaps and kinematic deviations from the smooth model, with velocity offsets on the order of $\sim 5~\rm{km/s}$. Again, the agreement between the linear model and simulation is excellent, with slight misalignments in $\delta z$ and $\delta v_x$.

 In Fig.~\ref{fig: Contrast_Amplitudes} we show the 1D density contrast of the stream, with the solid black curve in the top panel corresponding to the full simulation shown in Fig.~\ref{fig: ManyMajorImpacts}. The cyan plus points are from the linear model, which again show good agreement with the simulation. A unique aspect of our linearized stream model is that we can scale up and down the amplitude (i.e., mass) associated with each perturbation. Likewise, we can also adjust the scale radius of the subhalos impacting the stream perturbatively. All of these scalings enter our model through multiplicative factors, allowing us to rapidly sample a spectrum of subhalo masses and scale-radii as a post-processing step.

We illustrate a rescaling of the mass amplitudes associated with the subhalos in the top panel of Fig.~\ref{fig: Contrast_Amplitudes}. While we can scale up and down the subhalo masses independently, here we rescale the masses by a common multiplicative factor. As expected, larger subhalo masses create more significant density fluctuations, both in density deficits and enhancements. The former can be attributed to subhalos clearing out deeper gaps owing to their higher masses and therefore larger velocity kicks, while the latter effect of density enhancements can be attributed to the caustic stage, where particles boosted to higher and lower frequencies create density pileups along the tidal tails \citep{2015MNRAS.450.1136E}. 

The bottom panel of Fig.~\ref{fig: Contrast_Amplitudes} shows a common rescaling of the subhalo scale radii estimated perturbatively, at fixed subhalo mass. This panel clearly demonstrates that higher density subhalos produce larger density enhancements and deficits. It also demonstrates a mass-radius degeneracy, where one can achieve similar 1D density profiles by either rescaling the mass or the radius of the subhalo population. However, higher-order moments of the perturbed density distribution do display some differences. For instance, the gap around $\phi_1 \approx -18~\rm{deg}$ undergoes a slight shift in $\phi_1$ at larger subhalo masses, while decreasing the scale radius of the subhalo does not produce an appreciable shift.

Differences in density structure of the stream due to varying the mass of the subhalo versus its scale-radius can be understood from Eq.~\ref{eq: update_rule}. When increasing the mass of an individual subhalo, the gradient term $\nabla \Phi_\alpha$ is always larger in magnitude for all particles. Meanwhile, a small decrease to the scale radius of the subhalo only significantly increases the magnitude of the force for particles that suffer near or direct encounters with the subhalo. For instance, if we take a subhalo potential of the form $\Phi \propto M(r + r_s)^{-1}$, the derivative of its force with respect to $r_s$ is proportional to $M(r + r_s)^{-3}$. At large values of $r/r_s >> 1$, the subhalo's force is more sensitive to an increase in mass than a decrease in radius. While scale-radius deviations are degenerate with mass deviations for any single particle, using an ensemble of particles can help break degeneracies since not all of the particles share the same impact parameter with the subhalo. This means that many stars are needed to break the mass-radius degeneracy. We have also tested taking higher order moments of the density distribution, and find that these can be used to break mass-radius degeneracies, at least for subhalos on fixed orbits. We leave an exploration of higher order statistics and stream density structure to a future work. 

\section{Application to Streams in Realistic Potentials}\label{sec: applications} 
In this section we demonstrate our model in time-dependent potentials representative of the Milky Way environment. We consider three MW streams spanning the inner to outer halo. In \S\ref{sec: Pal5_and_Bar} we apply our perturbative model to a Pal 5 type stream in the presence of a rotating bar. In \S\ref{sec: OC_and_LMC} we consider the Orphan-Chenab (OC) stream, and linear perturbations to its tidal tails in the presence of the LMC. In \S\ref{sec: heating} we apply our method to a GD-1 like stream in the presence of many low-mass subhalos, as expected in CDM.

\subsection{Pal 5 and the Rotating Bar}\label{sec: Pal5_and_Bar}
Several previous works have argued that the surface density, width, and asymmetric tidal tails of Pal 5 are consistent with perturbations from the Galactic bar \citep{2017NatAs...1..633P,2017MNRAS.470...60E, 2020ApJ...889...70B}. While the cluster's pericenter is $\approx 7-8~\rm{kpc}$ from the Galactic center, the stream's orbit is prograde with the direction of the bar's rotation and therefore may have been seriously distorted if in resonance with the bar's pattern speed. In fact, bar-induced torques that fan out Pal 5's leading tail have been proposed as the mechanism behind the leading tail being several degrees shorter than its trailing tail \citep{2017NatAs...1..633P,2020ApJ...889...70B}.

\begin{figure*}
\centering\includegraphics[scale=.68]{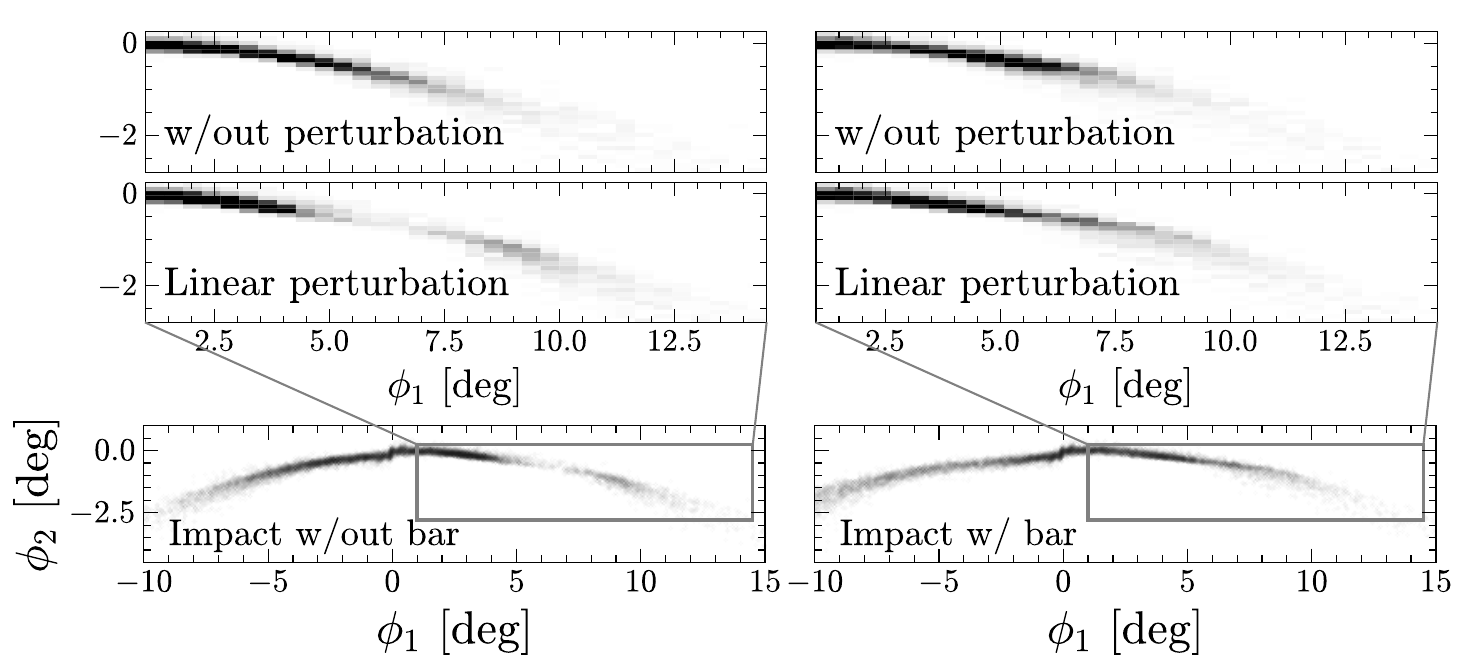}
    \caption{A Pal-5 like stream encounters the galactic bar and a $3\times 10^7~M_\odot$ dark matter subhalo. An advantage of our model is that we can account for the bar at non-linear order, and the subhalo at linear order. The left column shows the stream generated with the bar turned off, and a subhalo impact around $\phi_1 \approx 5~\rm{deg}$. The right column shows the same setup, but with the bar turned on. The bar increases the velocity dispersion of the stream, and redistributes density along both arms. The bar largely erases the gap cleared out by the subhalo impact, which occurred $> 1~\rm{Gyr}$ ago, corresponding to a few pericentric passages of the stream. }
    \label{fig: Pal5andBar}
\end{figure*} 

The truncation of the leading arm is not the only unusual feature of Pal 5's tidal tails. Both arms show significant density fluctuations, some of which can be attributed to epicycles as stars oscillate around their guiding centers, and a possible interaction with a massive dark matter subhalo \citep{2012MNRAS.420.2700K,2017MNRAS.470...60E}. The latter appears consistent with an appreciable under density along the trailing arm, and is not easily explained by perturbations from the bar.

To disentangle possible subhalo signatures in a stream affected by the Galactic bar, it is necessary to simultaneously model both effects. We consider two scenarios for Pal 5 suffering the same subhalo impact: first in a static, axisymmetric Milky Way potential, and then also incorporating a rotating bar  (Eq.~\ref{eq: bar}). The dynamical effect of the bar on the stream will be modeled at full non-linear order, by absorbing its potential in $H_{\rm base}$ (\S\ref{sec: method}). The pattern speed in the barred case is $\Omega_b = 38~\rm{km s^{-1} kpc^{-1}}$ \citep{2020ApJ...889...70B}. In both cases the final phase-space location of the progenitor is the same, and the orbit of the progenitor cluster does not change significantly. For the subhalo impact, we use a similar setup to \citet{2017MNRAS.470...60E} and model the impact of a Hernquist subhalo with mass $3\times 10^7 M_\odot$  and (CDM) scale-radius of $0.58~\rm{kpc}$. The impact is direct, and occurs $1.4~\rm{Gyr}$ ago. We treat the subhalo perturbatively in our model (i.e, the subhalo is treated as a $\Phi_\alpha$ potential in Eq.~\ref{eq: Hamiltonian}), and will compare to the equivalent non-perturbative simulation.

The result of this experiment is shown in Fig.~\ref{fig: Pal5andBar}, with the left and right columns showing models without and with a rotating bar, respectively. The depth, width, and overall morphology of the subhalo-induced gap at $\phi_1= 6~\rm{deg}$ differ between the unbarred and barred potentials. Specifically, in the unbarred potential, the subhalo impact removes approximately 80\% of the stars from a several-degree region across the stream. In contrast, the barred potential results in density variations with a shallower depth (density contrast $\approx 40\%$). The leading tail with $\phi_1 <0$ also differs between the two scenarios; in the barred potential, the stream's density is redistributed due to the scattering of orbital planes driven by the bar's torques.

\begin{figure}
\centering\includegraphics[scale=.6]{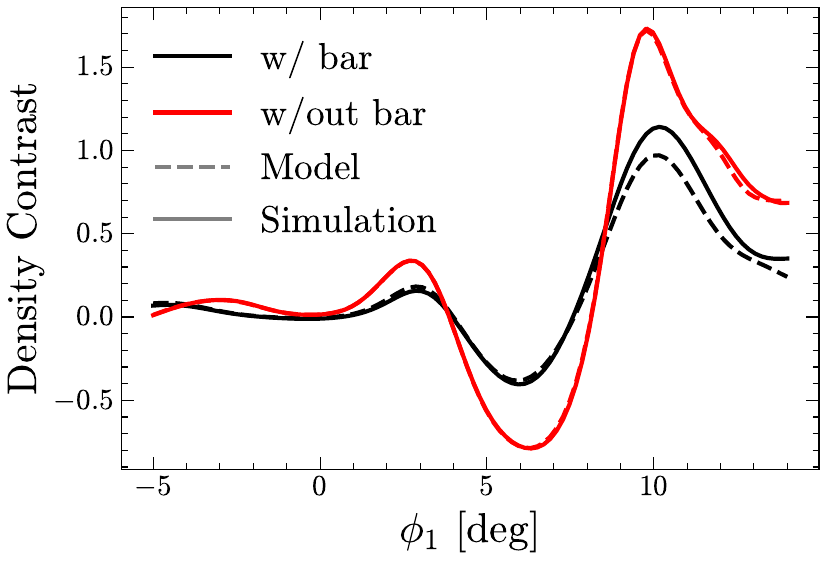}
    \caption{1D density contrast of the Pal-5 like streams shown in Fig.~\ref{fig: Pal5andBar}. In all cases, a subhalo impacts the stream clearing out a gap around $\approx 5~\rm{deg}$. The black curve shows the gap when the stream is generated with a bar, and red is without the bar. Solid and dashed lines correspond to the simulation and linear model, respectively. The depth and width of the gap change in the presence of the bar, and the linear model is able to characterize the gap in both scenarios.  }
    \label{fig: Pal5_Density}
\end{figure} 

A more quantitative comparison is provided in Fig.~\ref{fig: Pal5_Density}, where the unbarred and barred potentials are shown with red and black lines, respectively. The dashed line shows the result of our linear perturbation theory, while the solid lines show simulations. We find that the overall shape of the density contrast is similar between the barred and unbarred scenarios. The major difference is the amplitude of the signal. In our model Pal 5 experiences 4 pericentric passages after the subhalo impact, each of which increases the velocity dispersion of the stream and redistributes density along the tidal tails. The final velocity dispersion of the barred potential stream is $\approx 0.7~\rm{km/s}$ higher than the unbarred potential stream in the trailing arm. This explains the reduced density contrast in the barred potential, as kinematically hotter stellar systems obscure signatures of interactions with substructure due to their random thermal motions. Failure to account for the effect of the bar would lead to underestimating the subhalo mass, since the amplitude of the subhalo density signal is reduced by the stream's higher velocity dispersion.

There is a small disagreement ($\sim 10-15\%$) between the barred model and simulation around $\phi_1 \approx 10~\rm{deg}$. This is because the tidal field due to the bar is very sensitive to changes in test-particle orbit. Because a $3\times 10^7~M_\odot$ direct impact can produce large changes in an orbit, some particles experience an appreciably different tidal field such that higher order terms are needed in our expansion to exactly recover the correct dynamics. Still, only small disagreements in density structure even in this nonlinear regime are promising, since Pal 5 is more disturbed by the bar compared to all of the other known MW streams so far. 

While Fig.~\ref{fig: Pal5andBar} and Fig.~\ref{fig: Pal5_Density} show a single subhalo impact, we explored a several Gyr range of impact times. We find that when the subhalo impact occurs before Pal 5's last pericenter, the bar reduces the amplitude of the subhalo induced density signal. Our experiments indicate that for Pal 5, it is essential to incorporate the effects of the bar when inferring properties of potential subhalo interactions, since the bar can significantly change the characteristics of a density signal imprinted by a subhalo fly-by. Failure to account for a rotating bar can therefore lead to bias when inferring subhalo masses and radii.

\begin{figure*}
\centering\includegraphics[scale=.6]{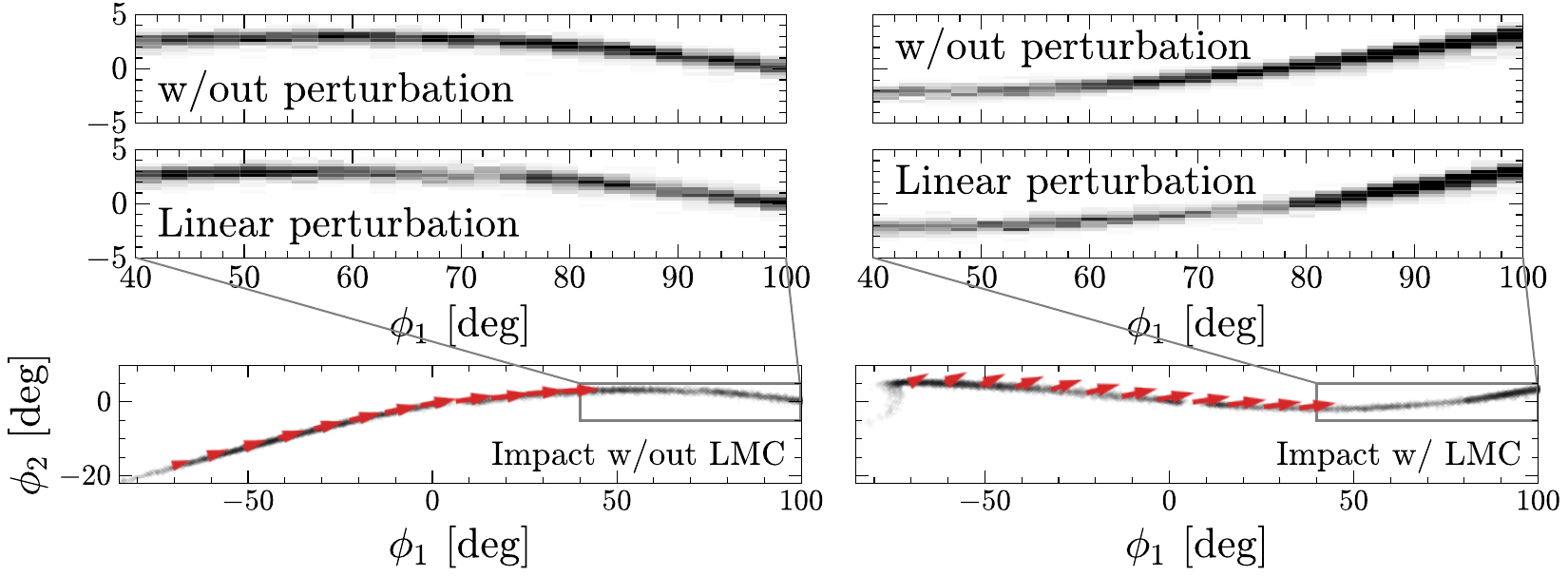}
    \caption{The LMC plays a crucial role in shaping outer halo streams, both in their track and density distribution. The left column shows an OC-like stream generated without the presence of the LMC. A single subhalo (mass of $6\times 10^7~M_\odot$) impacts the stream, producing a density break around $\phi_1 \approx 70~\rm{deg}$. Proper motion vectors are shown as red arrows. The right column shows the same stream (i.e., same final progenitor location) but with the LMC's potential turned on. The stream's track and proper motion vectors are misaligned due to the presence of the LMC. A subhalo impacts the stream in the same region again, with identical impact parameters to the scenario in the left (all taken relative to the stream). The LMC is incorporated in the base hamiltonian, whose equations of motion are solved at full non-linear order. Linear perturbation theory is then applied to model the subhalo's effect on the stream. Post-impact, the phase-mixing process is modified by the presence of the LMC due to the stream's changing orbit and the LMC's tidal field.}
    \label{fig: OC_Stream_phi12}
\end{figure*}

\begin{figure}
\centering\includegraphics[scale=.55]{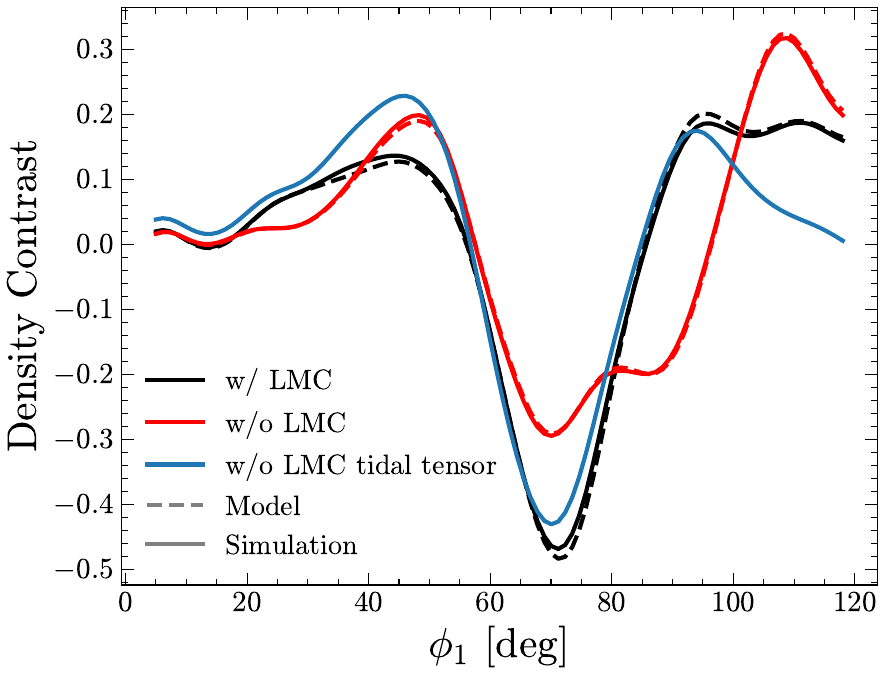}
    \caption{ 1D density contrast of the OC-like streams shown in Fig.~\ref{fig: OC_Stream_phi12}. A subhalo clears a gap around $\phi_1 \approx 70~\rm{deg}$, but the depth and width of the gap change as a function of modeling choices for the potential. Black and red curves show the density contrast when turning on and off the LMC, respectively. Solid black and red curves correspond to the simulation, while dashed curves show the prediction by our model. The blue curve shows how the density changes if the stream is on a fixed orbit, but with the LMC's tidal tensor turned off for particles that are perturbed by the subhalo. Differences in depth are small, but the caustic structure of the density distribution (i.e., outside of the gap) is modified by $10-15\%$ when turning off the LMC's tidal field. }
    \label{fig: OC_Stream_Density}
\end{figure}

\subsection{Orphan-Chenab Stream and the LMC}\label{sec: OC_and_LMC}
Outer halo streams are less likely to be affected by baryonic substructure such as the bar, giant molecular clouds, or even other globular clusters. This makes them excellent dark matter detectors. However, the outer Milky Way is currently being distorted by the infalling Large Milky/Magellanic Cloud (LMC) system, whose total mass is a sizeable 10\% that of the MW (see \citealt{2023Galax..11...59V} for an overview of the MW-LMC interaction). Here we explore how the LMC shapes the dynamics of outer halo streams and the evolution of stream gaps.

The effect of the LMC merger on the Milky Way environment is substantial. The 1:10 mass ratio merger displaces the MW's stellar disk \citep{2021NatAs...5..251P}, induces an outer halo response \citep{2021Natur.592..534C}, and is expected to generate a dark matter halo wake \citep{2019ApJ...884...51G}. The morphology and orbits of stellar streams are sensitive to these effects, and several signatures of the LMC's presence have been observed in MW streams (e.g., \citealt{2019MNRAS.485.4726K, 2021ApJ...923..149S, 2021MNRAS.506.2677E}).

A significant advantage of the method presented in this paper is that we can accommodate time-evolving, action non-conserving, and non-central potentials. We now illustrate this aspect of our work by focusing on the Orphan-Chenab (OC) stream. The OC stream exhibits a strong misalignment between its track and proper motion vectors, with the direction of the misalignment pointing towards the LMC's location. The precise location of the stream in the sky is difficult to recover without accounting for an LMC induced deflection \citep{2021MNRAS.506.2677E, 2023MNRAS.518..774L, 2023MNRAS.521.4936K}. Models of the OC stream evolving in the MW-LMC system indicate a closest approach with the LMC of around $6~\rm{kpc}$, $370~\rm{Myr}$ ago \citep{2023MNRAS.521.4936K}. 

For modeling the OC stream, we adopt the same dissolved progenitor location from \citet{2019MNRAS.487.2685E}, with a progenitor mass of  $10^6 M_\odot$ \citep{2019MNRAS.487.2685E, 2023ApJ...948..123H}. We evolve the OC stream for $4~\rm{Gyr}$ in two scenarios: one in our fiducial MW potential without the LMC, and one that includes the LMC (specified in \S\ref{sec: galaxy_model}). Our LMC model includes a simplified scenario of a static disk and halo with a time-evolving NFW potential representing the LMC on its orbit. The LMC is incorporated in our model at full non-linear order, by absorbing its potential into $H_{\rm base}$ (\S\ref{sec: method}). Employing more realistic MW potential models (e.g., using basis function expansions) is possible using our method. In both the static and time-evolving scenarios, we also simulate an encounter with a $6\times 10^7~M_\odot$ Hernquist subhalo. The subhalo impact is treated perturbatively using our model (i.e, the subhalo is treated as a $\Phi_\alpha$ potential in Eq.~\ref{eq: Hamiltonian}). The subhalo crosses the stream at a lookback time of $1.1~\rm{Gyr}$, roughly $0.7~\rm{Gyr}$ before the stream has its closest approach with the LMC. To ensure that we account for the change in orbit of the stream due to turning on and off the LMC potential, we utilize the impact sampling scheme discussed in \S\ref{sec: sampling_impacts}. We backwards integrate the particles around $\phi_1 = 75~\rm{deg}$ in the OC stream frame \citep{2019MNRAS.485.4726K} to determine the location of the subhalo at its closest approach. We choose $\phi_1 = 75~\rm{deg}$ since this is the approximate location of a tentative gap observed along the real OC stream with deep photometry from DECam \citep{2019MNRAS.485.4726K}. This experiment allows us to assess (a) how much the LMC changes the overall orbit of the stream, and (b) how the properties of a stream gap can change due to interactions with the LMC.

The result of this experiment is shown in Fig.~\ref{fig: OC_Stream_phi12}. The left column shows our OC stream model in a potential without the LMC, while the right column shows the model with the LMC included. Red arrows are the average proper motions in $\phi_1$ bins across the stream. Without the LMC (left), the stream's track and proper motions vectors are aligned. The stream with the LMC is very similar to previous OC models (e.g., \citealt{2023MNRAS.521.4936K}), with an abrupt change in the streams track at $\phi_1 < -50~\rm{deg}$. This shift in the stream's track is due to an interaction with the LMC, and is clearly not present in the model without the LMC. The concavity of the stream at $\phi_1 = 50~\rm{deg}$ differs between the two models. For the LMC model, the stream's curvature points in the direction of the LMC. This concavity is consistent with observations (e.g., \citealt{2019MNRAS.485.4726K}), and is very difficult (if not impossible) to produce in a static galaxy model with a central potential. The proper motion vectors along the stream with the LMC are strongly misaligned from the stream's track. The proper motions point towards the LMC's direction. This misalignment has been observed for the real OC stream \citep{2023MNRAS.521.4936K}.

The boxed window and inset panels in Fig.~\ref{fig: OC_Stream_phi12} highlight the region of the subhalo impact. Density maps are shown for the equivalent streams generated without a subhalo impact (top panels), while the perturbed stream density from our linear model is shown below. A gap is apparent in both cases. A more quantitative comparison between the gaps can be found in Fig.~\ref{fig: OC_Stream_Density}. The density contrast estimated via the model is shown by the dashed line, while the solid line shows the simulation. There is extremely good agreement between our model and the simulation. The gap carved out in the model without the LMC is shallower and wider compared to the gap in the LMC model, which appears $\approx 40\%$ deeper. The region of the stream with $\phi_1 >50~\rm{deg}$ is located at a similar average $\phi_2$ value and has a comparable distance distribution in both models, so the changes in gap properties are not due to projection effects, but due to the presence of the LMC.

After a subhalo perturbation, the LMC can change both the overall orbit of a stream, and the local tidal field probed by the stream. Both of these effects are included when comparing the black and red curves in Fig.~\ref{fig: OC_Stream_Density}. Here we disentangle the effect of the tidal field, and explore to what extent the LMC's tides are important in accounting for stream-gap evolution. We carry out this calculation, since it cannot be done with other existing frameworks for stream perturbations. The tidal field governs the divergence of nearby orbits after the subhalo perturbation is applied, and therefore the growth of a gap and its survival. To estimate the importance of the LMC's tidal field on the phase-mixing process, we now apply the same subhalo perturbation both with and without the LMC's tidal tensor using Eq.~\ref{eq: Delta_p_int}. Importantly, the base trajectory is the same in both cases, allowing us to isolate the effect of the LMC's tidal field on the stream from an overall change in orbit. The result of this experiment is represented by the blue curve in Fig.~\ref{fig: OC_Stream_Density}. Comparing the black (w/ LMC's tidal tensor) and blue curves (w/o LMC's tidal tensor), we see that the depth of the gap without the LMC's tidal field is slightly smaller, and the caustics (i.e., overdensities) outside the gap are more pronounced.

Using Eq.~\ref{eq: Delta_p_int}, we can provide a more quantitative assessment of how much the LMC's tidal field changes the process of phase-mixing after a subhalo perturbation. The difference between the linear-order correction terms when including and excluding the LMC's tidal tensor is
\begin{multline}\label{eq: tidal_diff}
    \left( \Delta \boldsymbol{p}\right)_{\mathrm{w/ }} - \left( \Delta \boldsymbol{p} \right)_{\mathrm{w/o}} =\\
    \epsilon\int\limits_{t_{\rm init}}^{t_f} dt\left[  \mathbf{T}_{\rm LMC}\left( t\right) \int\limits_{t_{\rm init}}^t dt^\prime \left\{ \left(\frac{d\boldsymbol{p}}{d\epsilon}\right)_{\rm w/o} - \left(\frac{d\boldsymbol{p}}{d\epsilon}\right)_{\rm w/}\right\} \right],
\end{multline}
where $\rm{w/}$ and $\rm{w/o}$ reference the inclusion and exclusion of the LMC's tidal tensor ($\mathbf{T}_{\rm LMC}$), respectively. All quantities within the integrands are evaluated at $\epsilon=0$. We compute Eq.~\ref{eq: tidal_diff} for all particles inside $\phi_1 \in [50,120]~\rm{deg}$, and find that the median value is $\sim 8~\rm{km/s}$ for the $6\times 10^7 M_\odot$ perturbation considered here. We can propagate this change in velocity to a change in position by integrating Eq.~\ref{eq: tidal_diff} again to obtain the accumulated change in position due to the LMC's tides. We obtain an average value of $\sim 1.3~\rm{kpc}$.
While these values are dependent on where along the stream the impact is centered, a $\sim 1~\rm{kpc}$ displacement is substantial. For outer halo GC streams that suffer a close passage with the LMC, we can anticipate that gaps created by subhalos for these features will be heavily modified by the LMC, since typical GC streams themselves span several $\rm{kpc}$, and have typical widths below the $100~\rm{pc}$ scale (e.g., \citealt{2022MNRAS.514.1757P}).

Eq.~\ref{eq: tidal_diff} also indicates that the LMC's tidal field is important for gap evolution in a subhalo mass-dependent manner ($\propto \epsilon$). Failure to account for the tidal effect of the LMC on gap evolution can bias inference of the subhalo mass, though the amount of bias is higher for the most massive subhalos, and therefore the most observable gaps. Streams that are less susceptible to the LMC's tidal field will similarly be less susceptible to this effect, though we expect that inference of Southern hemisphere streams where the LMC is located should account for the LMC's orbital and tidal effects when modeling gap evolution.

\subsection{GD-1 and Subhalo Heating}\label{sec: heating}
In this section we demonstrate how our method can be applied to rapidly generate many realizations of a stream for different subhalo mass functions. We focus on radial velocity dispersion as a key observable, since several prior works \citep{2002ApJ...570..656J, 2002MNRAS.332..915I,2011ApJ...731...58Y,2012ApJ...748...20C,2016ApJ...818..194N,2022MNRAS.513.3682D,2024arXiv240518522C} have argued that the heating (or scattering) of tidal tails can provide an indicator of the substructure environment. This section demonstrates the power of our method for statistical studies of streams and subhalo impacts, and a more in-depth analysis of the GD-1 stream and heating will be presented in a forthcoming work.

We consider the GD-1 like stream described in Appendix~\ref{app: prog_and_LMC}. For this stream, we generate a library of $\sim 10^4$ linear responses to subhalo impacts, enabling rapid adjustments to the strength of each perturbation as a post-processing step. The number of expected subhalo impacts is orders of magnitude less than $10^4$, so a library this large allows us to efficiently sample a wide range of combinations for subhalo orbits. We also obtain derivatives with respect to the subhalo scale-radius, which allows us to model departures from the CDM mass-radius relation. 

We sample linear responses using the impact sampling scheme outlined in \S\ref{sec: sampling_impacts}, and consider impacts from $0.5~\rm{Gyr}$ after the observable tails of the stream start forming up to the present day (a $2.5~\rm{Gyr}$ period). We exclude impacts in the earliest $0.5~\rm{Gyr}$, since the stream's tidal tails are very short during this period, resulting in a very small interaction cross-section. We have verified that including impacts from the first $0.5~\rm{Gyr}$ of this experiment does not change our results.

\begin{figure*}
\centering\includegraphics[scale=.53]{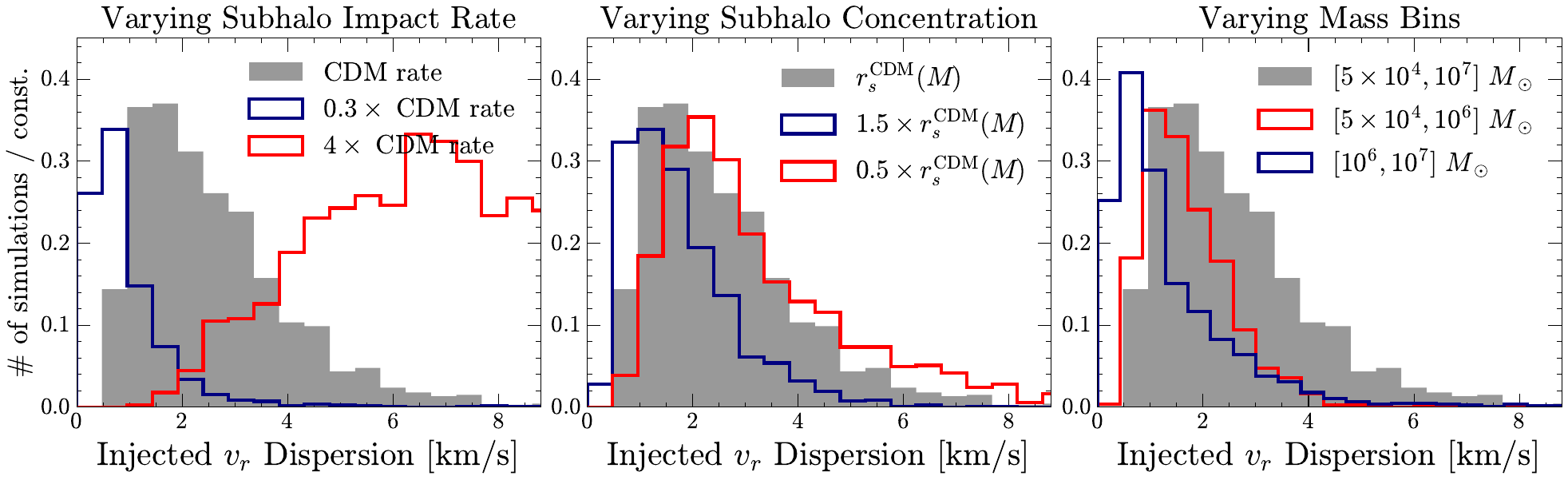}
    \caption{We generate thousands of simulations using our model as a function of different subhalo population parameters. Histograms represent the amount of injected velocity dispersion due to subhalo perturbations across the ensemble of simulations. 
    Gray histograms in all panels are the same, and represent the CDM expectation. Left: reducing the rate of impacts (blue) decreases heating, increasing the rate (red) drives heating of tidal tails. Middle: at the CDM impact rate, we vary the subhalo mass-radius relation. Higher density subhalos (red) produce kinematically hotter tails compared to lower density subhalos (blue). Right: here we isolate the effect of subhalos in a low mass bin (red) and a high mass bin (blue). All impacts are sampled at the CDM rate for each mass bin. The low mass bin sets the lower cutoff of the CDM (gray) histogram, while the high mass bin produces the high dispersion tail of the CDM distribution. This is because lower mass subhalos are more numerous, and therefore have a higher probability of several direct encounters with the stream for every model realization.   }
    \label{fig: Radial_Velocity_Dispersion_Injection}
\end{figure*}

\subsubsection{Kinematic Heating due to Subhalos}\label{sec: subhalo_heating}

The velocity dispersion of a stream is dependent on several factors. Internal cluster dynamics sets the intrinsic velocity dispersion of the stream, with higher mass globular clusters producing tidal tails with higher velocity dispersions. External factors include the orbital properties and phase of the stream (i.e., velocity dispersion can oscillate with the orbital phase) and its encounters with substructure. These factors combined set the baseline velocity dispersion. The orbit of a stream can be reliably modeled with our framework. While accounting for internal heating is in principle simple, as it is mainly dependent on the progenitor's mass, estimating the initial mass of a stream's progenitor can be challenging in practice.

Due to uncertainty in the GD-1 progenitor's mass and therefore the stream's intrinsic velocity dispersion, we first proceed in a cluster-mass agnostic sense, by computing the amount of velocity dispersion injected by the subhalo population. By injection, we mean how much additional heating subhalos should induce in the GD-1 stream. This is a similar calculation to estimating the diffusion coefficient generated by the substructure along the stream's trajectory \citep{2022MNRAS.513.3682D}. Using the library of velocity derivatives $\{(d\boldsymbol{p}/d\epsilon_\alpha, d^2\boldsymbol{p}/d\epsilon_\alpha dr_{s,\alpha})\}$ obtained for each particle, we first sample from this library according to the SHMF and generate model streams. We then take the variance of the perturbation velocities over the length of the stream, and add each subhalo's contribution to the variance in quadrature. The square root of the resulting variance is the velocity injection. We assume that velocity kicks between different subhalos are uncorrelated, and that the stream's unperturbed velocity is uncorrelated with the velocity kicks. The former is motivated since the velocity kicks are oriented over random directions, while the latter is true when averaging over the stream, since small-scale subhalo perturbations typically induce local velocity distortions rather than overall velocity shifts. Indeed, the covariance between stream star velocities and subhalo injected velocity kicks is very small in our simulations. Under these conditions, the squared velocity dispersion injection is simply the perturbed stream's velocity variance minus the unperturbed stream's velocity variance. Additionally, velocity injection as a statistic is less sensitive to reasonable variations of the progenitor mass around $10^4~M_\odot$. This is because, in this section, we are only using a cold GC stream as a tracer of the substructure environment; we are not yet making statements about the intrinsic velocity dispersion of the stream. Provided that the tracer stream follows the same orbit and occupies approximately the same phase-space volume, the number of subhalo impacts and the relative velocity of the stream and subhalos is not sensitive to 
reasonable variations in GC mass. For a dwarf galaxy stream, tidal tails occupy a larger spatial volume and could have different (higher) encounter rates with substructure.

Fig.~\ref{fig: Radial_Velocity_Dispersion_Injection} shows the injected radial velocity ($v_r$) dispersion for a GD-1 like stream due to subhalo impacts. Each histogram represents over $1000$ realizations of our model. We focus on the line-of-sight velocities as they are typically more precisely measured than tangential velocities. To determine the injected $v_r$ dispersion, we take the line-of-sight component of the velocity perturbation induced by subhalos, and compute the variance of this quantity over a $20~\rm{deg}$ region of the stream. The $20~\rm{deg}$ region is the same section of the stream with high-precision radial velocities from \citet{2020ApJ...892L..37B}. In each panel, we show distributions for the injected $v_r$ dispersion as a function of different subhalo population parameters. Each contribution to the histogram represents a single realization of our model (i.e., a single simulation). The histograms are generated by treating the number of subhalo impacts as a Poisson process with a mean impact rate determined from Eq.~\ref{eq: Num_impacts}. Additionally, we marginalize over many different subhalo orbits. Both of these factors contribute to the spread in injected velocity dispersion found across test-cases with fixed subhalo population parameters.

The left panel shows the CDM impact rate, while the blue and red histograms are for $0.3$ times and $4$ times the number of expected impacts, respectively. The subhalo mass range considered is $[5\times 10^4, 10^7]~M_\odot$. As expected, the injected velocity dispersion increases for more numerous subhalo impacts. Besides having a higher mean, the $4\times \rm{CDM}$ distribution is much wider than those from fewer impacts. This is due to two factors. First, the variance of a Poisson rate equals the rate itself, so we expect a $4\times$ spread in the number of impacts for a rate that is $4\times$ higher than the CDM value. Second, the $4\times\rm{CDM}$ case better samples the subhalo mass function, resulting in heavier tails of the injected velocity dispersion distribution on both ends.

The middle panel in Fig.~\ref{fig: Radial_Velocity_Dispersion_Injection} shows the result of varying the mass-concentration relation of the subhalos, all sampled at the CDM impact rate. As expected, higher density subhalos typically produce more velocity dispersion (red), while lower density subhalos reduce the high dispersion tail of the gray distribution. Comparing the left panel (rate) to the middle (concentration) suggests that velocity dispersions alone have a rate-concentration degeneracy, where similar shifts in the distributions can be achieved by changing either the Poisson rate of impact or the mass-radius relation. 

The right panel of Fig.~\ref{fig: Radial_Velocity_Dispersion_Injection} brackets the CDM distribution of velocity dispersion into two mass bins. The gray histogram shows the full mass range sampled, spanning several orders of magnitude. The red histogram shows the injected velocity dispersion across 1000s of simulations, only accounting for subhalos in the low-mass bin, $[5\times 10^4, 10^6]~M_\odot$. The blue histogram is generated with the same setup, though only accounting for subhalos in the high-mass decade, $[10^6,10^7]~M_\odot$. The average number of subhalos in each mass bin is determined from CDM expectations (Eq.~\ref{eq: Num_impacts}). We run the same number of simulations in each case. The blue histogram indicates that for many realization of the CDM mass function in the high-mass bin, it is not unusual to have roughly zero injected velocity dispersion. However, there is a high density tail extending to $\approx 5~\rm{km/s}$ of injected velocity dispersion. The red histogram indicates that all of the simulations we run for the low-mass subhalo bin produce a stream with an injected velocity dispersion of at least $\approx 0.5~\rm{km/s}$. Not a single realization produces a stream with zero injected velocity dispersion. Taking the blue and red histograms in combination, the low dispersion cutoff of the gray CDM histogram is set by the lowest mass range of subhalos, while the high dispersion tail is set by the highest mass range of subhalos. The reason the blue and red distributions differ significantly is because low-mass subhalos outnumber the high-mass subhalos. This means that for a given realization of the SHMF, there is a high probability of several low-mass subhalos having a direct encounter with the stream, whereas there are an insufficient number of high mass subhalos to consistently have direct encounters with the stream. Therefore, the minimum velocity dispersion measured across many GC streams should be dictated by the number of low-mass subhalos in the Galaxy. We discuss this point further in \S\ref{sec: discuss_heating}.

\subsubsection{The Velocity Dispersion of GD-1}
In this section we first calculate the radial velocity dispersion of the GD-1 stream itself, and then compare the dispersion to expectations from different dark matter models (\S\ref{sec: subhalo_heating}). To measure the radial velocity dispersion of GD-1, we use a sample of 43 spectroscopically confirmed GD-1 members from \citet{2020ApJ...892L..37B}. This sample extends over a 20~\rm{deg} segment of the stream, including 14 stars in the stream's spur component. We detrend the radial velocity curve as a function of $\phi_1$ by fitting a 3rd degree polynomial, and define radial velocity dispersion as the standard deviation of radial velocities around this polynomial. This approach is similar to the analysis in \citet{2021ApJ...911L..32G}. We find a radial velocity dispersion of $2.3 \pm 0.2~\rm{km/s}$, consistent with the velocity dispersion derived in \citet{2021ApJ...911L..32G} using the same data though a slightly different analysis procedure. The injected velocity dispersion predicted by our model in Fig.~\ref{fig: Radial_Velocity_Dispersion_Injection} is taken over the same $\phi_1$ range of the stream compared to the actual data. 

To estimate the intrinsic velocity dispersion of an unperturbed GD-1-like stream, we must adopt a mass for the progenitor and a dynamical age for the stream. We fix the age of the observable tails to $3~\rm{Gyr}$ since this value is consistent with the length of the observed stream and is preferred by direct $N-$body models of GD-1 \citep{2019MNRAS.485.5929W}. For the progenitor mass, we adopt a value of $0.5\times 10^4~M_\odot$, because this value produces a stream model that matches the observed GD-1 width and length. The radial velocity dispersion of the unperturbed resulting stream is $\approx 0.7~\rm{km/s}$, which is consistent with 
earlier models, including particle-spray and direct $N-$body representations of GD-1 \citep{2021ApJ...911L..32G,2019MNRAS.485.5929W}.
Assuming that the unperturbed star velocities are uncorrelated with velocity injections (discussed in \S\ref{sec: subhalo_heating}), we add the unperturbed velocity dispersion to the injected velocity dispersion in quadrature.  

\begin{figure}
\centering\includegraphics[scale=.55]{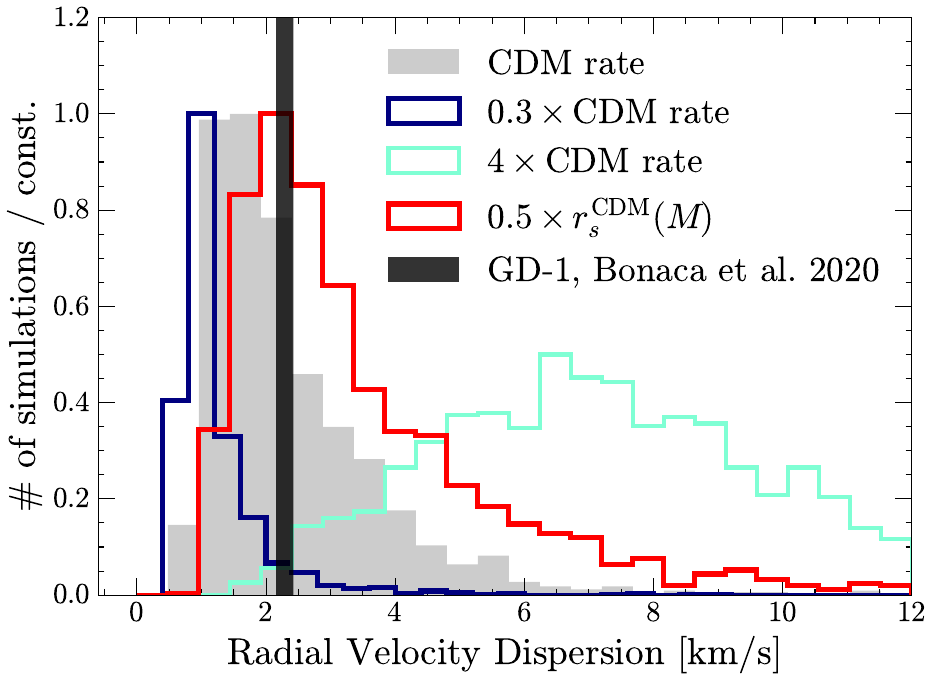}
    \caption{Predictions for the radial velocity dispersion of the GD-1 stream for different subhalo parameters. The measured dispersion for GD-1 is shown as a black vertical band representing a $1\sigma$ errorbar. The GD-1 velocity dispersion is consistent with expectations from CDM (gray) and a model with more compact subhalos (red), while it disfavors models with lower (dark blue) and higher (cyan) impact rates. }
    \label{fig: GD1_Heating_Constraint}
\end{figure} 

We compare the measured GD-1 radial velocity dispersion (vertical black bar) to forecasts from the previous section (histograms) in Fig.~\ref{fig: GD1_Heating_Constraint}. The CDM rate distribution (gray) is in good agreement with the actual GD-1 observations, with perhaps a slight preference for higher density subhalos at the CDM rate. There are physical reasons to expect denser subhalos, e.g., in SIDM (see, e.g., \citealt{2000PhRvL..84.3760S, 2024arXiv240919493Z}). A model with more numerous CDM subhalos overpredicts the GD-1 velocity dispersion (cyan), while a model with fewer CDM subhalos underpredicts the velocity dispersion (navy). The agreement between the observed GD-1 dispersion and the CDM forecast is a tantalizing demonstration of how stream kinematics could potentially be used to constrain the nature of dark matter. However, Fig.~\ref{fig: GD1_Heating_Constraint} assumes a fixed progenitor mass and stream age. In an upcoming work, we sample the degeneracy between the progenitor's mass, the stream's age, and its intrinsic velocity dispersion. We therefore defer a full statistical analysis of dark matter subhalos and GD-1's velocity dispersion to this future work.

\section{Discussion}\label{sec: discussion}
\subsection{Modeling Flexibility}
Our model is able to incorporate any differentiable function for the galactic potential and subhalo potentials, including time-dependent functions. Accounting for time-dependence in the Galactic potential is crucial for capturing orbital dynamics of the streams \citep{2022ApJ...939....2A, 2023MNRAS.518..774L, 2024arXiv241002574B}, while implementing time-dependent subhalo potentials allows us to extend our model to (e.g.) SIDM potentials with subhalo core-collapse \citep{2000PhRvL..84.3760S}, or semi-analytic treatments for subhalo tidal stripping (e.g., \citealt{2024JCAP...02..032Y}).

The flexibility of our approach is particularly important for modeling the effects of baryonic substructure on tidal tails. For example, the Galactic bar has a non-linear effect on certain inner halo streams, such as Pal-5 (\S\ref{sec: Pal5_and_Bar}), while the LMC plays a significant role in shaping outer halo streams, like the OC stream (\S\ref{sec: OC_and_LMC}). Although the bar and LMC themselves act as perturbations to these streams, they are not well approximated in the linear regime. 
Ultimately, we can include any known baryonic perturbation in the base model and treat smaller fluctuations in the potential perturbatively. 

For more massive subhalos where our linear approximation breaks down, we can include the non-linear effects of a massive subhalo in $\Phi_{\rm base}$, while solving for smaller-scale perturbations in the series of perturbing potentials, $\Phi_\alpha$. The former involves solving for the parameters of a single massive subhalo flyby (e.g., \citealt{2019ApJ...880...38B,2024arXiv240402953H}), while the latter entails pre-computing a library of linear responses around the perturbed stream (\S\ref{sec: heating}). This hybrid approach combines the accuracy needed for modeling the impacts of more massive subhalos, and the speed necessary for modeling the effects of less massive, but more numerous subhalos.

We have formulated our model in terms of generic release functions for particles leaving the stream (\S\ref{sec: boundary_conditions}). Because of this, our model does not depend on particle-spray based prescriptions for stream formation, but can also be applied to more accurate stream models, e.g., direct $N-$body simulations. Provided that we can track when and where a particle leaves the progenitor cluster, we are free to apply our method to the resulting tidal tails using initial conditions generated from $N-$body models (e.g., \citealt{2014ApJ...788..181N, 2019MNRAS.485.5929W, 2021NatAs...5..957G,2023ApJ...946..104W}). 

\subsection{Stellar Stream Velocity Dispersion}\label{sec: discuss_heating}
Observable gaps in streams are only expected for the higher mass decades of a subhalo population, with subhalos above a mass of $\gtrsim 10^6~M_\odot$ producing observable density fluctuations along kinematically cold GC streams \citep{2011ApJ...731...58Y,2016MNRAS.463..102E, 2019arXiv190201055D}. For a GD-1 like stream, one expects $\sim 1$ appreciable gap to form given its orbit and approximate dynamical age \citep{2011ApJ...731...58Y, 2016MNRAS.463..102E, 2023arXiv230915998A}. On the other hand, for a CDM subhalo population GD-1 should encounter $\lesssim 100$ subhalos in the mass range of $5\times 10^4-10^6 M_\odot$ \citep{2011ApJ...731...58Y, 2016MNRAS.463..102E, 2017MNRAS.466..628B, 2023arXiv230915998A}. While not all subhalos will clear out measurable gaps, the large number of lower mass subhalos are expected to increase the overall velocity dispersion of the stream \citep{2002ApJ...570..656J,2002MNRAS.332..915I,2012ApJ...748...20C,2022MNRAS.513.3682D, 2024arXiv240518522C}.

In \S\ref{sec: heating} we demonstrated how our model can be used to construct probability distributions for a stream's velocity dispersion under different assumptions for the number, mass, and central concentration of subhalos. We find that for GD-1, if there is an absence of subhalos below $10^6~M_\odot$, it would not be unusual for the stream to have avoided subhalo induced heating entirely, since more massive subhalos are fewer in number compared to the lower mass subhalos. Conversely, if there is a plethora of low-mass subhalos below $10^6~M_\odot$, then we find that every realization of our stream model has a higher velocity dispersion. Applying this reasoning to streams with similar impact rates to GD-1 (i.e., inner halo streams), we expect substantial kinematic heating due to subhalos for the population of observed MW streams. This implies that measuring the velocity dispersion of stream populations could provide a test of dark matter subhalos down to masses of roughly $5\times 10^4~M_\odot$, since we expect the velocity distribution of all streams in the inner halo to be inflated by low-mass subhalos. There are already several MW streams with measured radial velocity dispersions \citep{2020ApJ...892L..37B,2022ApJ...928...30L, 2024arXiv240706336V}. Radial velocity measurements can be used to test for deviations around CDM as we show in this work, or alternatives to CDM like warm dark matter \citep{2024arXiv240518522C}.

\subsection{Accelerating Inference of the Subhalo Mass Function}
An advantage of our perturbative model is that by linearizing the effect of a subhalo on the stream of interest, we are free to algebraically adjust the presence, absence, and strength of the perturbation as a post-processing step. Provided that a large library of derivatives have been pre-computed,  inference of the SHMF essentially reduces to constrained linear regression (or a spectral unmixing problem), for which there is an expansive literature of tools to solve (e.g., \citealt{911111, Keshava2003ASO, bioucasdias:hal-00760787,  8075417}). 

Because linear models are extremely computationally efficient, our approach allows for fitting the properties of individual subhalos, as well as constraining their population statistics. Derivatives for the subhalo library can be precomputed, representing the most computationally expensive step of the process. In Appendix~\ref{app: efficiency} we discuss the wall clock time of this step, and find that it is competitive with \texttt{streampepperdf} (an action-angle based model for perturbed streams) from \texttt{Galpy} \citep{2015ApJS..216...29B, 2017MNRAS.466..628B}, when run on a GPU.

\subsection{Limitations}
Our method relies on tracer particles, so structure below the inter-particle spacing will always be suppressed. This is the main limitation of our approach, though in principle it is possible for us to generate very dense stream models, and parallelize the computation of derivatives across several GPUs (see Appendix~\ref{app: efficiency}). In practice, we have found that fitting all streams with a KDE and choosing a motivated smoothing scale set by cross-validation techniques provides an efficient means to estimate a smooth stream model, but we still must not interpret this model below the smoothing scale. For a GD-1 like stream, we find a smoothing scale of roughly $0.1~\rm{deg}$ when using 12000 particles.

In order to obtain an accurate representation of tidal tail density, it is important to generate accurate release conditions for stars that become unbound from the progenitor. Our model is compatible with any differentiable release function of the progenitor's trajectory (\S\ref{sec: boundary_conditions}). For simplicity, in this work we adopted the particle spray implementation from \citet{2015MNRAS.452..301F} to generate initial conditions for stream particles. This release function is capable of reproducing the detailed features of streams seen in $N-$body simulations \citep{2015MNRAS.452..301F}, but the success of this approach can depend on the orbit of the progenitor \citep{2024arXiv240801496C}. Crucially, our model is not built off of this particle spray implementation. Instead, our approach is expressed in terms of generic release functions, so that we can apply our method to other release functions that take into account the progenitor's orbit (e.g., \citealt{2023MNRAS.525.3662A, 2024arXiv240801496C}). Furthermore, we could also apply our approach to $N-$body streams directly, since once a star-particle becomes unbound from the progenitor it can be safely treated as a test particle evolving in the Galactic potential. We intend to explore an application of our method to direct $N-$body streams in a future work.

Finally, we apply perturbation theory to every particle in the stream for a single subhalo. In most cases, only particles that are closest to the subhalo are imparted with appreciable velocity kicks. A future implementation could achieve greater speed by only applying perturbation theory to regions of the stream that have a close impact with each subhalo.

\subsection{Comparison to Previous Works}
Several works have focused on fitting individual subhalo impacts from (e.g.) stream gaps \citep{2019ApJ...880...38B, 2017MNRAS.470...60E, 2024arXiv240402953H}, which is possible with our method, though fewer works have developed a general method (short of direct $N$-body simulations) that can be applied efficiently to the regime of many small-scale impacts. The regime of many small-scale impacts is studied in  \citet{2017MNRAS.466..628B} and \citet{2022MNRAS.513.3682D}. We draw comparisons to these works below.

In \citet{2017MNRAS.466..628B}, a linear perturbation theory for streams is developed in action-angle coordinates. Subhalo encounters are treated impulsively (i.e., instantaneously), allowing a single velocity kick to be converted to frequency-angle kicks, and finally mapped forward in time and transformed to observable coordinates. Their method models the stream as one-dimensional, so all perturbations are projected along the stream's dominant expansion direction in (i.e.) angle-angle space. This dimensionality reduction has the substantial benefit of computational efficiency, though there are higher order statistics that could be lost in the density of the stream perpendicular to its elongated axis (i.e., the $\phi_2$ dimension). A major advantage of the 1D approximation is that a density function for the stream can be derived, allowing one to efficiently compute the structure of the stream down to arbitrarily small-scales. Our method is based on tracer particles, so we are limited to scales of the stream above the inter-particle spacing. 

An advantage of our work compared to the method of \citet{2017MNRAS.466..628B} is that we do not require transformation to action-angle coordinates. This means that we can adopt any time-dependent, aspherical potential, provided that it is differentiable. Therefore, we may include the effects of baryonic substructure like the galactic bar, the LMC, giant molecular clouds, etc. in our base model, and apply perturbation theory around the resulting nonlinear simulation. We have also simplified the combinatorics of stream and subhalo interactions. In \citet{2017MNRAS.466..628B}, impacts are undone sequentially by simulating many combinations of subhalo encounters. Because the $\mathcal{O}(\epsilon^2)$ interaction terms in Eq.~\ref{eq: expansion_series} can be ignored in the absence of very massive direct impacts, using our method one can sample different sets of subhalo orbits and impacts by algebraically rescaling the perturbation strength of each subhalo. 
Additionally, our stream model is multidimensional, and can be used to study higher order statistics associated with (e.g.) the width and $\phi_2$ density of the stream. We also self-consistently model the stream progenitor, such that it too can be perturbed by subhalo fly-bys. Finally, we do not rely on the impulse approximation, allowing us to, in principle, break degeneracies between the subhalo mass, radius, and velocity from stream observables \citep{2024arXiv240402953H}.

We now draw a comparison to the work of \citet{2022MNRAS.513.3682D}. In their work, the Fokker-Planck equation is employed to model the evolution of phase-space density in a potential entirely sourced by substructure (i.e., no smooth galaxy component), characterized as a stochastic background force. The work we have developed here is similar to a monte-carlo version of their approach (i.e., we solve a Langevin-type equation, rather than the Fokker-Planck equation). For instance, in \S\ref{sec: heating} we sample a physically motivated stochastic background force, by assuming $N>> 1$ perturbing potentials in Eq.~\ref{eq: Hamiltonian} and resampling the perturbation strengths to obtain random realizations of the background force. Similar to \citet{2022MNRAS.513.3682D}, we can take correlation functions of the particle distribution along a stream, and connect the resulting power-spectrum directly to the perturbation strengths using Eq.~\ref{eq: linearized_qp}. The advantage of our method is that we do not need to approximate orbital dynamics of the stream, and we can use arbitrary models for the background and substructure potentials.

\section{Summary and Conclusion}\label{sec: conclusion}
We have developed a new model to describe stellar streams in the presence of substructure. Our approach can model the evolution of streams in arbitrary time-dependent potentials, and incorporate the effect of small-scale fluctuations in the potential perturbatively. We have validated our approach against simulations, and demonstrated how the phase-mixing of a perturbed stream can differ between static and time-evolving potentials. We have shown that known baryonic perturbers, such as the LMC and the bar, can also be incorporated in our model at full non-linear order.

In the regime of many subhalo impacts, our model can be used to compute the linear response of a stream to a large precomputed library of subhalo orbits. Once these responses are calculated, the strength, presence, and absence of each subhalo perturbation can then be re-scaled in a post-processing step, allowing us to sample from statistical distributions for the number, mass, and scale-radius of subhalo populations. Our method is built within the $\texttt{Jax}$ Python framework, and utilizes automatic differentiation to seamlessly integrate arbitrary user-defined potentials and generic stream models. The method runs on GPUs for a substantial speedup in computation.

We demonstrated and validated our method for streams that have a close-encounter with the galactic bar (e.g., Pal-5) or with the LMC (e.g., the OC stream). The dynamical effects of known baryonic substructures were incorporated in the non-linear model, while perturbations due to subhalos were applied at linear order in the subhalo mass and radius. We show that modeling these two regimes in tandem is crucial for an unbiased inference of subhalo properties from stream gaps and kinematics. Our model also makes predictions for the velocity dispersion of streams as a function of subhalo properties. We find that the velocity dispersion of GD-1 is consistent with a CDM subhalo population, though is perhaps better fit by more compact subhalos. We also find that low mass subhalos ($\sim 10^5~M_\odot$) should dictate the minimum velocity dispersion measured across many different streams.

Upcoming photometric and spectroscopic surveys are expected to measure streams at unprecedented depth and radial velocity precision. Our work will provide a pathway to model stream observables as a function of subhalo properties and statistics, while maintaining flexibility in both the stream and potential models. This flexibility allows us to incorporate the effects of (e.g.) the LMC on the orbital dynamics of streams, as well as the subsequent gap and velocity evolution induced by subhalo perturbations. This represents a crucial step in modeling the Milky Way environment across orders of magnitude in mass scales, from the stellar halo down to subhalos.

\section*{Acknowledgments}
JN is supported by a National Science Foundation Graduate Research
Fellowship, Grant No. DGE-2039656. Any opinions, findings, and conclusions
or recommendations expressed in this material are those of the author(s) and
do not necessarily reflect the views of the National Science Foundation. 
KVJ acknowledges support from  the Simons Foundation, Award ID: 1018465.  
We thank M. Sten Delos, Peter Melchior, Jeremy Goodman, Shirley Ho, Shaunak Modak, Sarah Pearson, Ebtihal Abdelaziz, Maureen Iplenski, and the Carnegie Stream Team for helpful comments and discussions. We are pleased to acknowledge that the work reported on in this paper was substantially performed using the Princeton Research Computing resources at Princeton University which is a consortium of groups led by the Princeton Institute for Computational Science and Engineering (PICSciE) and Office of Information Technology's Research Computing.

\software{Jax \citep{jax2018github}, Diffrax \citep{kidger2021on}, Astropy \citep{astropy:2018}, Gala \citep{gala}}

\appendix

\section{Stream Progenitors and LMC Location}\label{app: prog_and_LMC}
\textbf{Pal-5:} We use the convenient tabulation from \citet{2017NatAs...1..633P} for the cluster's location, derived from \citet{2003AJ....126.2385O,2016ApJ...833...31B,2015ApJ...811..123F}. We adopt a progenitor mass of $2\times 10^4~M_\odot$ \citep{2016MNRAS.460.2711T, 2020ApJ...889...70B}.

\textbf{GD-1:} To estimate a progenitor location for GD-1, we use $\sim 1000$ probable stream members from \citet{2023arXiv231116960S} and take the average phase-space coordinate in a $\sim 2~\rm{deg}$ band of the stream centered on $\phi_1 = 20~\rm{deg}$. This is the tentative location of the dissolved stream progenitor \citep{2019ApJ...880...38B}. In \S\ref{sec: heating} we use a progenitor mass of $0.5\times 10^4$ to match the observed width of the stream. This value is consistent with earlier models from \citet{2019MNRAS.485.5929W, 2021ApJ...911L..32G}.

\textbf{Orphan-Chenab:} We use table A1, first column from \citet{2019MNRAS.487.2685E}. We adopt a progenitor mass of $10^6~M_\odot$ \citep{2019MNRAS.487.2685E, 2023ApJ...948..123H}

\textbf{LMC location:} The ICRS location we adopt for the LMC is $\alpha = 81.28~\rm{deg}, \ \delta = -69.78~\rm{deg}$, \ $r_{\rm helio} = 49.6~\rm{kpc}, \ \mu_{\alpha*} = 1.858~\rm{mas/yr}, \ \mu_{\delta} = 0.385~\rm{mas/yr}, \ v_{\rm los} = 262.2~\rm{km/s}$. The on-sky location and proper motions are from \citet{2021A&A...649A...7G}, the distance is from \citet{2019Natur.567..200P}, and the line-of-sight velocity is from \citet{2002AJ....124.2639V}.

\section{$N-$Body Comparison}\label{app: N_body}
Throughout this work we have adopted the particle-spray technique from \citet{2015MNRAS.452..301F} for determining the release function introduced in \S\ref{sec: boundary_conditions}. Crucially, our work is not dependent on this particular stream model, but can be applied to arbitrary differentiable release functions. Still, the particle-spray technique is an appealing approach for its straightforward implementation, and has been calibrated against $N-$body simulations in \citet{2015MNRAS.452..301F}.

\begin{figure}
\centering\includegraphics[scale=.45]{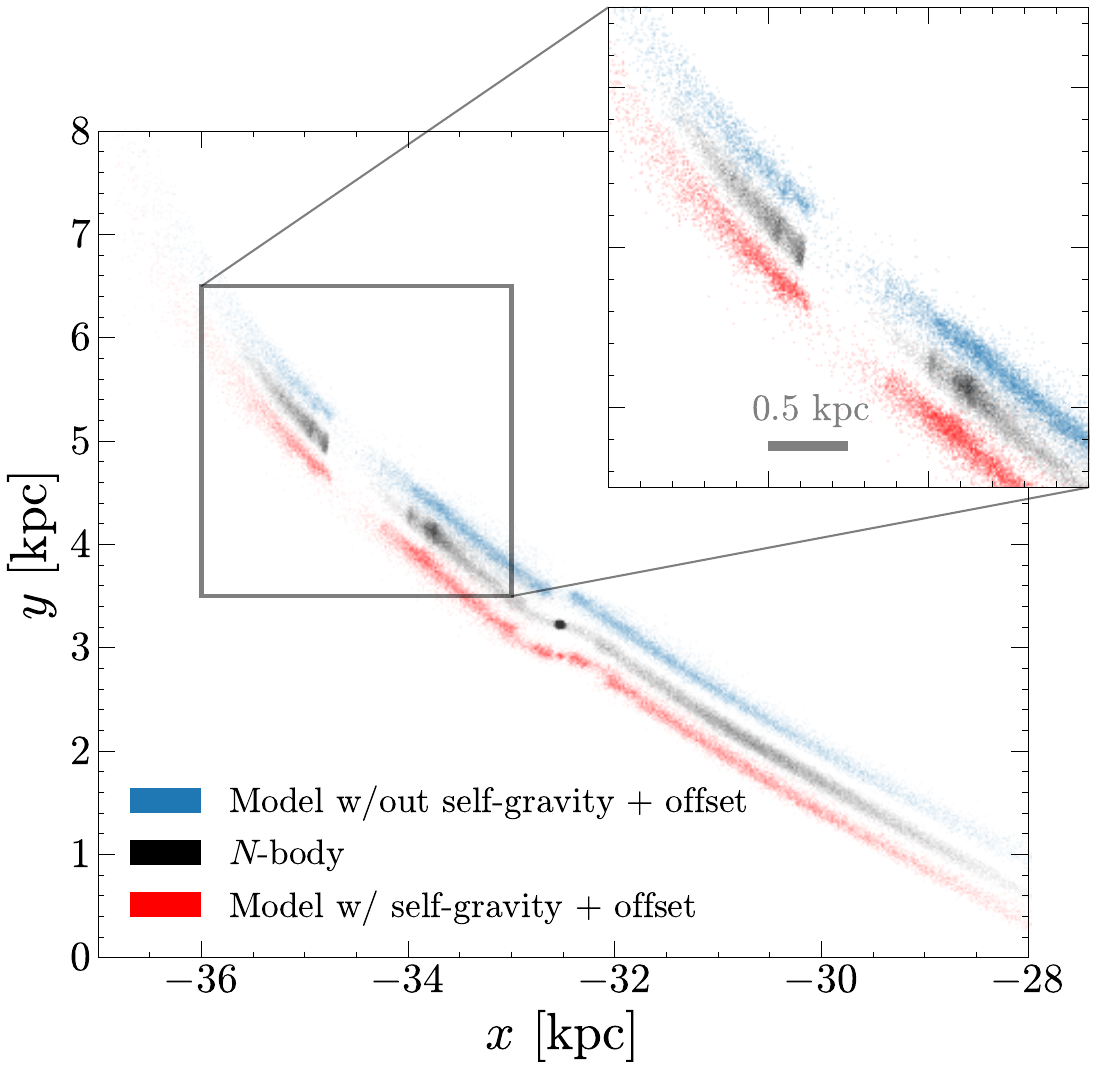}
    \caption{Black: $N-$body stream model with a $10^7~M_\odot$ subhalo impact. Blue: particle-spray model without self-gravity of the progenitor and the same subhalo impact. Red: particle-spray model with self-gravity of the progenitor included and the subhalo impact. An offset is applied to the blue and red streams for visual comparison.}
    \label{fig: NbodyComparison}
\end{figure}

Here we compare a particle-spray stream with our linear perturbation formalism against the equivalent stream generated in a $N-$body simulation. A $10^7~M_\odot$ subhalo impacts the test stream $1~\rm{Gyr}$ ago. The $N-$body result generated from a cluster with $10^5$ particles is shown by the black points in Fig.~\ref{fig: NbodyComparison}. The particle-spray stream (with the subhalo impact) is in blue, and the particle-spray stream with self-gravity of the progenitor is shown in red. The particle-spray streams have a constant offset added to aid in visual comparison.

Interestingly, the particle-spray model without self-gravity does a better job at reproducing the $N-$body stream, outside of the progenitor's location. We believe this is because the release function developed in \citet{2015MNRAS.452..301F} is calibrated against $N-$body simulations with self-gravity of the progenitor already incorporated. Therefore, the red stream overestimates the effect of self-gravity. While our approach can include more generic models for streams including self-gravity of the progenitor, we adopt the particle-spray approach for now since it provides a good match to the $N-$body model, by construction. In a future work, we will adopt more accurate release functions motivated by $N-$body models (e.g., \citealt{2023MNRAS.525.3662A, 2024arXiv240801496C}).

\section{Efficiency of Method}\label{app: efficiency}
\begin{figure}
\centering\includegraphics[scale=.58]{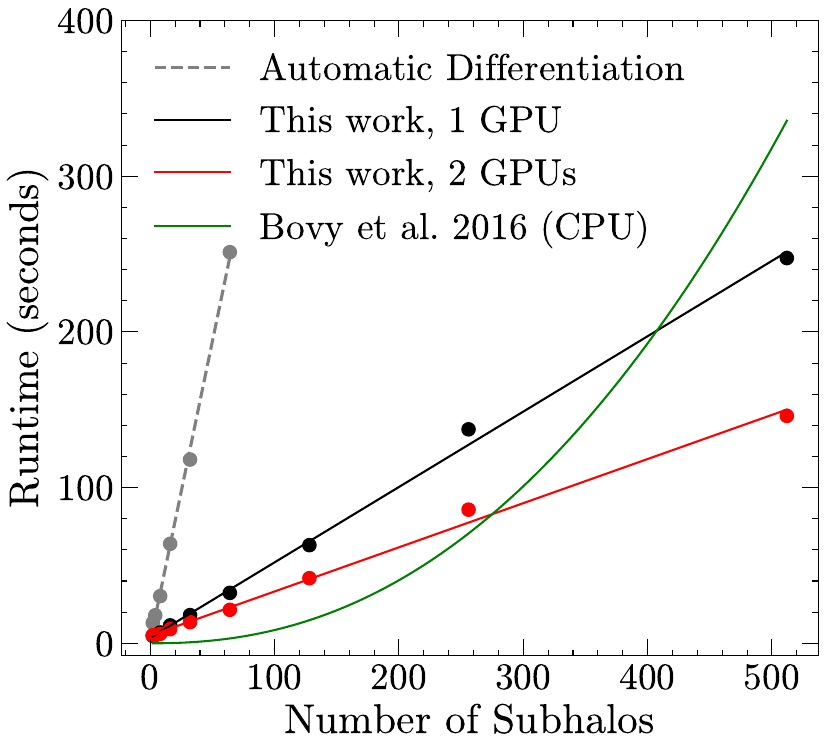}
    \caption{Runtime comparison for different methods and hardware configurations as a function of the number of subhalos. We apply perturbation theory to the mass and radius of subhalos. We obtain the perturbative corrections using automatic differentiation (gray; with GPU), and the analytic treatment developed in this work using a single GPU (black) and two GPUs (red). While the simulation output of the gray and black/red lines are identical, automatic differentiation takes significantly more compute compared to the forward-mode differentiation of Hamilton's equations that we have developed. Our code is straightforward to parallelize further across more GPUs. The green curve is the scaling found in \citet{2017MNRAS.466..628B}. }
    \label{fig: Wallclock_time}
\end{figure} 

In Fig.~\ref{fig: Wallclock_time} we show the amount of time it takes to apply perturbation theory to a Pal-5 like stream consisting of $12000$ particles, evolving for $\sim 5~\rm{Gyr}$. Because we apply perturbation theory in the mass and radius of the subhalos, each particle has an associated 12D vector consisting of mass and radius derivatives in phase-space, per subhalo. When we apply perturbation theory using only automatic differentiation (i.e., neglecting the analytic derivations we carried out in \S\ref{sec: method}), the efficiency of the method scales poorly with the number of subhalos (gray line). We then apply our method, using the forward mode differentiation of Hamilton's equations developed in \S\ref{sec: method}, and show the runtime for a single GPU, and parallelization across 2 GPUs (black and red, respectively). For the 2 GPU case, we split the stream into two segments (leading and trailing), running computations for each on their own GPU in parallel.

In principle, we can continue breaking the stream down to smaller sub-parts using several GPUs. In green, we show the scaling from \citet{2017MNRAS.466..628B}, which is based on action-angle methods for generating perturbed streams. Importantly, because our code is vectorized, our method scales linearly with the number of subhalo provided that one has enough GPU memory. The method from \citet{2017MNRAS.466..628B} is currently implemented on CPUs, and relies on brute force simulation of different subhalo combinations. While this is a more accurate approach in principle, we have shown throughout this work that the effect of subhalos with masses $\lesssim 10^7~M_\odot$ can be successfully modeled by neglecting interaction terms. This also contributes to the linear scaling seen in Fig.~\ref{fig: Wallclock_time} for our method.

\section{Relation to Canonical Perturbation Theory in Actions and Angles}\label{app: canonical_theory}
Here we discuss how the perturbative formalism developed in \S\ref{sec: method} is related to canonical perturbation theory in actions and angles. In general, we can write $\mathbf{J} = \mathbf{J}\left(\boldsymbol{q}, \boldsymbol{p} , t\right)$ and $\boldsymbol{\theta} = \boldsymbol{\theta}\left(\boldsymbol{q}, \boldsymbol{p} , t\right)$. If $H_{\rm base}$ is integrable, then it can be defined entirely in terms of the actions of the base (unperturbed) system, $H_{\rm base}\left(\mathbf{J}_0\right)$. For the problem $H_{\rm base} + \epsilon H_1$, for small $\epsilon$ we can express the perturbing Hamiltonian in terms of the base actions $\mathbf{J}_0$ and angles $\boldsymbol{\theta}_0$. That is, 
\begin{equation}
    H\left(\boldsymbol{\theta}_0, \mathbf{J}_0\right) = H_{\rm base}\left(\mathbf{J}_0\right) + \epsilon H_1\left(\boldsymbol{\theta}_0, \mathbf{J}_0\right).
\end{equation}
The next step is to perform a canonical transformation, so that the new actions and angles for the full perturbed system are functions of the base actions and angles. Namely,
\begin{equation}
\begin{split}
    \mathbf{J}^\prime\left(t\right) &= \mathbf{J}_0 + \epsilon\mathbf{J}_1\left(\mathbf{J}_0, t\right) + \epsilon^2\mathbf{J}_2\left(\mathbf{J}_0, t\right) +...\\
    \boldsymbol{\theta}^\prime\left(t\right) &= \boldsymbol{\theta}_0(t) + \epsilon\boldsymbol{\theta}_1(\mathbf{J}_0, t) + \epsilon^2\boldsymbol{\theta}_2(\mathbf{J}_0, t) +...
    \end{split}
\end{equation}

From this expression, we have
\begin{equation}
    \mathbf{J}_1 = \frac{d\mathbf{J}^\prime}{d\epsilon}\Big\vert_{\epsilon=0} = \left[\frac{d\mathbf{J}^\prime}{d\boldsymbol{q}} \frac{d\boldsymbol{q}}{d\epsilon} + \frac{d\mathbf{J}^\prime}{d\boldsymbol{p}}\frac{d\boldsymbol{p}}{d\epsilon}\right]_{\epsilon=0}.
\end{equation}
Evaluating the bracketed quantity at $\epsilon=0$ yields
\begin{equation}
    \mathbf{J}_1 = \frac{d\mathbf{J}_0}{d\boldsymbol{q}} \frac{d\boldsymbol{q}}{d\epsilon} + \frac{d\mathbf{J}_0}{d\boldsymbol{p}} \frac{d\boldsymbol{p}}{d\epsilon}.
\end{equation}
Similarly for $\boldsymbol{\theta}_1$, we have
\begin{equation}
    \boldsymbol{\theta}_1 = \frac{d\boldsymbol{\theta}^\prime}{d\epsilon}\Big\vert_{\epsilon=0} = \frac{d\boldsymbol{\theta}_0}{d\boldsymbol{q}}\frac{d\boldsymbol{q}}{d\epsilon} + \frac{d\boldsymbol{\theta}_0}{d\boldsymbol{p}}\frac{d\boldsymbol{p}}{d\epsilon}.
\end{equation}

We can express these relations in matrix form, giving us
\begin{equation}
    \begin{pmatrix}
        \boldsymbol{\theta}_1 \\
        \mathbf{J}_1
    \end{pmatrix}=
    \begin{pmatrix}
        \frac{d\boldsymbol{\theta}_0}{d\boldsymbol{q}} & \frac{d\boldsymbol{\theta}_0}{d\boldsymbol{p}} \\
        \frac{d\mathbf{J}_0}{d\boldsymbol{q}} & \frac{d\mathbf{J}_0}{d\boldsymbol{p}}
    \end{pmatrix}
    \begin{pmatrix}
        \frac{d\boldsymbol{q}}{d\epsilon}\\
        \frac{d\boldsymbol{p}}{d\epsilon}
    \end{pmatrix}.
\end{equation}  
The linear order corrections to the base EOM that we use in the paper are related to the same first order corrections in canonical perturbation theory, up to a linear transformation. Crucially, we never need to transform to action-angle coordinates, allowing us to easily implement general forms for the galactic potential.

\section{Example with the Simple Harmonic Oscillator}\label{app: SHO}

Here we provide an analytic example of our method. The results of this appendix are shown in the main text, \S\ref{sec: conceptual_example}. We consider a cubic perturbation to the 1D simple harmonic oscillator, with a full Hamiltonian of the form
\begin{equation}\label{eq: SHO_ham}
\begin{split}
    H\left(q,p, t\right) &= H_{\rm base}\left(q,p\right) + \epsilon \Phi_{1}\left(q,p, t\right) \\
    &= \frac{p^2}{2} + \frac{1}{2} \Omega^2 q^2 + \frac{\epsilon}{3} q^3 W(t),
\end{split}
\end{equation}
where $W(t)$ is a window function in time that evaluates to $1$ in the interval $t \in [t_{\rm min}, t_{\rm max}]$, and $0$ otherwise (i.e., like a subhalo fly-by the perturbation is transient). A solution to the EOM for the base Hamiltonian is $q_{\rm base}(t) = q_0 \cos\left(\Omega t\right)$.

Eq.~\ref{eq: update_rule} provides a prescription for obtaining the leading order perturbative correction to the dynamics of a particle subject to the perturbation $\epsilon\Phi_1$.  Specializing Eq.~\ref{eq: update_rule} to the Hamiltonian from Eq.~\ref{eq: SHO_ham}, the equation to solve is
\begin{equation}\label{eq: specialized_pert_sho}
    \frac{d}{dt}\begin{pmatrix}
        d{{q}}/d\epsilon  \\
        d{{p}}/d\epsilon
    \end{pmatrix}_{ {\epsilon}=0} = \begin{pmatrix}
        dp/d\epsilon\\
        -q^2 W(t) - \Omega^2 \frac{dq}{d\epsilon}
    \end{pmatrix}_{\epsilon=0},
\end{equation}
where the tidal tensor of the base potential is $\Omega^2$. Eq.~\ref{eq: specialized_pert_sho} is a coupled first order ODE. Computationally, this ODE is very easy to solve, given initial conditions for $dq/d\epsilon$ and $dp/d\epsilon$. Analytic solutions are more easily obtained by converting Eq.~\ref{eq: specialized_pert_sho} to a single, second-order ODE, using the first vector element $d\dot{q}/d\epsilon = dp/d\epsilon$. Taking another time-derivative gives
\begin{equation}\label{eq: sho_2ndODE}
\begin{split}
    \frac{d\ddot{q}}{d\epsilon}\Big\vert_{\epsilon=0} &= -\left[q^2 W(t) + \Omega^2 \frac{dq}{d\epsilon}\right]_{\epsilon=0} \\
    &= -q_0^2\cos^2\left(\Omega t\right) W(t) - \Omega^2\frac{dq}{d\epsilon}\Big\vert_{\epsilon=0},
\end{split}
\end{equation}
where $q\vert_{\epsilon=0}$ is just the solution to the EOM at zeroth order in the perturbation, $q_0\cos{(\Omega t)}$. 

If we can solve the time-dependence in Eq.~\ref{eq: sho_2ndODE}, we will have obtained the leading order correction to the motion. We will assume that the perturbation turns on at $t_{\rm min} = 0$, and will evolve an ensemble of particles up to that point (starting from $t<0$). We will assume the initial conditions $dq/d\epsilon =0$ and $d\dot{q}/d\epsilon = 0$ at the initial time. Solutions to Eq.~\ref{eq: sho_2ndODE} are sinusoidal. The effect of the perturbation is illustrated in Fig.~\ref{fig: method_workflow}, and discussed in \S\ref{sec: conceptual_example}.

\section{Boundary Conditions for Structural Parameters of the Perturbations}\label{app: structural_boundary_conds}
In \S\ref{sec: varying_structural_params}, we derive an integro-differential equation for the leading order response of an orbit to variations in the structural parameters of the perturbation. In this appendix, we discuss the boundary conditions to solve this equation (Eq.~\ref{eq: update_rule_theta}) numerically.

To solve Eq.~\ref{eq: update_rule_theta}, we must initialize our integration with the proper values for ($\partial \boldsymbol{q}_{\alpha 1}/\partial\boldsymbol{\theta}_\alpha, \partial \boldsymbol{p}_{\alpha 1}/\partial\boldsymbol{\theta}_\alpha)$. First, we note that the initial value for $\boldsymbol{q}_{\alpha 1} = d\boldsymbol{q}/d\epsilon_\alpha$ and $\boldsymbol{p}_{\alpha 1} = d\boldsymbol{p}/d\epsilon_\alpha$ is given by Eq.~\ref{eq: init_cond_general}. Therefore, to obtain the initial condition for Eq.~\ref{eq: update_rule_theta}, we only need to differentiate Eq.~\ref{eq: init_cond_general} with respect to $\boldsymbol{\theta}_\alpha$ on both sides. The result of this differentiation is
\begin{small}
\begin{multline}\label{eq: init_cond_theta}
    \begin{pmatrix}
        \partial \boldsymbol{q}_{\alpha 1}/\partial\boldsymbol{\theta}_\alpha \\
        \partial \boldsymbol{p}_{\alpha 1}/\partial\boldsymbol{\theta}_\alpha 
    \end{pmatrix}_{(t_{\rm rel},\boldsymbol{\epsilon}=0)} = \\\frac{\partial\mathbf{F}_{\rm rel}}{\partial\boldsymbol{w}}\Big\vert_{\boldsymbol{w}_{\rm prog}(t_{\rm rel}, \boldsymbol{\epsilon}=0)} 
    \begin{pmatrix}
        \partial\boldsymbol{q}_{\alpha 1, \rm prog}/\partial\boldsymbol{\theta}_\alpha \\
        \partial\boldsymbol{p}_{\alpha 1, \rm prog}/\partial\boldsymbol{\theta}_\alpha
    \end{pmatrix}_{(t_{\rm rel}, \boldsymbol{\epsilon}=0)},
\end{multline}\end{small}
where the derivatives are evaluated at $\boldsymbol{\theta}_\alpha = \boldsymbol{\theta}^*_\alpha$, and $\mathbf{F}_{\rm rel}$ does not depend on $\boldsymbol{\theta}_\alpha$ since it is only a function of the base potential, $\Phi_{\rm base}$. 

Exactly the same as in \S\ref{sec: boundary_conditions}, we have offloaded the problem of finding generic initial conditions for every particle, to only finding a boundary condition for the progenitor. Provided that an initial state for the progenitor can be chosen, then we can use the Jacobian $\partial\mathbf{F}_{\rm rel}/\partial \boldsymbol{w}$ in Eq.~\ref{eq: init_cond_theta} to transform the progenitor's derivatives at release time to an initial condition for each particle at its respective release time. 

To set a boundary condition for the progenitor, we impose the same constraint that is used in \S\ref{sec: boundary_conditions} and require the the phase-space position of the progenitor at the final time ($t_f$) is independent of any perturbation parameters (including $\epsilon_\alpha$ and $\boldsymbol{\theta}_\alpha$). Mathematically, we require
\begin{equation}\label{eq: prog_ic_theta}
    \frac{\partial \left(\boldsymbol{q}_{\alpha 1, \rm{prog}},  \boldsymbol{p}_{\alpha 1,\rm{prog}} \right)}{\partial \boldsymbol{\theta}_\alpha }\Big\vert_{t_f, \boldsymbol{\theta}= \boldsymbol{\theta}_*,\boldsymbol{\epsilon}=0} = 0.
\end{equation}
With Eq.~\ref{eq: prog_ic_theta}, we have the necessary information to integrate the progenitor's $\boldsymbol{\theta}_\alpha$ derivatives through time, and therefore we can obtain initial conditions at leading order in the perturbations for every particle.

\begin{figure*}
\centering\includegraphics[scale=.6]{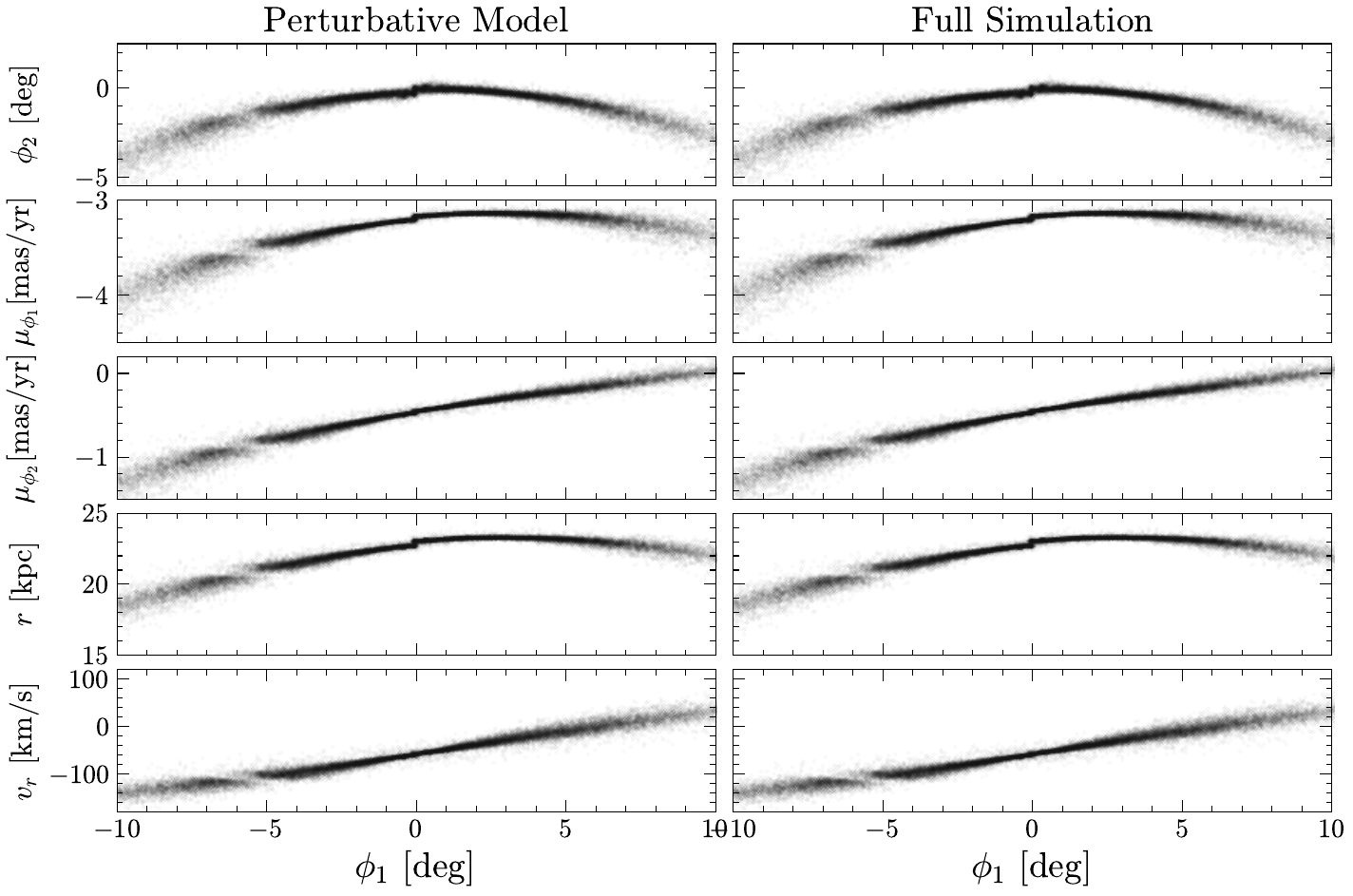}
    \caption{Mock-observations across all phase-space dimensions of the Pal-5 like stream shown in Fig.~\ref{fig: single_impact}. The response of the stream to the subhalo is calculated using our perturbative model (left) and the simulation (right). Across all phase-space dimensions, the perturbative model appears nearly identical to the equivalent stream produced without approximation of the stream-subhalo interaction.  }
    \label{fig: Kinematic_View_Single_Impact}
\end{figure*}

\section{Additional Validation}\label{app: additional_validation}
Here we provide additional validation for the streams generated throughout this work. In Fig.~\ref{fig: Kinematic_View_Single_Impact} we show the Pal 5 like stream studied in \S\ref{sec: single_impacts}, projected in heliocentric coordinates. The linear model with the subhalo perturbation is shown in the left column, while the simulation is in the right column. Our model clearly captures the distinctive kinematic effect of a subhalo fly-by, with a positive and negative velocity kick along the stream on either side of the subhalo's impact location ($\phi_1 \approx -5.5~\rm{deg}$). 

\begin{figure*}
\centering\includegraphics[scale=.58]{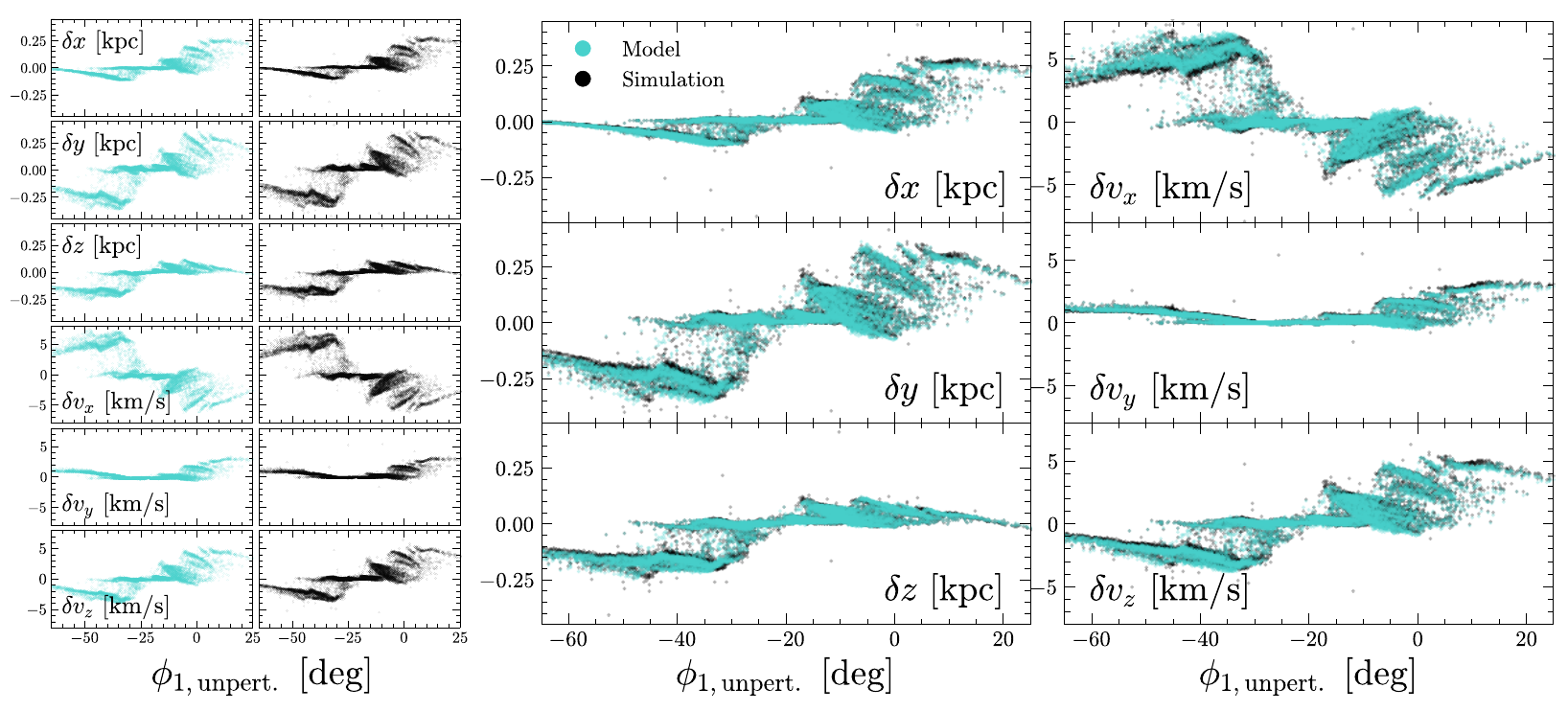}
    \caption{Same as Fig.~\ref{fig: delta_Kinematic_single}, but for the CDM expected number of impacts for a GD-1 like stream.  }
    \label{fig: CDM_overlaid}
\end{figure*}

In Fig.~\ref{fig: CDM_overlaid} we compare, on a per-particle basis, the difference between the perturbed and unperturbed stream discussed in \S\ref{sec: many_CDM_impacts} (impacts at the CDM rate). The model and simulation show excellent agreement in each phase-space dimension. Nested feathers are due to the stream encountering multiple subhalos along the same tidal tail, and are also amplified by perturbations to the progenitor's orbit which create larger offsets in $\delta(\boldsymbol{x},\boldsymbol{v})$.

\section{Connection to Automatic Differentiation}\label{app: autodiff}
Automatic differentiation (AD) provides a means to compute exact derivatives of complicated functions. At a basic level, AD works by constructing a graph of all mathematical operations applied to an object, and applying the chain rule through the graph (i.e., even complicated functions can be decomposed as a composition of simpler functions whose derivatives are well known). 

Our particle-spray stream model is implemented in the \texttt{Jax} python library, which supports AD, GPU acceleration, vectorized, and parallelized computations. In \texttt{Jax}-based python code, computing the derivative of a function is extremely straightforward. For instance, the derivative of a function $\texttt{f(x)}$ is obtained by calling $\texttt{grad(f)(x)}$ once the proper libraries are imported.

For the subject of this paper, the function $\texttt{f}$ is a differential equation solver, which consists of a velocity and acceleration term (from Hamilton's equations), a set of initial conditions, and model parameters. The initial conditions are evolved using an ODE solver:
\begin{equation}
    \left(\boldsymbol{q}_f, \boldsymbol{p}_f \right) = \texttt{ODESolve}\left[ \dot{\boldsymbol{q}} , \dot{\boldsymbol{p}} |  \boldsymbol{q}_{\rm init}, \boldsymbol{p}_{\rm init}, t_i, t_f, {\epsilon} = 0\right],
\end{equation}
where $\boldsymbol{q}_f$ and $\boldsymbol{p}_f$ represent the phase-space position of a particle at time $t_f$, and $(\boldsymbol{q}_{\rm init}, \boldsymbol{p}_{\rm init})$ are initial conditions at time $t_i$.

In this work all models are implemented in the \texttt{Jax} library, and for orbit integration we make significant use of the \texttt{diffrax} package, which provides a suite of \texttt{Jax}-based differential equation solvers. Because the integrator itself is differentiable, obtaining the leading order corrections to the EOM (at linear order in ${\epsilon}$) is easy as
\begin{multline}\label{eq: grad_OdeSolve}
    \left(\frac{d\boldsymbol{q}_f}{d{\epsilon}}, \frac{d\boldsymbol{p}_f}{d{\epsilon}} \right)_{\boldsymbol{\epsilon}=0} \\= \texttt{grad}\left( \texttt{ODESolve} \right)\left[ \dot{\boldsymbol{q}} , \dot{\boldsymbol{p}} | \boldsymbol{q}_{\rm init}, \boldsymbol{p}_{\rm init}, t_i, t_f, {\epsilon} = 0\right].
\end{multline}

Eq.~\ref{eq: grad_OdeSolve} is essentially the pseudo-code version of Eq.~\ref{eq: update_rule}, though in practice one can use Eq.~\ref{eq: grad_OdeSolve} in a differentiable programming environment  without having to analytically derive the terms that show up in the perturbative analysis. Implementing Eq.~\ref{eq: grad_OdeSolve} amounts to simulating a stream with subhalos that have zero mass, and then differentiating the final snapshot of the simulation with respect to the mass parameters. The resulting derivatives that are returned are the first order corrections to the EOM, because the dynamical system is Hamiltonian.

In practice, obtaining the perturbative correction terms through Eq.~\ref{eq: grad_OdeSolve} seems appealing, since one can construct a base-stream model without having to derive its derivatives. However, there is a cost to Eq.~\ref{eq: grad_OdeSolve}: AD requires substantial memory usage, and does not cleanly reveal the terms that are being calculated in Eq.~\ref{eq: update_rule} (i.e., the tidal tensor). The latter is a theoretical consideration, but the former represents a relatively severe limitation. In particular, for each new perturbation one needs to call Eq.~\ref{eq: grad_OdeSolve}. Calling this function executes two primary tasks: one is to integrate the base stream model without perturbations, and the other is to differentiate each step of the base stream model with respect to the perturbation parameter. For different perturbation parameters, the same base stream model is recomputed, and the gradients of each step are recomputed. For this reason, direct application of Eq.~\ref{eq: grad_OdeSolve} scales poorly with the number of perturbations.

\bibliography{thebib}
\end{document}